\documentstyle[prb,eqsecnum,aps,epsf]{revtex}
\font\bba=msbm10 scaled 1200
\font\bbb=msbm8 
\font\bbc=msbm6 
\newfam\bbfam
\textfont\bbfam=\bba
\scriptfont\bbfam=\bbb
\scriptscriptfont\bbfam=\bbc
\def\bb{\fam\bbfam\bba}
\def\R{{\bb R}}

\def\tr{\text{ tr}}
\begin{document}
\draft
\title
{ Sine-Gordon Theory  for the Equation of State of Classical Hard-Core Coulomb
systems. III Loopwise Expansion }
\author{Jean-Michel Caillol \thanks{e-mail: Jean-Michel.Caillol@th.u-psud.fr}}
\address{LPT - CNRS (UMR 8627) \\
         Bat. 210,
         Universit\'e de Paris Sud \\
         F-91405 Orsay Cedex, France}
         
\date{\today}
\maketitle
\begin{abstract}
We present an exact field theoretical representation of an ionic solution
made  of charged hard spheres. The action of the field theory is obtained 
by performing a Hubbard-Stratonovich transform of the 
configurational Boltzmann factor.
It is shown that the Stillinger-Lovett sum rules
are satisfied if and only if all the field correlation functions  are
short range functions. 
The  mean field, Gaussian and two-loops approximations of the theory
are derived and discussed.  The mean field approximation for the free
energy constitutes an exact lower bound for the exact free energy, while the
mean field pressure is an exact upper bound.
The one-loop order approximation is shown to be
identical with the random phase approximation of the theory of liquids. 
Finally, at the two-loop order and in the pecular case of the restricted
primitive model, one recovers results obtained
in the framework of the mode expansion theory.
\end{abstract}
\section{Introduction}
Various ionic systems including electrolyte solutions, molten salts, and 
colloids
can be studied with a good approximation
in the framework of the so-called primitive model (PM) which
consists in a mixture of $M$ species of charged hard spheres (HS) which differ
by their respective charges and (or) diameters.\cite{Hansen} 
Of special interest is the restricted primitive model (RPM) where $M=2$,
the hard spheres have all the same diameter, and  the cations and anions
bear opposite charges $\pm q$. In many instances, we shall also consider
the special primitive model (SPM), where the number $M$
of species as well as the charges are arbitrary but all the ions have the
 same diameter $\sigma$. 

In the two first parts of this work, published
some years ago and hereafter referred to as I and II, 
we have established  an exact field theoretical representation 
of the RPM.\cite{Cai-Raim,Raim-Cai} The action of this field theory, which is
obtained by applying the Kac-Siegert-Stratonovich-Hubbard-Edwards (KSSHE)
\cite{Kac,Siegert,Strato,Hubbard1,Hubbard2,Edwards} 
transform to the Coulomb potential, looks like the sine-Gordon
action
to which it reduces in the limit of point-like ions,\cite{Samuel,Brydges,Orland}
hence the slightly abusive title of this series of papers. 
Nowadays we prefer the acronym KSSHE to christen the action.
The extended sine-Gordon action
derived in paper I for the RPM is obtained here for a general PM. The
regularization of the Coulomb potential which is required to define properly the
KSSHE transform is  obtained by a smearing of the charge over the HS volume. A
more general treatment where a part of the Coulomb interaction is incorporated
in the reference system is discussed in the review of Brydges and Martin.
\cite{Brydges} 

The developments of refs. I and II are based on a cumulant expansion of the
grand partition function  reorganized in ascending powers of either the 
fugacity or the inverse temperature.
In this way one can obtain the exact low fugacity and
high temperature expansions of the pressure and the free energy of the RPM.
Of course the  expressions obtained in that manner are already known
from the theory of liquids and were
derived years ago in the framework of Mayer graph
expansions.\cite{Mayer,Stell-Lebo}
In the present paper we proceed differently. 
After having obtained the KSSHE action for the
general PM (see sec.\ \ref{KSSHE}), we reorganize the cumulant expansion by
grouping some classes of Feynman diagrams. The resulting loopwise expansion
is explicitly computed up to order two in the number of loops.
 In the case of the RPM, 
the two-loop order free energy turns out to coincide with an expression derived more than
thirty years ago by Chandler and Andersen 
\cite{Chand-Ander} in the framework of the so-called mode expansion
theory.\cite{Yukno,McMillan} Reorganizing the loop-expansion 
in ascending powers
of the inverse temperature gives back the high-temperature expansions
of paper\ II and ref.\cite{Stell-Lebo}

Our paper is organized in the following way. In next section\ II we show
how to construct a well-defined KSSHE transform
by regularizing the Coulomb potential at short distances by means of a smearing of
the charges inside the volumes of the HS. In section\ III we establish the general
relations between the charge correlation functions and the correlations of 
the KSSHE field. From the known asymptotic behavior of the former one 
can deduce
that of the latter. The conclusion, which is detailed at length in section\ IV, is
that the n-body correlations of the KSSHE field are short-ranged functions;
stated otherwise, the KSSHE field is a non-critical field. The so-called
Stillinger-Lovett sum rules, both for the homogeneous \cite{SL} and the 
inhomogeneous 
fluid \cite{Carnie} emerge as a consequence of this behavior.
In section\ V the mean field (MF) level of the theory is studied in detail. 
The MF free
energy $\beta {\cal A}_{MF}$ is shown to be a strictly convex functional of 
the $M$ partial densities 
and to constitute a rigorous lower bound of
the  exact free energy. The former property excludes a fluid-fluid transition at
the MF level while the latter serves to define an optimized MF free
energy by maximizing $\beta {\cal A}_{MF}$ with respect 
to the variations of the
smearing functions.  An explicit expression of 
the optimized $\beta {\cal A}_{MF}$ is obtained in the case of an homogeneous
fluid.
From the MF solution for the inhomogeneous system we also  deduce 
the expressions of the n-body correlation and vertex functions of the
homogeneous
system in the Gaussian approximation.\cite{Ma} This Gaussian approximation
is discussed in section\ VI and shown to be
equivalent to the random phase approximation (RPA)
of the theory of liquids.\cite{Hansen,Chand-RPA}
Finally, a two-loop order calculation is performed in section\ VII. 
The 
resulting
expression for the the free energy of the RPM is shown to be 
identical with that obtained by  Chandler and Andersen 
\cite{Chand-Ander} in the framework of the first version of the mode expansion
theory. Conclusions are drawn in section\ VIII.
\section{The KSSHE Transform}
\label{KSSHE}
\subsection{The model}
\label{model}
We shall consider only the three dimensional (3D) version of the  
(PM), 
i.e. a mixture of $M$ species of charged
hard spheres.\cite{Hansen}
The ions of the species $\alpha$ ($\alpha=1, \ldots, M $) are characterized by
their diameter $\sigma_{\alpha}$ and their electric charge $q_{\alpha}$. The 
molecular structure of the solvent is ignored and it is treated as a
continuum, the dielectric constant of which has been absorbed in the 
definition of the charges $q_{\alpha}$.  The solution is  
made of {\em both} positive  {\em and} negative ions so that
the electroneutrality in the bulk can be satisfied 
without adding any unphysical neutralizing background to the system.
The particles occupy a domain $\Omega \subset \R^3$ of volume $\Omega$
 of the ordinary 
space with free boundary conditions. Only  configurations $\omega \equiv 
(N_{1};\vec{r}_{1}^{1}, \ldots, \vec{r}^{1}_{N_{1}} \vert \ldots \vert
N_{\alpha };\vec{r}_{1}^{\alpha}, \ldots, \vec{r}^{\alpha}_{N_{\alpha}} \vert 
N_{M};\vec{r}_{1}^{M}, \ldots, \vec{r}^{M}_{N_{M}} )$ 
($\vec{r}^{\alpha}_{i_{\alpha}} \in \Omega $ ) 
without overlaps of  the
spheres - i.e. such that  $\| \vec{r}^{\alpha}_{i_{\alpha}}
- \vec{r}^{\beta}_{i_{\beta}}\|  \geq (\sigma_{\alpha} 
+\sigma_{\beta})/2 $ -  
do contribute to the partition or grand partition functions. In such a  
configuration,
the charge $q_{\alpha}$ of each ion  can be smeared 
out inside 
its volume according to a spherically symmetric distribution 
$q_{\alpha} \tau_{\alpha}(r)$ 
without altering the configurational energy  
as a consequence of Gauss theorem.  The distribution $\tau_{\alpha}(r)$ is 
{\em a priori} arbitrary, 
provided it  satisfies the following properties :
\begin{mathletters}
    \label{tau}
    \begin{eqnarray}
\tau_{\alpha}(r) &=& 0 \; \ \text{ if } \;  r \ge  \overline{\sigma}_{\alpha} \equiv 
\sigma_{\alpha} /2 \; , \\
\int d^{3}\vec{r} \;\tau_{\alpha}(r)  &=& 1 \label{wb}\; .
     \end{eqnarray}            
\end{mathletters} 

\noindent   
The electrostatic interaction energy of two  charge distributions 
$\tau_{\alpha}$ and $\tau_{\beta}$ the centers of which are 
located at the points $\vec{r}_{1}$ and 
$\vec{r}_{2}$ of $\Omega$ respectively 
will be noted  $w_{\alpha,\beta}(1,2) $. It reads as
  \begin{eqnarray}
      \label{w}
   w_{\alpha,\beta}(1,2)&=& \int d^{3} r_{1'}  \int d^{3} r_{2'} \; 
   \tau_{\alpha}(\| \vec{r}_{1} - \vec{r}_{1'}\|) v_{c}( \| 
   \vec{r}_{1'}-\vec{r}_{2'} \|)
   \tau_{\beta}(\| \vec{r}_{2'} - \vec{r}_{2}\|)  \;   , \nonumber \\
   &\equiv& \tau_{\alpha}(1,1') \;  v_{c}(1',2') \; \tau_{\beta}(2',2) \; ,
   \end{eqnarray}   
where  $v_{c}(r)=1/r$ is the Coulomb potential. Note that in this 
paper, summation over repeated, either discrete or continuous indices 
will always be meant (except if explicitly stated otherwise). As a 
consequence of eqs.\ (\ref{tau}) and of Gauss theorem $w_{\alpha,\beta}(1,2)=
1/r_{12}$ for 
$r_{12} \ge \overline{\sigma}_{\alpha} + 
\overline{\sigma}_{\beta} $. Note that the Fourier transform 
$\widetilde{w}_{\alpha,\beta}(k)$ of the interaction takes the simple 
form
\begin{equation}
  \widetilde{w}_{\alpha,\beta}(k) = \frac{4 \pi}{k^2} 
  \widetilde{\tau}_{\alpha}(k)
   \widetilde{\tau}_{\beta}(k) \; ,
 \end{equation} 
which diverges for $k \to 0$ as $4 \pi /k^2$ since  
$\widetilde{\tau}_{\alpha}(0)=1$, as follows from eq.\ (\ref{wb}). 
Finally we shall denote by 
\begin{equation}
    \label{vab}
 v_{\alpha,\beta}(1,2) = q_{\alpha} q_{\beta} \; w_{\alpha,\beta}(1,2)
 \end{equation} 
the pair interaction of two ions.

The electrostatic potential 
energy of the configuration $\omega$ times the inverse temperature 
$\beta = 1/kT$ can be written as
\begin{equation}
    \label{U}
    \beta U_{el}(\omega) 
    = \frac{\beta}{2} \;  \widehat{\rho}_{C}(1) \;
    v_{c}(1,2) \; \widehat{\rho}_{C}(2) - N_{\alpha}
    \nu_{\alpha}^{S}
   \; ,
\end{equation}     
where $\widehat{\rho}_{C}(1)$ is the microscopic 
charge density in the configuration $\omega$ at the point $\vec{r}_{1}$
and $ \nu_{\alpha}^{S}$ 
is the self-energy of the charge distribution $q_{\alpha} \tau_{\alpha}(r)$. 
In general, for a sufficiently regular distribution 
$\tau_{\alpha}(r)$, the self-energy 
\begin{equation}  
    \label{nuS}
    \nu_{\alpha}^{S}=
\frac{\beta q_{\alpha}^2}{2} \; w_{\alpha,\alpha}(0) = \frac{\beta 
q_{\alpha}^2}{2} \;
\int d^{\vee}k \; 
\frac{4 \pi }{k^{2}}
\widetilde{\tau}_{\alpha}(k) ^{2}
\; ,
  \end{equation}
where $d^{\vee}k \equiv d^{3}\vec{k}/(2 \pi)^{3}$, is a well-defined 
positive and finite quantity. Of course $\nu_{\alpha}^{S}$ diverges for
point-like charges which makes the KSSHE transform, to be introduced
in next section, an ill-defined object in that case.
The  microscopic smeared charge density $\widehat{\rho}_{C}(\vec{r} )$ 
which enters eq.\ (\ref{U}) reads
\begin{equation}  
    \label{roc}
  \widehat{\rho}_{C}(1) = q_{\alpha} \; 
  \tau_{\alpha}(1,1') \;  \widehat{\rho}_{\alpha}(1')  
  \; ,
  \end{equation}
  where 
\begin{equation}  
   \widehat{\rho}_{\alpha}(1)  =
   \sum_{i_{\alpha}=1}^{N_{\alpha}} \; \delta^3( \vec{r}_{1}
   -\vec{r}_{i_{\alpha}}^{\alpha}) 
\end{equation}
is the microscopic number density of the species $\alpha$ at the point
$\vec{r}_{1}$. 

It will prove convenient to make use of Dirac's notations for matrix 
elements and scalar products and to rewrite the 
energy\ (\ref{U}) as
\begin{eqnarray} 
 \frac{1}{2}  \widehat{\rho}_{C}(1) \;
    v_{c}(1,2) \; \widehat{\rho}_{C}(2)  &=&
    \frac{1}{2} \; \left < \widehat{\rho}_{C} \vert v_{c} \vert 
    \widehat{\rho}_{C} \right > \; , \nonumber \\
    &=& \left < \widehat{\rho}_{C} \vert \widehat{V} \right > \; ,
\end{eqnarray}    
 where     $\widehat{V}(1) \equiv \widehat{\rho}_{C}(1')
 \; v_{c}(1',1)  $
 denotes the microscopic 
 electric potential at the point $\vec{r}_{1}$  in the configuration 
 $\omega$. Of course  $\widehat{V}$ is solution of the $3D$ Poisson equation, i.e.
\begin{equation} 
    \Delta_{1} \;  \widehat{V}(1) = -4 \pi \widehat{\rho}_{C}(1) \; .
  \end{equation}  
\subsection{The KSSHE transform of the Boltzmann factor}
\label{Boltz}
The Boltzmann factor in the configuration $\omega$ is equal to
\begin{equation}
    \label{Boltz1}
    \exp \left( - \beta U \left(\omega \right)\right) = 
     \exp \left( - \beta U_{HS} \left(\omega \right) \right) \; 
     \times
     \exp \left( - \beta U_{el} \left(\omega \right) \right) \; ,
  \end{equation} 
where $U_{HS} \left(\omega \right)$ denotes the contribution of the hard 
cores to the configurational energy. We perform now a KSSHE 
transform in order to rewrite eq.\ (\ref{Boltz1}) as
\cite{Cai-Raim,Raim-Cai,Kac,Siegert,Strato,Hubbard1,Hubbard2,Edwards,Samuel,Brydges,Orland}
\begin{equation}
    \label{Boltz2}
    \exp \left( - \beta U \left(\omega \right) \right) = 
  \exp \left( - \beta U_{HS} \left(\omega \right) \right) \; 
   \exp \left(  N_{\alpha} \; \nu_{\alpha}^{S} \right)
   \left<  \exp \left( i  \beta^{1/2}   \left<   \widehat{\rho}_{C}  
   \vert  \varphi    \right>    \right)  \right>_{v_{c}} \; ,
  \end{equation}    
where the brackets $\left< \ldots \right>_{v_{c}}$ denote Gaussian 
averages over the real scalar field $\varphi(\vec{r})$, i.e.
\begin{eqnarray}
\label{moyv}
\left< \ldots  \right>_{ v_{c}}&\equiv&
{\cal N}_{v_{c}}^{-1} \int {\cal D} \varphi
\ldots \exp\left(-\frac{1}{2} \left< \varphi \vert  v_{c}^{-1}
\vert \varphi \right>\right) \; , \nonumber \\
{\cal N}_{v_{c}} &\equiv& \int {\cal D } \varphi
\exp\left(-\frac{1}{2} \left< \varphi \vert  v_{c}^{-1}
\vert \varphi \right>\right) \; ,
\end{eqnarray}
where
\begin{equation}
    \label{vmoins1}
 v_{c}^{-1}(1,2) = - \frac{1}{4 \pi} \Delta_{1}\delta (1,2) 
 \end{equation} 
is the inverse of the positive operator $v_{c}(1,2)$. Therefore one 
has, after an integration by parts
\begin{equation}
    \label{Nv}
    {\cal N}_{v_{c}}=
    \int {\cal D } \varphi \; 
    \exp \left( 
    - \frac{1}{8 \pi} \int_{\Omega} d^{3} \vec{r}\; ( \vec{\nabla} 
    \varphi )^{2}
    \right)
\end{equation}     
The functional integrals which enter eqs\ (\ref{moyv}) and\ (\ref{Nv})
can be given a precise meaning when grounded
perfect conductor boundary conditions (BC) are adopted;
periodic BC's also work if only neutral configurations are considered,
we refer the reader to the literature for more details.
\cite{Cai-Raim,Brydges,Ma,Dowrick,Parisi} 
It will be convenient to write
\begin{equation}
    \label{Phi0}
  \left<   \widehat{\rho}_{C}  \vert  \varphi    \right>=
  \left<   \widehat{\rho}_{\alpha}  \vert  \phi_{\alpha}   
  \right> \; ,
\end{equation}     
where the smeared field $\phi_{\alpha}  $ is defined as 
\begin{equation}
    \label{Phi}
 \phi_{\alpha} (1) \equiv \beta^{1/2} \; q_{\alpha} 
 \tau_{\alpha}(1,1') \; \varphi(1') \; .
 \end{equation}
 The field $ i \phi_{\alpha} (1) $ may thus be seen as an external one-body 
 potential acting on the particles of the species $\alpha$; indeed,  one can 
 rewrite the Boltmann factor\ (\ref{Boltz2}) under the form
\begin{eqnarray}
    \label{Boltz3}
    \exp \left( - \beta U \left(\omega \right) \right)& = &
  \exp \left( - \beta U_{HS} \left(\omega \right) \right) \; 
   \exp \left(  N_{\alpha} \; \nu_{\alpha}^{S} \right)
   \langle  \exp \left(  i  \langle  \widehat{\rho}_{\alpha}  
   \vert  \phi_{\alpha}
      \rangle    \right)  \rangle_{v_{c}} \; , \nonumber \\ 
      &=&  \exp \left( - \beta U_{HS} \left(\omega \right) \right) \;
      \exp \left(  N_{\alpha} \; \nu_{\alpha}^{S} \right)
      \langle  \exp \left(
      \sum_{\alpha=1}^{M} \sum_{i_{\alpha}=1}^{N_{\alpha}}
      i \phi_{\alpha}\left( \vec{r}_{i_{\alpha}}^{\alpha}\right)
         \right)  \rangle_{v_{c}}
      \; .
 \end{eqnarray}   
\subsection{The Physical meaning of the auxiliary field }
In a given configurational $\omega$ let us define an action
\begin{equation}
    \label{h}
    h [\varphi ] =\frac{1}{2} \left< \varphi \vert 
    v_{c}^{_{-1}} \vert \varphi \right> - i \; \left< \phi_{\alpha} 
    \vert  \widehat{\rho}_{\alpha}  \right> \; ,
 \end{equation}      
 and a partition function
 \begin{equation}
     \label{z}
     z\left( \omega \right) = \left< \exp\left(  - h 
     \left[\varphi\right] \right) \right>_{v_{c}} \; .
  \end{equation}     
(Henceforth we shall specify the arguments of functionals by means of 
brackets, the variables of ordinary functions being enclosed as usual by  parenthesis.)
The saddle point of the functional $h [\varphi ]$ is obtained 
by solving the equation
\begin{equation}
    \label{statio0}
    \left.\frac{\delta h}{\delta \varphi(\vec{r})}\right |_{\overline{\varphi}}
=0 \; ,
 \end{equation}  
 which can be recast under the form of the Poisson equation
 \begin{equation}
     \label{statio1}
     \Delta_{1} \overline{\varphi}(1)= - 4 \pi i \beta^{1/2}
  \widehat{\rho}_{C} (1) \; ,
 \end{equation}  
 the solution of which is of course 
 \begin{equation}  
 \overline{\varphi}(1)= i \beta^{1/2} \widehat{V}(1) \; .
 \end{equation}   
 Therefore, at the saddle point,  the field $\overline{\varphi}$ can 
 be identified with the microscopic electric potential in the configuration
$\omega$, up to an imaginary 
 multiplicative constant.
 Moreover, it is easy to show that the value of $h[\varphi]$ is 
 nothing but  the energy of the configuration, i.e.
\begin{equation}  
    h[ \overline{\varphi}] = \frac{\beta}{2} \left<
    \widehat{\rho}_{C} \vert v_{c} \vert  \widehat{\rho}_{C}
    \right> \; .
\end{equation} 
Let us make now the change of variables 
$ \varphi = \overline{\varphi} + \delta \varphi$ 
where $\delta \varphi$ is a real scalar field. It follows from the 
stationarity condition\ (\ref{statio0}) that
\begin{equation} 
    h[\varphi]=  h[ \overline{\varphi}] +\frac{1}{2}
    \left< \delta \varphi \vert v_{c}^{-1} \vert \delta \varphi 
    \right> \; ,
\end{equation} 
which confirms that  $h[ \overline{\varphi}] $ is indeed a minimum of 
the functional $h[\varphi]$ since $v_{c}^{-1}$ is a positive 
operator. For a more complicated Hamiltonian than $h[\varphi]$, the approximation
consisting in truncating its functional Taylor expansion  about the 
saddle point at the second order level is called the Gaussian
approximation. \cite{Ma} 
This approximation is obviously exact for $z(\omega)$ because
$h[\varphi]$ is a quadratic form. A direct calculation indeed confirms  that  
\begin{eqnarray}
z(\omega) &=& \exp( -h[ \overline{\varphi}]) \;
\frac{\int {\cal D} \delta \varphi
 \exp\left(-\frac{1}{2} \left< \delta \varphi \vert  v_{c}^{-1}
\vert \delta \varphi \right>\right)}{\int {\cal D} \varphi
\exp\left(-\frac{1}{2} \left< \varphi \vert  v_{c}^{-1}
\vert \varphi \right>\right)} \; , \nonumber \\
&=&
\exp \left(    -\frac{\beta}{2} \langle
    \widehat{\rho}_{C} \vert v_{c} \vert  \widehat{\rho}_{C}
    \rangle \right) \; .
\end{eqnarray}
\subsection{The KSSHE transform of the grand partition function}
\label{csi}
 Henceforward we shall 
work in the grand canonical (GC) ensemble. We denote by $\mu_{\alpha}$ 
the chemical potential of the species $\alpha$ and by $\psi_{\alpha}(\vec{r})$ 
the external potential with which the particles of the species
 $\alpha$ interact eventually.
According to a terminology due to J. Percus,\cite{Percus} we shall define the local
chemical potential $\nu_{\alpha}(\vec{r})$ as 
$\beta(\mu_{\alpha} -\psi_{\alpha}(\vec{r}))$.
With these notations, the GC partition function of the system takes the form
\begin{equation}
\label{Xi}
\Xi[ \{\nu_{\alpha} \}]=\sum_{N_{1}=0}^{\infty} \frac{1}{N_{1}!}
\ldots
\sum_{N_{M}=0}^{\infty} \frac{1}{N_{M}!}
\int_{\Omega}d^{3}\vec{r}_{1}^{1}
\ldots d^{3}\vec{r}_{N_{M}}^{M} \exp(-\beta
U(\omega)) \prod_{\alpha=1}^{M} \prod_{i_{\alpha}=1}^{N_{\alpha}}
\exp\left(\nu_{\alpha}\left( \vec{r}_{i_{\alpha}}^{\alpha}\right)\right) \; .
\end{equation}
 Grand canonical averages of dynamic 
variables ${\cal A}(\omega)$ will be noted $\left< {\cal A}(\omega) 
\right>_{GC}$. Inserting
the expression\ (\ref{Boltz3}) of the Boltmann factor 
in eq.\ (\ref{Xi}) one obtains the 
KSSHE representation of $\Xi$ 
\begin{equation}
    \label{basic}
  \Xi[ \{\nu_{\alpha} \}]=  \left<
   \Xi_{HS}\left[ \{\overline{\nu}_{\alpha} +i \phi_{\alpha}\}\right]
   \right>_{v_{c}}  \; ,  
\end{equation}  
where $\overline{\nu}_{\alpha} = \nu_{\alpha} +\nu_{\alpha}^{S}$ and
$\Xi_{HS}\left[ \{\overline{\nu}_{\alpha} +i 
\phi_{\alpha}\}\right]$ is the GC partition function of a mixture of bare
hard spheres in the presence of the local chemical potentials
 $\overline{\nu}_{\alpha} +i 
\phi_{\alpha}$. The above result generalizes to the case of the PM 
the result obtained in paper\ I for the restricted 
primitive model. It is also possible to incorporate  a part of the Coulomb
interaction in the reference potential which yields a more general expression
than eq.\ (\ref{basic}) as detailed in the review of Brydges and Martin (cf 
eq.\ (2.29) of ref \cite{Brydges}). However, in the liquid domain, the
thermodynamics and correlations of this reference system are, by contrast with
those of the HS fluid, little known in general. 
Relations similar to eq.\ (\ref{basic}) have also
been obtained and  discussed for neutral fluids.\cite{Siegert,Hubbard2,Brilliantov,Caillol}

To make some contact with statistical field theory we  introduce the
effective Hamiltonian (or action)
\begin{equation}
\label{H}
{\cal H}[\varphi]  = \frac{1}{2} \left< 
\varphi \vert  v_{c}^{-1} \vert \varphi \right>  - 
\log \Xi_{HS}\left[ \{\overline{\nu}_{\alpha} +i \phi_{\alpha}\}\right] \; ,
\end{equation}
which allows us to recast $\Xi$ under the form
\begin{equation}
\label{Xinewlook}
\Xi[ \{\nu_{\alpha} \}]= {\cal N}_{v_{c}}^{-1} \int {\cal D} \varphi
\; \exp(-{\cal H}[\varphi])
\: .
\end{equation}

It will be important in the sequel 
to distinguish carefully, besides the GC averages  $<\ldots>_{GC}$,
between two types of statistical field
averages :
the already defined  $<\ldots>_{ v_{c}}$
and the  $<\ldots>_{{\cal H}}$ that we define as
\begin{equation}
<A[\varphi]>_{{\cal H}} \equiv \frac{\int {\cal D} \varphi \;
 \exp(-{\cal H}[\varphi])A[\varphi]}{\int {\cal D} \varphi
 \; \exp(-{\cal H}[\varphi])} \; .
\end{equation}
With these definitions in mind one notes that for an arbitrary functional 
${\cal A}[\varphi]$ one has the relation
\begin{equation}
\label{means}
<A[\varphi]>_{{\cal H}}= \frac{\; \left<A[\varphi] \;  
\Xi_{HS}\left[ \{\overline{\nu}_{\alpha} +i \phi_{\alpha}\}\right] 
\right>_{v_{c}}}{\left< \;
\Xi_{HS}\left[ \{\overline{\nu}_{\alpha} +i 
\phi_{\alpha}\}\right]\right>_{ v_{c}}} \; .
\end{equation}
\section{Correlation functions}
\label{corre}
\subsection{Zoology}
\label{def-corre}
The ordinary and truncated (or connected) density correlation  functions will
be
defined in this paper as \cite{Stell1,Stell2} 
\begin{eqnarray}
\label{defcorre}
G^{(n)}_{\alpha_{1} \ldots \alpha_{n} } [\{\nu_{\alpha} \}](1, \ldots, n) &=
&\left< \prod_{1=1}^{n} \widehat{\rho}_{\alpha_{i}}
        ( i ) \right>_{GC} \; ,\nonumber \\
        &=&
\Xi[\{\nu_{\alpha} \}]^{-1}\frac{\delta^{n} \;\Xi[\{\nu_{\alpha} \}]}
{\delta \nu_{\alpha_{1}}(1) \ldots \delta \nu_{\alpha_{n}}(n)}           \; ,\nonumber \\
G^{(n)\; T}_{\alpha_{1} \ldots \alpha_{n} }[\{\nu_{\alpha} \}](1, \ldots, n) &=& 
\frac{\delta^{n} \log \Xi[\{\nu_{\alpha} \}]}
{\delta \nu_{\alpha_{1}}(1) \ldots \delta \nu_{\alpha_{n}}(n)} \; .
\end{eqnarray}
Our notation emphasizes the fact that the  $G^{(n)}_{\alpha_{1} \ldots \alpha_{n} }$ 
(truncated or not) are functionals of the local chemical potentials 
 $\nu_{\alpha}(\vec{r})$ and
functions of the coordinates $(1,\ldots, n) \equiv (\vec{r}_{1},\ldots,
\vec{r}_{n})$. Note however that, in the remainder of the paper,
we shall frequently omit to 
quote the functional dependence of $G^{(n)}_{\alpha_{1} \ldots \alpha_{n} }$
upon the  $\nu_{\alpha}$ 
when no ambiguity is possible.
In standard textbooks of liquid theory\cite{Hansen} the n-body correlation
functions are more frequently defined as functional
derivatives of $\Xi$ or $\log \Xi$  with respect to the activities
 $z_{\alpha}= 
\exp(\nu_{\alpha})$  rather than with respect to
the local chemical potentials. This yields differences involving delta functions. For
instance for $n=2$ and for a homogeneous system one has 
\begin{eqnarray}
\label{}
G^{(2)}_{\alpha \beta}(1,2) &=&
\rho_{\alpha }  \rho_{\beta  }  g_{\alpha \beta}(r_{12})+ \rho_{\alpha }
\delta_{\alpha,\beta}\; \delta(1,2) 
\; , \nonumber \\
G^{(2)\; T}_{\alpha \beta} (1,2) &=& 
\rho_{\alpha }  \rho_{\beta  } h_{\alpha \beta}(r_{12}) + \rho_{\alpha } 
\delta_{\alpha,\beta}\;  \delta(1,2) \; ,
\end{eqnarray}
where  $\rho_{\alpha } $ is the equilibrium number density of the species $\alpha$
and $g_{\alpha \beta}(r)$ the usual pair 
distribution function; finally $h_{\alpha \beta}=g_{\alpha \beta}-1$.

The charge correlations will play an important role in subsequent 
developments. They are defined as 
\begin{equation}
\label{defcorreC}
G^{(n)}_{C} (1, \ldots, n) =
\left< \prod_{1=1}^{n} \widehat{\rho}_{C}
        ( i ) \right>_{GC} \; .
\end{equation}
It follows from the definition\ (\ref{roc}) of the smeared density of 
charge $\widehat{\rho}_{C}$ that eq.\ (\ref{defcorreC}) can be 
rewritten alternatively
\begin{equation}
\label{defcorreC2}
G^{(n)}_{C} (1, \ldots, n) =
q_{\alpha_{1}} \ldots q_{\alpha_{n }} \; 
\tau_{\alpha_{1}} (1,1') \ldots \tau_{\alpha_{n}} (n,n')  \;
G^{(n)}_{\alpha_{1} \ldots \alpha_{n} } (1', \ldots, n')
 \; .
\end{equation}
Clearly the operator 
\begin{equation}
\label{theta}
\Theta(1) \equiv i \beta^{1/2} q_{\alpha} \; \tau_{\alpha} (1,1')
\; \frac{\delta}{\delta \nu_{\alpha}(1')} \; 
\end{equation}
is the generator of the charge correlations for we have clearly
\begin{equation}
\label{defcorreC3}
i^{n} \beta^{n/2} \; G^{(n)}_{C} (1, \ldots, n) =
\Xi^{-1} \Theta(1) \ldots \Theta(n) \Xi
\; .
\end{equation}
The truncated charge correlations can thus be defined according to
\begin{equation}
\label{defcorreCT}
i^{n} \beta^{n/2} \; G^{(n) \;T}_{C} (1, \ldots, n) =
 \Theta(1) \ldots \Theta(n) \log  \Xi 
\; .
\end{equation}
On the one hand 
\begin{equation}
\label{defcorreCT2}
G^{(n)\; T}_{C} (1, \ldots, n) =
q_{\alpha_{1}} \ldots q_{\alpha_{n }} \; 
\tau_{\alpha_{1}} (1,1') \ldots \tau_{\alpha_{n}} (n,n')  \;
G^{(n)\; T}_{\alpha_{1} \ldots \alpha_{n} } (1', \ldots, n')
 \; ,
\end{equation}
and, in the other hand \cite{Dowrick,Stell1,Stell2}
\begin{equation}
\label{defcorreCT3}
G^{(n)\; T}_{C} (1, \ldots, n) =
G^{(n) \; T}_{C} ( 1,\ldots,n)
- \sum \prod_{m<n}G^{(m)\; T}_{C} (i_{1},\ldots,i_{m})  \; ,
\end{equation}
where the sum of products is carried out over all possible partitions of 
the set $(1,\ldots,n)$ into subsets of cardinality $m<n$. The functions 
$G^{(n)}_{C}$ (resp. $G^{(n)\; T}_{C}$) for different values of $n$ are not
independent; they are related by a hierarchy of equations most conveniently
written with the help of the operator $\Theta$ defined at eq.\ (\ref{theta}). 
The
hierarchies for the $G^{(n)}_{C}$ and the $G^{(n)\; T}_{C}$ are derived in 
appendix\ \ref{apA}.

In the field theoretical representation of the PM
the field correlation functions  play a key role. 
They are defined as
\begin{mathletters}
    \begin{eqnarray}
G^{(n)}_{\varphi} (1, \ldots, n)& =&
\left< \varphi(1) \ldots \varphi(n)  \right>_{{\cal H}} \label{gphi}
\; , \\
G^{(n)\; T}_{\varphi} (1, \ldots, n)& =&
G^{(n) \; T}_{\varphi}( 1,\ldots,n)
- \sum \prod_{m<n}G^{(m)\; T}_{\varphi}(i_{1},\ldots,i_{m})  \; .
     \end{eqnarray}     
 \end{mathletters}
 
 \noindent
Of course the  $G^{(n)}_{\varphi}$, as the charge correlation 
functions, are functionals of the local chemical potentials.
The hierarchies for the $G^{(n)}_{\varphi}$ and the $G^{(n)\; T}_{\varphi}$
are derived in appendix\ \ref{apA}.

\subsection{Relations between the charge and field correlation 
functions}
\label{gcgphi} 
\subsubsection{The density and charge correlation functions as statistical field 
averages}
\label{rel1} 
It follows from the definition\ (\ref{defcorre}) of 
$ G^{(n)}_{\alpha_{1} \ldots \alpha_{n} }$ and from
the KSSHE 
representation\ (\ref{basic}) of the grand partition function that we 
have
\begin{eqnarray}
    G^{(n)}_{\alpha_{1} \ldots \alpha_{n} }\left[ \{ \nu_{\alpha}  \right \}] 
    (1, \ldots, n) &=
& \Xi^{-1} {\cal N}_{v_{c}}^{-1} 
\int {\cal{D}} \varphi \; \exp \left( -\frac{1}{2}
 \left< 
\varphi \vert  v_{c}^{-1} \vert \varphi \right>  \right)
\frac{\delta^{n} \;\Xi_{HS}\left[ \{\overline{\nu}_{\alpha} +i \phi_{\alpha}\}\right] }
{\delta \nu_{\alpha_{1}}(1) \ldots \delta \nu_{\alpha_{n}}(n)}  \; , 
\nonumber \\
&=&
\Xi^{-1} \left< \Xi_{HS} \; 
 G^{(n)}_{HS, \; \alpha_{1} \ldots \alpha_{n} } \left[ \{\overline{\nu}_{\alpha} +i 
\phi_{\alpha}\} \right]  (1, \ldots, n) 
 \right>_{v_{c}} \; ,
 \end{eqnarray}
 where $G^{(n)}_{HS, \; \alpha_{1} \ldots \alpha_{n} } \left[ \{\overline{\nu}_{\alpha} +i 
\phi_{\alpha}\} \right]  (1, \ldots, n) $ denotes the density correlation 
function of the reference HS fluid in the presence of the local chemical 
potentials $\{\overline{\nu}_{\alpha} +i 
\phi_{\alpha}\}$.
 Thence, making use of eq.\ (\ref{means})
 \begin{equation}
     \label{klug}
    G^{(n)}_{\alpha_{1} \ldots \alpha_{n} }\left[ \{ \nu_{\alpha}  \right \}]
    (1, \ldots, n)=  
    \left<   G^{(n)}_{HS,\; \alpha_{1} \ldots \alpha_{n} }
    \left[ \{\overline{\nu}_{\alpha} +i 
\phi_{\alpha}\} \right]  (1, \ldots, n) 
    \right>_{{\cal H}} \; .
  \end{equation} 
 Eq.\ (\ref{klug}), which extends to ionic mixtures a relation that we 
 derived elsewhere for simple non-charged fluids,\cite{Caillol} 
 although aesthetic is not very useful 
 since the hard sphere correlations $G^{(n)}_{HS, \;  \alpha_{1} \ldots 
 \alpha_{n} }$ are complicated functionals of the field $\varphi$. 
 However the case $n=1$ is of some interest. In that case  eq.\ 
 (\ref{klug}) says that
 \begin{equation}
  \label{klug1}
  \rho_{\beta}\left[ \{ \nu_{\alpha}  \right \}]  (1)=
  \left<   \rho_{HS, \; \beta}
    \left[ \{\overline{\nu}_{\alpha} +i 
\phi_{\alpha}\} \right]  (1) \right>_{{\cal H}} \; ,
  \end{equation} 
  
 It follows readily from the expression\ (\ref{defcorreC2}) of   the 
 charge correlation function that we also have
\begin{equation}
     \label{klu}
     G^{(n)}_{C}\left[ \{ \nu_{\alpha}  \right \}]
    (1, \ldots, n)=  
    \left<   G^{(n)}_{HS,\; C }
    \left[ \{\overline{\nu}_{\alpha} +i 
\phi_{\alpha}\} \right]  (1, \ldots, n) 
    \right>_{{\cal H}} \; ,
  \end{equation} 
where
\begin{equation}
    \label{bidule}
G^{(n)}_{HS, \; C} (1, \ldots, n) =
q_{\alpha_{1}} \ldots q_{\alpha_{n }} \; 
\tau_{\alpha_{1}} (1,1') \ldots \tau_{\alpha_{n}} (n,n')  \;
G^{(n)}_{HS, \; \alpha_{1} \ldots \alpha_{n} } (1', \ldots, n') \; .
 \end{equation} 
Specializing  eq.\ (\ref{klu}) for $n=1$ we note that 
 \begin{equation}
  \label{klugc}
  \rho_{C}\left[ \{ \nu_{\alpha}  \right \}] (1)=
  \left<   \rho_{HS, \; C}
    \left[ \{\overline{\nu}_{\alpha} +i 
\phi_{\alpha}\} \right]  (1) \right>_{{\cal H}} \; , 
  \end{equation} 
  where
 \begin{equation}
  \label{rhoc} 
  \rho_{HS, \; C}(1)=q_{\alpha}\tau_{\alpha}(1,1') \rho_{HS, \; \alpha}(1') \; .
  \end{equation}   
\subsubsection{Relations between $G^{(n) }_{C}$ and  $G^{(n) 
}_{\varphi}$}
\label{tomi}
It follows readily from the expression\ (\ref{defcorreCT}) of 
$ G^{(n)}_{C}$ and from
the KSSHE 
representation\ (\ref{basic}) of the grand partition function that 
\begin{eqnarray}
    \label{Gc1}
 i^{n} \beta^{n/2}   G^{(n)}_{C}\left[ \{ \nu_{\alpha}  \right \}]  (1, \ldots, n) &=&
 \Xi^{-1}\left[ \{ \nu_{\alpha}  \right \}] \;  {\cal N}_{v_{c}}^{-1} 
\int {\cal{D}} \varphi \; \exp \left( -\frac{1}{2}
 \left< 
\varphi \vert  v_{c}^{-1} \vert \varphi \right>  \right) \times 
\nonumber \\
& \times &
\Theta(1) \ldots \Theta(n) \; \Xi_{HS}
\left[ \{\overline{\nu}_{\alpha} +i \phi_{\alpha}\}\right] 
\; .
\end{eqnarray}
At this point we make the remark that
 \begin{eqnarray}
       \label{rela}
     \frac{\delta \;  \Xi_{HS}\left[ \{\overline{\nu}_{\alpha} +i 
\phi_{\alpha}\}\right]}{\delta \; \varphi(1)}&=&
\frac{\delta \;  i \phi_{\alpha}(1')}
{ \delta \varphi(1)} \; 
 \frac{\delta \;  \Xi_{HS}\left[ \{\overline{\nu}_{\alpha} +i 
\phi_{\alpha}\}\right]}{\delta \; \nu_{\alpha}(1')} \; 
 \; , \nonumber \\
 &=&\Theta(1) \; \Xi_{HS}\left[ \{\overline{\nu}_{\alpha} +i 
\phi_{\alpha}\}\right]   \; .
\end{eqnarray} 
The relation\ (\ref{rela}) enables us to replace the operators $\Theta(i)$ 
which occur  the right hand side (RHS)
of eq.\ (\ref{Gc1}) by 
functional derivatives with respect to the field $\varphi$. 
Then, performing $n$ functional integrations by parts\cite{Parisi,Caillol}
yields
\begin{equation}
    \label{gene}
    i^{n} \beta^{n/2}  G^{(n)}_{C}
    (1, \ldots, n) = (-)^{n}
   \left<  \frac{\delta^{n} \exp \left( -\frac{1}{2}
 \left< 
\varphi \vert  v_{c}^{-1} \vert \varphi \right>  \right)}
{\delta \varphi(1) \ldots \delta 
    \varphi(n)}  \right>_{{\cal H}}
\end{equation}    
The relation\ (\ref{gene}) can be used to obtain an explicit representation of
$G^{(n)}_{C}$  
in terms of the field correlations as long as $n$ is not too large. 
Let us first consider the case $n=1$ in which  Eq.\ (\ref{gene}) takes the 
simple form
\begin{equation}
    \label{neq1}
    \Delta_{1} <\varphi(1) >_{{\cal H}}= - 4 \pi i \beta^{1/2}  
    \rho_{C}(1) \; .
\end{equation}  
Once again (cf eq.\ (\ref{statio1})) we obtain the Poisson equation, 
the solution of which is of course
\begin{equation}
     <\varphi(1) >_{{\cal H}}= i \beta^{1/2} V(1) \; ,
\end{equation}  
where $V(1)$ is the GC average of the configurational electric potential, i.e. 
$V(1)= <\widehat{V}(1) >_{GC}$.

 In the case $n=2$ eq.\ (\ref{gene}) says that
\begin{equation}
  \label{neq2a}
 \beta G^{(2)}_{C}(1, 2)= \frac{-1}{4 \pi} \Delta_{1} \delta(1,2)
 -  \frac{1}{(4 \pi)^{2}} \Delta_{1} \Delta_{2} G^{(2)}_{\varphi}(1, 
 2) \; ,
\end{equation}  
or, by reverting the equation
\begin{equation}
  \label{neq2b}
 G^{(2)}_{\varphi}(1, 2) = v_{c}(1,2) - \beta  v_{c}(1,1')
  G^{(2)}_{C}(1', 2') v_{c}(2',2) \; .
\end{equation}  
Eqs.\ (\ref{neq2a}) and\ (\ref{neq2b}) extend to  
electrolyte solutions relations  obtained recently for  neutral 
fluids. \cite{Caillol} Equations of this type were also derived by Ciach 
and Stell  in the framework of a heuristic field theory 
of the RPM.\cite{Ciach-Stell}

By combining eqs.\ (\ref{neq1}) and\ (\ref{neq2a}) one can show easily 
that the truncated two-body charge correlation function satisfies to
 a similar relation, i.e.
\begin{equation}
  \label{neq2T}
 \beta G^{(2)\; T}_{C}(1, 2)= \frac{-1}{4 \pi} \Delta_{1} \delta(1,2)
 -  \frac{1}{(4 \pi)^{2}} \Delta_{1} \Delta_{2} G^{(2) \; T}_{\varphi}(1, 
 2) \; .
\end{equation}
In the case $n=3$ eq.\ (\ref{gene}) yields an awkward expression 
for $G^{(3)}_{C}$. 
However the truncated 3-body charge correlation function  takes the  simple 
form
\begin{equation}
  \label{neq3}
  i \beta^{3/2} G^{(3) \; T}_{C}(1, 2,3) =
  \frac{1}{(4 \pi)^{3}} \Delta_{1} \Delta_{2}  \Delta_{3} G^{(3) \; T}_{\varphi}(1, 
 2,3) \; ,
\end{equation}
which can be obtained by brute force calculation. 
The above result suggests that there are simple relations between $G^{(n)\; T}_{C}$ 
and $G^{(n)\; T}_{\varphi}$ for  values of $n \geq 3$. 
Indeed, let us apply the operator $\Theta(2) \ldots \Theta(3)$ ($n\geq 
3$) to  both sides of eq.\ (\ref{neq1}). Then, making use of the 
hierarchy equations satisfied by the $G^{(n) \; T}_{C}$ and the
$ G^{(n) \; T}_{\varphi}$ (cf eqs.\ (\ref{hie1}) and\ (\ref{hie4}) of appendix\ A)
one gets immediately the aesthetic generic formula
\begin{equation}
  \label{neqn}
  i^{n} \beta^{n/2} G^{(n) \; T}_{C}(1,\ldots,n) =
  \frac{(-1)^{n}}{(4 \pi)^{n}} \Delta_{1} \ldots  \Delta_{n}
  G^{(n) \; T}_{\varphi}(1, ,\ldots,n)  \; \; (\forall n \geq 3) \; .
\end{equation} 
\subsubsection{Correlations of the electric potential}
\label{corre-pot}
It is obvious that 
\begin{eqnarray}
    \left< \widehat{V}(1 ) \ldots \widehat{V}(n) \right>_{GC}&=&
    v_{c}(1,1') \ldots v_{c}(n,n ') \; G^{(n) }_{C}(1',\ldots,n') \; , 
    \nonumber \\
    \left< \widehat{V}(1) \ldots \widehat{V}(n) \right>_{GC}^{T} &=&
    v_{c}(1,1') \ldots v_{c}(n,n ') \;  G^{(n) \; T}_{C}(1',\ldots,n') \; .
\end{eqnarray}    
Combining the above relations with those obtained in sec.\ (\ref{tomi}) 
one gets
\begin{mathletters}
    \label{correVexact}
\begin{eqnarray}
    i \beta^{1/2}  \left< \widehat{V}(1 ) \right>_{GC} &=& 
    \left< \varphi(1)\right>_{{\cal H}} \; , \\
    \beta \left< \widehat{V}(1) \widehat{V}(2 ) \right>_{GC}^{T}  &=& 
    v_{c}(1,2) -  G^{(2) \; T}_{\varphi}(1,2)   \; , \label{correVexactb}\\
 i^{n} \beta^{n/2}    \left< \widehat{V}(1) 
 \ldots \widehat{V}(n) \right>_{GC}^{T} &=&  
  G^{(n) \; T}_{\varphi}(1, ,\ldots,n)  \; \; (\forall n \geq 3) \; .
\end{eqnarray}  
\end{mathletters}

\noindent
What  can be learned from the above relations is the subject of  next 
section.

\section{Stillinger-Lovett sum rules}
\label{SL}
A salient property of 3D ionic liquids is the screening effect. To 
paraphrase Ph. Martin, this type of fluid " in thermal equilibrium 
does not tolerate any charge 
inhomogeneity over more than a few intermolecular 
distances".\cite{Martin} Even at 
the liquid-vapor critical point where the correlation length 
associated with the fluctuations of density  diverges it is believed, and has
been checked by means of
numerical simulations in the case of the RPM,\cite{Caillol-Weis} 
that the correlation length 
associated with the fluctuations of charge remains 
finite. From the 
existence of screening it is possible to deduce sum rules for the 
charge  correlation functions 
for both homogeneous and 
inhomogeneous systems, the so-called Stillinger-Lovett (SL) sum 
rules.\cite{SL,Carnie,Janco1} As pointed out by B. Jancovici, these 
rules may be rederived under the sole assumption that the system 
behaves macroscopically as a conductor in the sense that " the laws of 
macroscopic electrostatics are assumed to be obeyed for length scales 
large compared to the microscopic characteristic lengths of the 
model."\cite{Janco2}
We examine below how the SL rules can be deduced from simple 
hypothesis on the behavior of 
the KSSHE field correlation functions.

Let us first consider a homogeneous system. In this case 
$<\varphi>_{{\cal H}}$ is a constant as well as are the 
densities $\rho_{\alpha}$ and the smeared density of charge $\rho_{C}$.
It follows then from eq.\ (\ref{neq1}) that 
the smeared density $\rho_{C}=0$. This is nothing but the usual 
electroneutrality condition since 
$\rho_{C}=\rho_{\alpha}q_{\alpha}\widetilde{\tau}_{\alpha}(0)$ 
can be identified with the usual local charge density for
$\widetilde{\tau}_{\alpha}(0)=1$  (property\ (\ref{wb}) of the 
smearing function $\tau_{\alpha}$). Therefore, in the framework of our 
formalism, the electroneutrality condition
\begin{equation}
    \label{SL0}
    \rho_{\alpha}q_{\alpha}  =0
\end{equation} 
is automatically satisfied for an arbitrary set of chemical potentials 
$\{\nu_{\alpha}\}$, a well known property of ionic fluids.\cite{Brydges}
Note that eq.\ (\ref{SL0}) implies that there are only $M-1$ independent
chemical potentials.

The correlations of the electric potential has been studied by 
various approaches and asymptotically one has 
\begin{mathletters}
    \label{correVasympt}
\begin{eqnarray}
    \beta \left< \widehat{V}(1) \; \widehat{V}(2) \right>_{GC}^{T}  &=& 
    v_{c}(1,2)   \; , \\
   \left< \widehat{V}(1) 
 \ldots \widehat{V}(n) \right>_{GC}^{T} &=&  0 \; .
\end{eqnarray}  
\end{mathletters}

\noindent
It must be stressed that these expressions are valid for relative 
distances large compared to the microscopic characteristic lengths of the 
system and if the correlations decay fast enough or, equivalently, if
the system behaves as a macroscopic conductor.\cite{Brydges,Martin,Janco2,Lebo-Martin}
 The comparison of these asymptotic behaviors with the exact 
relations\ (\ref{correVexact}) derived in sec.\ (\ref{corre-pot}) 
entails that the truncated KSSHE field correlation functions $G^{(n) \; 
T}_{\varphi}$ are short range functions; stated otherwise, 
the KSSHE field is a non-critical field. Conversely, this property being
taken as given, we show now that one can infer the SL rules.

We consider now a non-homogeneous system and we take the Laplacian of 
eq.\ (\ref{correVexactb}). We get
\begin{equation}
    \label{carni0}
  \beta \left< \widehat{\rho}_{C}(1 ) \widehat{V}(2) \right>_{GC}^{T}=  
  \delta(1,2) +
  \frac{1}{4 \pi} \; \Delta_{1} G^{(2) \; T}_{\varphi}(1,2) \; .
\end{equation}
Let us integrate eq.\ (\ref{carni0}) over $\vec{r}_{1}$. With the 
hypothesis that $G^{(2) \; T}_{\varphi}(1,2)$ is short range the 
integration of the Laplacian gives zero by an application of Green's 
theorem. Therefore
\begin{equation}
    \label{carni}
    \beta \int d(1) \left< \widehat{\rho}_{C}(1 ) \widehat{V}(2) 
    \right>_{GC}^{T}=1 
\end{equation}    
which is the Carnie-Chan sum rule.\cite{Carnie} In the case of a 
homogeneous system, the Carnie-Chan sum rule is equivalent to the SL 
sum rule.\cite{Janco2} Let us retrieve these sum rules in our framework. 
We start with eq.\ (\ref{neq2T}) for a homogeneous system. In this 
case 
$G^{(2) \; T}_{C}(1,2) \equiv G^{(2) \; 
T}_{C}(\vec{r}=\vec{r}_{1} - \vec{r}_{2}))$ and we have
\begin{equation}
    \label{neq2Tbis}
    -4 \pi \beta G^{(2) \; T}_{C}(\vec{r}) =
    \Delta \delta(\vec{r}) -\frac{1}{4 \pi}\Delta \Delta G^{(2) \; 
    T}_{\varphi}(\vec{r}) \; .
\end{equation} 
With the hypothesis that $G^{(2) \; T}_{\varphi}(1,2)$ is a short range
function the 
integration over $\vec{r}$ gives zero, i.e.
\begin{equation}
    \label{SL1}
    \int d^{3}\vec{r} \; G^{(2) \; T}_{C}(\vec{r}) =0 \; 
\end{equation} 
which is the first SL rule.\cite{SL}
Similarly, after integration by parts
\begin{equation}
    \label{SL20}
  4 \pi \beta\int d^{3}\vec{r}\;  r^{2} G^{(2) \; T}_{C}(\vec{r})=
  -6 + \frac{6}{4 \pi} \int d^{3}\vec{r}\; \Delta G^{(2) \; 
  T}_{\varphi}(\vec{r}) \; .
\end{equation} 
If $G^{(2) \; T}_{\varphi}(\vec{r})$ is a short range function then
the integral in 
the RHS  vanishes and we are left with the second SL 
sum rule

\begin{equation}
    \label{SL2}
    \frac{2 \pi \beta}{3} \int d^{3}\vec{r}\;  r^{2} G^{(2) \; 
    T}_{C}(\vec{r})=-1 \; .
\end{equation}   
The SL rules are more conveniently written in terms of the pair 
distributions $h_{\alpha,\beta}$ introduced at eq.\ (\ref{})
\begin{mathletters}
 \label{slab}
\begin{eqnarray}
    \rho_{\gamma} q_{\gamma} \widetilde{h}_{\alpha,\gamma}(k=0) 
    &=&-q_{\alpha} \label{sla} \\
    \frac{2 \pi \beta}{3} \; q_{\alpha}q_{\gamma}
    \rho_{\alpha}\rho_{\gamma} \int d^{3}\vec{r} \; r^{2} 
    h_{\alpha,\gamma}(r)&=&-1\label{slb}  \; 
\end{eqnarray} 
\end{mathletters}

\noindent
Although eqs.\ (\ref{slb}) and\ (\ref{SL2}) are equivalent, eq.\ 
(\ref{sla}) which says that the cloud of charges which surrounds a given ion 
of species $\alpha$ has a total charge $-q_{\alpha}$ implies eq.\ (\ref{SL1}) 
but is more precise. It can nevertheless be derived directly with methods similar 
to those used in this section. The proof is given in appendix\ \ref{apB}.

We want to precise that the results derived in this section are not valid for
$2D$ systems which can undergo a Kosterlitz-Thouless transition.\cite{KT}
In the
low-temperature KT phase of a $2D$ PM, the SL rules are violated and 
the sine-Gordon field should exhibit long range correlations with an algebraic
decay. In this case, $\varphi$ is a critical field and
$<\varphi>$ is related to the order parameter of the KT transition.
\cite{Levin}
\section{The mean field theory}
\label{MF}
\subsection{Mean field equations}
\label{MFeq}

We define the MF level  or saddle point approximation of the theory
by the equation
\begin{equation}
    \label{MFdef}
\Xi_{MF}[\{\nu_{\alpha} \}]\equiv \exp\left(-{\cal
H}\left(\overline{\varphi}\right)\right) \; ,
\end{equation}
where, at $\varphi=\overline{\varphi}$, the action ${\cal H}$ 
is stationary.
It follows  from the expression \ (\ref{H}) of ${\cal H}$ 
that the stationarity condition 
\begin{equation}
    \label{statio}
\left.\frac{\delta {\cal H}}{\delta \varphi(\vec{r})}
\right|_{\overline{\varphi}}=0 \; 
\end{equation}
may be rewritten as 
\begin{mathletters}
\begin{eqnarray}
    \label{eqMF}
    \Delta_{1} \overline{\varphi}(1)&=&
  -4 \pi   i \beta^{1/2} \; \rho_{HS,\;  C}
    \left[ \{\overline{\nu}_{\alpha} +i 
\overline{\phi}_{\alpha}\} \right]  (1) \; , \\
\overline{\phi}_{\alpha}(1)&=& q_{\alpha} \tau_{\alpha}(1,1') \;
\overline{\varphi}(1') \; .
\end{eqnarray} 
\end{mathletters}

\noindent
Two comments are in place here. 
Firstly, it can be pointed out that  eq.\ (\ref{eqMF}) is very 
similar to the well known  Poisson-Boltzmann equation.\cite{Outhwaite}
Secondly, it is instructive to compare  eq.\ (\ref{eqMF}) with the following 
exact relation, easily 
deduced from eqs.\ (\ref{klugc}) and\ (\ref{neq1})
\begin{equation}
    \label{exactr}
    \Delta_{1} \langle \varphi (1) \rangle_{{\cal H}}=
    -4 \pi   i \beta^{1/2} \;   
  \langle
 \rho_{HS,\;  C}
    \left[ \{\overline{\nu}_{\alpha} +i 
\phi_{\alpha}\} \right]  (1)  
 \rangle_{{\cal H}} \; .
\end{equation}    
One notes that eq.\ (\ref{eqMF})
could have been guessed from the exact eq.\ 
(\ref{exactr}) with the usual assumption of an absence of field 
correlations at the MF level.
\subsection{The mean field grand canonical free energy}
\label{MFfree}
The MF grand potential  is easily obtained by 
substituting $\overline{\varphi}$ in eq.\ (\ref{MFdef}) with the result 
\begin{eqnarray}
    \label{MFlXi}
    \log \Xi_{MF}[\{\nu_{\alpha} \}]&=&
    \log \Xi_{HS}[\{\overline{\nu}_{\alpha} +i 
\overline{\phi}_{\alpha}\}] - \frac{1}{2}\left<  \overline{\varphi} \;\vert
  v_{c}^{-1}
\vert \;\overline{\varphi}\right>
\nonumber \\
      &=&\log \Xi_{HS}[\{\overline{\nu}_{\alpha} +i 
\overline{\phi}_{\alpha}\}]  +
\frac{\beta}{2} \left<  \rho_{HS,\;  C}
    \left[ \{\overline{\nu}_{\alpha} +i 
\overline{\phi}_{\alpha}\} \right]  \vert v_{c} \vert \rho_{HS, \; C}
    \left[ \{\overline{\nu}_{\alpha} +i 
\overline{\phi}_{\alpha}\} \right] 
\right>
\end{eqnarray}  
The MF number density of species $\gamma$ is obtained by taking the 
functional derivative of $ \log \Xi_{MF}[\{\nu_{\alpha} \}]$ with 
respect to the local chemical potential $\nu_{\gamma}$. On 
the one hand we have
\begin{eqnarray}
    \label{roMFa}
    \frac{ \delta\log \Xi_{HS}[\{\overline{\nu}_{\alpha} +i 
\overline{\phi}_{\alpha}\}] }{\delta \nu_{\gamma}(1)}&=&
\rho_{HS, \; \gamma}[\{\overline{\nu}_{\alpha} +i 
\overline{\phi}_{\alpha}\}](1) + \rho_{HS,  \; \beta}[\{\overline{\nu}_{\alpha} +i 
\overline{\phi}_{\alpha}\}](1')
 \frac{ \delta i \overline{\phi}_{\beta}(1')}{\delta 
 \nu_{\gamma}(1)} \nonumber \\
 &=&\rho_{HS, \;  \gamma}[\{\overline{\nu}_{\alpha} +i 
\overline{\phi}_{\alpha}\}](1) + i \beta^{1/2} \;
 \rho_{HS, \; C}[\{\overline{\nu}_{\alpha} +i 
\overline{\phi}_{\alpha}\}](1')
 \frac{ \delta \overline{\varphi}(1')}{\delta \nu_{\gamma}(1)}
\; ,
\end{eqnarray}
and, in the other hand 
\begin{equation}
    \label{roMFb}
   \frac{\delta }{ \delta \nu_{\gamma}(1)} \frac{1}{2}
   <\overline{\varphi} \vert v_{c}^{-1} \vert \overline{\varphi}>=-   
   \frac{ \delta \overline{\varphi}(1')}{\delta \nu_{\gamma}(1)}
   \frac{1}{4 \pi}\;\Delta_{1'}\overline{\varphi}(1') \; .
\end{equation}   
Substracting eqs.\ (\ref{roMFa}) and \ (\ref{roMFb}) and making use 
of the stationarity condition\ (\ref{eqMF}) gives us 
\begin{equation}
    \label{roMF}
    \rho_{MF,\; \gamma}(1) = \rho_{HS, \; \gamma}
   [\{\overline{\nu}_{\alpha} +i 
\overline{\phi}_{\alpha}\}](1) \; ,
 \end{equation} 
from which we infer the expression of the MF charge density :
\begin{equation}
    \label{rocMF}
    \rho_{MF, \; C}(1) = \rho_{HS,\;  C}
   [\{\overline{\nu}_{\alpha} +i 
\overline{\phi}_{\alpha}\}](1) \; .
 \end{equation} 
In the case of a homogeneous system the MF KSSHE field  
$\overline{\varphi}$ is uniform and $\overline{\phi}_{\alpha}$ 
reduces to $\overline{\phi}_{\alpha}=\beta^{1/2} q_{\alpha} 
\overline{\varphi}$ since $\widetilde{\tau}_{\alpha}(0)=1$. Moreover, 
in this case,
the MF equation\ (\ref{eqMF}) takes the form of the charge
 neutrality condition
\begin{equation}
    \label{neutraMF}
    \rho_{MF, \; C}\equiv q_{\alpha}\; \rho_{MF, \;\alpha}=0 \; ,
 \end{equation}   
 where $\rho_{MF, \;\alpha}= \rho_{HS, \; \alpha}[\{ 
 \overline{\nu}_{\alpha} + i \beta^{1/2} q_{\alpha}\overline{\varphi} \}]$.
  Therefore
 $\overline{\varphi}$ is an imaginary number which adjusts itself in 
 such a way that the charge neutrality condition\ (\ref{neutraMF}) is 
 satisfied. For instance for a RPM with $\nu_1=\nu_2$ one has
 $\overline{\varphi}=0$, while for a binary  SPM ($M=2$) one finds,
with the same hypothesis on the chemical potentials,
that $\overline{\varphi}=-i \sqrt{\beta}
 \log(\vert q_2/q_1 \vert)/(q_1-q_2)$. 

 The MF grand canonical free energy ${\cal A}_{MF}$
 is defined as the Legendre transform of 
 $ \log \Xi_{MF}[\{\nu_{\alpha} \}]$ with respect to the local 
 chemical potentials $\{\nu_{\alpha} \}$.\cite{Percus,Stell1,Caillol2} 
 Therefore one has
 \begin{equation}
     \beta {\cal A}_{MF}=
     <\rho_{MF, \; \alpha}\vert \nu_{\alpha}> - 
     \log \Xi_{MF}[\{\nu_{\alpha} \}] \; .
 \end{equation} 
 When expressed in terms of its natural variables $\{ 
 \rho_{MF, \; \alpha}\}$, the functional  $\beta {\cal A}_{MF}[\{ 
 \rho_{MF, \; \alpha}\}]$ reads as
  \begin{equation}
      \label{AMF}
 \beta{\cal A}_{MF}[\{ \rho_{MF, \; \alpha}\}]  =\beta
  {\cal A}_{HS}[\{ \rho_{MF, \; \alpha}\}] -
  <\rho_{MF, \; \alpha} \vert \nu_{\alpha}^{S} > 
  +
  \frac{\beta}{2} \langle \rho_{MF, \; C} \vert v_{c} \vert \rho_{MF, \; C} 
  \rangle \; ,
\end{equation} 
as a short calculation will show. 
The functional  ${\cal A}_{HS}[\{ \rho_{\alpha}\}] $ which appears
in the RHS of eq.\ 
(\ref{AMF}) is the exact GC free energy of an inhomogeneous hard spheres mixture. As 
it is well known it is a convex functional of the densities  
$\{\rho_{\alpha}\}$. \cite{Caillol2,Chayes} On the other hand the quadratic 
form $< \rho_{C} \vert v_{c} \vert \rho_{C} >$ is definite positive, 
therefore (strictly) convex. The last contibrution to ${\cal A}_{MF}$, 
i.e. the linear term $  <\rho_{\alpha} \vert \nu_{\alpha} > $, will 
not change our conclusion that  ${\cal A}_{MF}[\{ \rho_{\alpha}\}] $ 
is a convex functional of the densities $\{ \rho_{\alpha}\}$.
Thence one has, for all
$0\leq \lambda \leq 1$
 \begin{equation}
     \label{ine}
 {\cal A}_{MF}\left[\left\{\lambda \rho_{\alpha}
 +(1-\lambda) \rho_{\alpha}^{'}\right\}
 \right]
  \leq
 \lambda \;  {\cal A}_{MF}\left[\left\{ \rho_{\alpha} \right\}\right]
 + (1-  \lambda ) \; {\cal A}_{MF}\left[\{\rho_{\alpha}^{'}\}\right] \; .
\end{equation}   
   
Since ${\cal A}_{MF}[\{ \rho_{\alpha}\}]$ is convex then its 
Legendre transform exists and is also convex. Of course, 
the Legendre transform being involutive, it 
must be identified with $ \log 
\Xi_{MF}[\{\nu_{\alpha} \}]$. Therefore one has\cite{Caillol2}
\begin{mathletters}
\begin{eqnarray}
 \label{Leg1} 
 \beta {\cal A}_{MF}[\{ \rho_{\alpha}\}] & =& \sup_{\{\nu_{\alpha} \}} 
 \left( \langle \rho_{\alpha}  \vert \nu_{\alpha} \rangle - 
 \log \Xi_{MF} [\{\nu_{\alpha} \}] \right) \; ,  \\
 \label{Leg2} 
 \log \Xi_{MF} [\{\nu_{\alpha} \}] &=& \sup_{\{ \rho_{\alpha} \}} 
 \left( \langle \rho_{\alpha}  \vert \nu_{\alpha} \rangle -  
 \beta {\cal A}_{MF} \left[\{ \rho_{\alpha}\}\right] \right)
 \; .
\end{eqnarray}
\end{mathletters}
\noindent
Some comments are in order.
Firstly, assuming that a mixture of hard spheres cannot undergo a liquid-liquid 
transition (liquid-vapor or dimixion)\cite{Hansen} one can safely 
conclude that ${\cal A}_{HS}[\{ \rho_{\alpha}\}]$ is a strictly convex 
functional in the fluid region. As the sum of two strictly convex 
functionals (cf eq.\ (\ref{AMF}) ) the MF free energy functional
${\cal A}_{MF}[\{ \rho_{\alpha}\}]$ is also strictly convex in 
the fluid region as is the Poisson-Boltzmann functional.\cite{Trizac}
In eq.\ (\ref{ine}) the inequality can thus be 
replaced by a strict inequality for $ 0 < \lambda < 1 $.
An important consequence is that the solution the the MF 
equation\ (\ref{statio}) is 
unique for a given set of chemical potential $\{\nu_{\alpha} \}$ which rules out
the possibility of a fluid-fluid transition of the PM at the MF level.
Secondly, the charge neutrality condition\ 
(\ref{neutraMF}) implies that for a homogeneous system the Helmoltz 
free energy per unit volume takes the (too much) simple form
\begin{equation}
    \label{fMF}
    \beta f_{MF}(\{ \rho_{\alpha}\})\equiv
    \frac{\beta{\cal A}_{MF}}{\Omega}=   \beta f_{HS}(\{ \rho_{\alpha}\})
   - \rho_{\alpha}  \nu_{\alpha}^{S} \; .
\end{equation} 
where the charge neutrality condition $ \rho_{\alpha} q_{\alpha}=0$ has 
been imposed  (otherwise $\beta f_{MF}$ diverges to $+\infty$).
\subsection{Mean field correlation functions}
\label{MFcorre}
\subsubsection{General case}
\label{mongeneral}
It is the place here to recall that a necessary and sufficient condition
for the convexity of $\log \Xi_{MF}[\{\nu_{\alpha} \}] $
and ${\cal A}_{MF}[\{ \rho_{\alpha}\}] $  is that their second order 
functional derivatives are
positive operators, i.e. \cite{Percus,Caillol2,Chayes}
\begin{eqnarray}
\label{GetC}
\left< \delta \nu_{\alpha} \; \vert \; \frac{\delta^{(2)}
\log \Xi_{MF}}{\delta \nu_{\alpha}(1)  
\delta \nu_{\beta}(2)}\;  \vert \; \delta \nu_{\beta} \right> 
& >&  0  \;, \nonumber \\
\left< \delta \rho_{\alpha} \; \vert\;  \frac{\delta^{(2)} 
\beta {\cal A}_{MF}}
{\delta \rho_{\alpha}(1) 
\delta \rho_{\beta}(2)} \;  \vert \; 
 \delta \rho_{\beta} \right> 
& > &  0 \;.
\end{eqnarray} 
We stress that the inequalities\ (\ref{GetC}) are strict in the fluid phase because the two 
functionals  $\log \Xi_{MF}[\{\nu_{\alpha} \}] $ 
and ${\cal A}_{MF}[\{ \rho_{\alpha}\}] $ are both strictly convex. 
The second order derivatives of $\log \Xi_{MF}$ with respect to the 
local chemical potentials are the density correlation functions at 
the MF level: 

\begin{equation}
    \label{GMF}
    G^{ T }_{MF, \; \alpha \beta}[\{\nu_{\alpha} \}](1,2) \equiv
 \frac{\delta^{(2)}
\log \Xi_{MF}}{\delta \nu_{\alpha}(1)  
\delta \nu_{\beta}(2)}   
 \end{equation}
Note that, since we  consider only two-point functions in 
this section, we have further economized the notation by dropping the subscript
$(2)$ in the definition of $ G^{ T }_{MF, \; \alpha \beta}$. 
The second order derivatives of $\beta {\cal A}_{MF}$ with respect to the 
densities are related to the the two-points direct correlation functions.
\cite{Percus,Stell2,Caillol2} Define
\begin{equation}
    \label{CMF}
    \widehat{C}_{MF, \; \alpha \beta}[\{\rho_{\alpha} \}](1,2) \equiv - \; 
 \frac{\delta^{(2)} 
\beta {\cal A}_{MF}}
{\delta \rho_{\alpha}(1)  
\delta \rho_{\beta}(2)}   \; .   
 \end{equation} 
 In the terminology of statistical field theory 
 $\widehat{C}_{MF, \; \alpha \beta}$ is minus the two-point proper vertex;
  it is related to the usual direct 
 correlation function $ c_{\alpha \beta}$ of the theory of liquids 
 by the relation \cite{Caillol2}
 \begin{equation}
  \widehat{C}_{MF, \; \alpha \beta}(1,2)  = c_{MF, \; \alpha 
    \beta}(1,2) - \frac{1}{\rho_{\alpha}(1)}
    \delta_{\alpha,\beta} \; \delta(1,2) \; .
  \end{equation} 
The strict convexity of the functionals $\log \Xi_{MF}[\{\nu_{\alpha} 
\}]$ and $\beta {\cal A}_{MF}[\{\rho_{\alpha} \}]$ guarantees the 
existence and unicity of the functions  $ G^{T }_{MF, \; \alpha \beta}$ 
and $\widehat{C}_{MF, \; \alpha \beta}$. Moreover the operators 
$ G^{ T }_{MF, \; \alpha \beta}$ and $-\widehat{C}_{MF, \; \alpha \beta}$ 
are both strictly positive and  $-\widehat{C}_{MF, \; \alpha \beta}$ is the
inverse of $ G^{ T }_{MF, \; \alpha \beta}$, i.e.

\begin{equation}
\label{OZ1}
- \widehat{C}_{MF, \; \alpha \beta}(1,2) \;
G^{T }_{MF, \;  \beta \gamma }(2,3) =  \delta_{\alpha, \gamma} \; 
\delta(1,3) \; ,
 \end{equation} 
which, when reexpressed in terms of the functions $h_{\alpha \beta}$ and 
$c_{\alpha \beta}$, takes the familiar form of the Orstein-Zernike 
equation\cite{Hansen}

\begin{equation}
\label{OZ2}
h_{MF, \; \alpha \beta}(1,2) =c_{MF, \; \alpha \beta}(1,2) +
h_{MF, \; \alpha \gamma}(1,3) \rho_{\gamma}(3) c_{MF, \; \gamma\beta}(3,2)
\; .
 \end{equation}
 $\widehat{C}_{MF, \; \alpha \beta}(1,2) $ is obtained readily from the 
 expression\ (\ref{AMF}) of the MF free energy functional  $\beta {\cal 
 A}_{MF}[\{\rho_{\alpha} \}]$ with the simple result
\begin{equation}
    \label{CMF2}
\widehat{C}_{MF, \; \alpha \beta}    (1,2)  =
     \widehat{C}_{HS, \;  \alpha \beta}(1,2) - \beta q_{\alpha} q_{\beta} 
     w_{\alpha \beta}(1,2) \; ,
 \end{equation} 
where $ \widehat{C}_{ HS, \; \alpha \beta}(1,2) $ is minus the exact two-point 
proper vertex of the reference HS fluid at the mean field densities $\{ 
\rho_{MF, \;  \alpha} \} $. Eq.\ (\ref{CMF2}) implies that for a 
homogeneous system and for $r_{12}\geq (\overline{\sigma}_{\alpha} + 
\overline{\sigma}_{\beta})$ the direct MF correlation functions 
reads as 
\begin{equation}
    \label{cMF}
    c_{MF, \; \alpha \beta}(r_{12})= c_{HS, \; \alpha \beta}(r_{12}) - \beta
    q_{\alpha} q_{\beta}/ r_{12}  \; ,
\end{equation}   
which is nothing but the RPA closure of the theory of
liquids.\cite{Hansen,Chand-RPA} 

The calculation of $G^{MF, \; T }_{ \beta \gamma }(1,2)$ is more 
involved. Our 
starting point will be the equation
\begin{eqnarray}
  G^{ T }_{MF, \;  \alpha \beta }(1,2) &=&
  \frac{\delta \rho_{MF, \; \alpha}[\{\nu_{\gamma}\}](1)}{\delta \nu_{\beta}(2)}
  \nonumber \\
  &=&\frac{\delta \rho_{HS, \; \alpha}
  [\{\overline{\nu}_{\gamma} + i \overline{\phi}_{\gamma}\}]
  (1)}{\delta \nu_{\beta}(2)} \; .
\end{eqnarray}    
Clearly $\rho_{HS, \; \alpha}
  [\{\overline{\nu}_{\gamma} + i \overline{\phi}_{\gamma}\}]$ depends 
  upon $\nu_{\beta}$ directly but also through the smeared mean 
  fields $\overline{\phi}_{\gamma}$; therefore one has
\begin{eqnarray}
  G^{T }_{MF, \; \alpha \beta  }(1,2) &=&  
  \left. \frac{\delta \rho_{HS, \; \alpha} [\{\overline{\nu}_{\gamma} 
  + i \overline{\phi}_{\gamma}\}](1)}{\delta 
  \nu_{\beta}(2)}\right \vert_{\overline{\varphi}} +
  \left. \frac{\delta \rho_{HS, \; \alpha}
  [\{\overline{\nu}_{\gamma} + i \overline{\phi}_{\gamma}\}]
  (1)}{\delta \overline{\varphi}(3)}\right \vert_{\nu_{\gamma}}
 \frac{\delta  \overline{\varphi}(3)}{\delta \nu_{\beta}(2)} 
 \nonumber \\
 &=&G^{ T }_{HS, \; \alpha \beta  }(1,2) + i \beta^{1/2}
 G^{T }_{MF, \; \alpha \gamma }(1,4) q_{\gamma} \tau_{\gamma}(4,3)
 \frac{\delta  \overline{\varphi}(3)}{\delta \nu_{\beta}(2)} \; . 
\end{eqnarray}     
Now, remarking that the stationarity condition\ (\ref{eqMF}) 
implies that
\begin{equation}
  \frac{\delta  \overline{\varphi}(3)}{\delta \nu_{\beta}(2)}=
  i \beta^{1/2} v_{c}(3,5) q_{\delta} \tau_{\delta} (5,6) 
  G^{ T }_{MF, \; \delta \beta}(6,2) \; ,
 \end{equation}   
one obtains
\begin{eqnarray}
    \label{zombi}
 G^{ T }_{MF, \; \alpha \beta  }(1,2)&=&G^{T }_{HS, \; \alpha \beta  
 }(1,2) \nonumber \\
 &- &\beta G^{T }_{HS, \; \alpha \gamma }(1,4) q_{\gamma} 
 \tau_{\gamma}(4,3) v_{c}(3,5) q_{\delta}\tau_{\delta}(5,6) 
G^{T }_{MF, \; \delta \beta  }(6,2)  \; ,
\end{eqnarray}      
which can be rewritten under a matricial form as
\begin{equation}
    \label{GMF1}
  \underline{G}^{ T }_{MF}(1,2)=
  \underline{G}^{T }_{HS}(1,2) - \beta \; 
  \underline{G}^{T }_{HS}(1,3) \; \underline{v}(3,4) \; 
  \underline{G}^{T }_{MF}(4,2) \; ,
 \end{equation}  
 where $ \underline{G}^{T }_{MF(HS)}(1,2)$ denotes the matrix of 
 elements $G^{T }_{MF(HS), \; \alpha \beta}(1,2)$ and
$ \underline{v}(1,2) $ that of elements $v_{\alpha \beta}(1,2)$
 (cf eq.\ (\ref{vab}))
The formal solution of eq.\ (\ref{GMF1}) is then
\begin{equation}
    \label{GMF2}
    \underline{G}^{T }_{MF}(1,2)=
    \left(  \underline{1}  + \beta \;  \underline{G}^{ T }_{HS} * 
    \underline{v} \right)^{-1} 
    \; * \; 
     \underline{G}^{ T }_{HS} \; ,
 \end{equation}  
where $ \underline{1}_{\alpha \beta}(1,2)=\delta_{\alpha \beta} 
\delta(1,2)$ is the unit operator and  the star "*" denotes a convolution.
 One easily checks that  $G^{T }_{MF, \; \alpha \beta}$ and 
$\widehat{C}_{MF, \; \alpha \beta}$ as given by eqs.\ (\ref{GMF2}) 
and\ (\ref{CMF2}) respectively do satisfy the OZ equation\ (\ref{OZ1}).

In the case of a homogeneous system eq.\ (\ref{GMF2}) can be 
simplified considerably. Let us work in Fourier space where we have

\begin{equation}
    \widetilde{ \underline{G}}^{ T }_{MF} (k)=
    \left(\underline{U} -\widetilde{\underline{P}}\left( k \right) 
    \right)^{-1} . \;
   \widetilde{ \underline{G}}^{T }_{HS} (k) \; ,
 \end{equation}  
 where $\underline{U}_{\alpha \beta}=\delta_{\alpha \beta}$ is the 
 unit matrix of rank $M \times M$ and
 $\widetilde{\underline{P}}( k )$ denotes the matrix of 
 elements
 \begin{equation}
     \label{matP}
     \widetilde{P}_{\alpha \beta}= - \beta \widetilde{v}_{c}(k) 
     \widetilde{G}^{T}_{HS, \; \alpha \gamma}(k) q_{\gamma}q_{\beta}
     \widetilde{\tau}_{\gamma}(k)  \widetilde{\tau}_{\beta}(k) 
     \; \; (\text{no sum over } \beta)
     \;.
  \end{equation}
 $\widetilde{\underline{P}}( k )$  has the remarkable property that
 \begin{equation}
     \label{prop}
     \widetilde{\underline{P}}^{2}(k)= -\beta \widetilde{v}_{c}(k) 
      \widetilde{G}^{T}_{HS, \; C}(k)  \widetilde{\underline{P}}(k) \; ,
 \end{equation}   
 which leads us to search the inverse of 
 $\underline{U} - \widetilde{\underline{P}}(k) $ under the form 
 $(\underline{U} - \widetilde{\underline{P}}(k))^{-1}= \underline{U} +\widetilde{\mu}(k) 
 \widetilde{\underline{P}}(k)$. The 
 identity $(\underline{U} + \widetilde{\mu}(k) \widetilde{\underline{P}}(k)) .
 (\underline{U} - 
\widetilde{ \underline{P}}(k))=\underline{1}$ 
when  combined with the property\ (\ref{prop}) implies that
 \begin{equation}
     \label{mu}
 \mu(k)= \frac{1}{1+ \beta \widetilde{v}_{c}(k) 
 \widetilde{G}^{T}_{HS, \; C}(k)} \; .
 \end{equation} 
Therefore 
 \begin{equation}
     \label{prop2}
   \left( \underline{U} - \widetilde{\underline{P}}(k)\right)^{-1} =
     \underline{U} + \frac{\widetilde{\underline{P}}(k) }{1+ \beta \widetilde{v}_{c}(k) 
 \widetilde{G}^{T}_{HS, \; C}(k)} \; ,
 \end{equation}    
 from which a simple expression for
 $ \widetilde{ G}^{T }_{MF, \; \alpha \beta} (k)$ is easily obtained 
 \begin{mathletters}
     \label{sy}
 \begin{eqnarray}
  \widetilde{ G}^{T }_{MF, \; \alpha \beta} (k)&=&
  \widetilde{ G}^{T }_{HS, \; \alpha \beta} (k) -
  \frac{\beta \widetilde{v}_{c}(k) \Gamma_{\alpha}(k)
  \Gamma_{\beta}(k)}{ 1 +\beta \widetilde{v}_{c}(k)
 \widetilde{ G}^{T }_{HS, \; C} (k) } \; ,\\
 \Gamma_{\alpha}(k)&=& \widetilde{ G}^{T }_{HS, \; \alpha \gamma} (k)
 q_{\gamma} \widetilde{\tau}_{\gamma}(k) \; .
  \end{eqnarray}   
  \end{mathletters}

\noindent  
  The expressions of the Fourier transforms of the
  usual pair correlations $h_{MF, \; \alpha \beta}$ 
  can then be deduced from eqs.\ (\ref{sy}). One finds
  \begin{equation}
      \label{lesh}
   \widetilde{ h}_{MF, \; \alpha \beta} (k)=
  \widetilde{ h}_{HS, \; \alpha \beta} (k) 
  -\frac{\Gamma_{\alpha}(k)}{\rho_{\alpha}}
  \frac{\Gamma_{\beta}(k)}{\rho_{\beta}}
  \frac{\beta \widetilde{v}_{c}(k)} 
  { 1 +\beta \widetilde{v}_{c}(k)
 \widetilde{ G}^{T }_{HS, \; C} (k) } \; .
\end{equation}  

 An important comment is in place here. As well 
known, the fact that the direct correlation functions\ (\ref{cMF}) 
behave as the Coulomb potential at large distances is sufficient to 
ensure that the SL rules are satisfied. \cite{Hansen,Blum} 
Therefore, the  $ h_{MF, \; \alpha 
\beta}(r)$, the Fourier transforms of which are given above, 
automatically satisfy to the SL sum rules\ (\ref{slab}).

 The two-point MF charge correlation  is obtained by  
 taking the 
 convolution of the two members of eq.\ (\ref{zombi}) with $q_{\alpha}
 q_{\beta} \tau_{\alpha}(1,1') \tau_{\beta}(2,2') $ which gives the equation 
 \begin{equation}
 G^{T }_{MF, \; C}(1,2) = G^{T }_{HS, \; C}(1,2)
 -\beta  G^{T }_{HS, \; C}(1,1') \; v_{c}(1',2') \; G^{T 
 }_{MF, \; C}(2',2) \; ,
 \end{equation}   
 the formal solution of which is 
 \begin{equation}
     \label{GCMF}
  G^{ T }_{MF, \; C}=\left( 1 + \beta   G^{T }_{HS, \; C}* v_{c}\right)^{-1}*
  G^{ T }_{HS, \; C} \; .
\end{equation}  
Therefore, for a homogeneous system, the Fourier transform of
$G^{T }_{MF, \; C}(r)$ has the simple expression
\begin{equation}
     \label{GCMF2}
 \widetilde{ G}^{T }_{MF, \; C} (k)=
 \frac{k^{2} \widetilde{ G}^{T }_{HS, \; C} (k)}{
 k^{2} + 4 \pi \beta \widetilde{ G}^{T }_{HS, \; C} (k)} \; .
 \end{equation}
 
 Finally the truncated pair correlation function of the KSSHE field 
 is obtained by using eq.\ (\ref{neq2T}) which gives us
 \begin{eqnarray}
    \label{GphiMF}  
   G^{T }_{MF, \; \varphi}(1,2) &=& v_{c}(1,2) -
   \beta v_{c}(1,1')  \;  G^{T }_{MF, \; C}(1',2')\;
    v_{c}(2',2)  \nonumber 
   \\
   &=&
  \left( 1 + \beta   G^{T }_{HS, \; C}* v_{c}\right)^{-1}(1,1') \;
  v_{c}(1',2) \; .
 \end{eqnarray}  
For an homogeneous fluid, one thus has in Fourier space 
\begin{equation}
  \label{GphiMF2}   
   \widetilde{G}^{ T }_{MF, \; \varphi}(k)=
      \frac{4 \pi}{
 k^{2} + 4 \pi \beta \widetilde{ G}^{T }_{HS, \; C} (k)} \; .
 \end{equation} 
\subsubsection{Application to the special primitive model}
\label{SPM}
In the SPM all the spheres have the same diameter 
$\sigma_{\alpha}=\sigma$ and we also assume that all the smearing 
distributions $\tau_{\alpha}$ are identical with the same $\tau$. 
With  these assumptions it is shown in app. \ref{apD} that the 
functions $G_{HS, \; C}^{(n) \; T}$ are nearly decoupled of the 
density correlations $G_{HS }^{(n) \; T}$ of the HS reference fluid.
For instance one has for $n=2$
\begin{equation}
\label{GHSC2}
    \widetilde{G}_{HS, \; C}^{T}(k)=\rho_{MF, \; \alpha} 
    q_{\alpha}^{2} \; 
    \widetilde{\tau}^{2}(k) \; ,
\end{equation}    
which entails considerable simplifications. Indeed the MF pair 
correlation functions take the simple form
\begin{equation}
    \label{tttt}
     \widetilde{h}_{MF, \; \alpha \beta}(k)=
       \widetilde{h}_{HS}(k)
       - \frac{4 \pi \beta q_{\alpha}q_{\beta}  \widetilde{\tau}^{2}(k) 
       }{k^{2} + \kappa_{MF}^{2}  \widetilde{\tau}^{2}(k) } \; ,
\end{equation}    
where $ \kappa_{MF}^{2}=4 \pi \beta \rho_{MF, \; \alpha} 
q_{\alpha}^{2}$ is the squared Debye number and $h_{HS}(r)$ the 
truncated pair correlation function of the HS fluid at the density 
$\rho_{MF}= \sum \rho_{MF, \; \alpha}$. 
Eq.\ (\ref{tttt}) when combined with the electroneutrality condition 
implies that the correlations of the (total) density  are equal to those 
of the HS reference fluid which rules out a liquid-vapor transition of the SPM 
at the MF level. Moreover the 
charge and KSSHE field correlations of the SPM  appear to be 
completely decoupled from the density fluctuations  since one finds
\begin{mathletters}
    \begin{eqnarray}
 \widetilde{G}_{MF, \; C}^{T}(k)&=& \frac{1}{4 \pi \beta}     
\frac{\kappa_{MF}^{2}  \widetilde{\tau}^{2}(k) k^{2}}{
 k^{2}+\widetilde{\tau}^{2}(k) \kappa_{MF}^{2} }
 \; , \\
 \widetilde{G}_{MF, \; \varphi}^{T}(k)&=&  
\frac{4 \pi}{ k^{2}+\widetilde{\tau}^{2}(k) \kappa_{MF}^{2} }   \; .
    \end{eqnarray}      
\end{mathletters}

\noindent
Note that for point charge distributions (i.e. 
$\+\widetilde{\tau}(k)=1$) the above functions reduce to the Yukawa 
potential, more precisely $$G_{MF, \; 
\varphi}^{T}(r)=\exp(-\kappa_{MF}\; r)/r  \; ,$$ and 
$$4 \pi \beta G_{MF, \; C}^{T}(r)=\kappa_{MF}^{2}\delta^{(3)}(\vec{r}) -\kappa_{MF}^{4}
\exp(-\kappa_{MF}\; r)/4 \pi r \; .$$.
\subsection{The mean field  free energy as an exact lower bound of the 
free energy }
\label{Bound} 
It can be shown, under quite general conditions,
that the logarithm of the 
grand-partition function $\log \Xi[\{ \mu_{\alpha} \}]$ is a convex functional of 
the local chemical  potentials even before the passage to the 
thermodynamic limit. Similarly, the exact
Kohn-Sham free energy $\beta {\cal A}[\{ \rho_{\alpha} \}]$
is  a convex functional of the local 
densities.
Moreover  
 $\log \Xi [\{ \mu_{\alpha} \}]$ and  ${\cal A}[\{ \rho_{\alpha} \}]$ 
 constitute a pair of Legendre transforms which can be expressed as
 \cite{Percus,Caillol2,Chayes}
  
\begin{mathletters}
\begin{eqnarray}
 \label{Lege1} 
 \beta {\cal A}[\{ \rho_{\alpha}\}] & =& \sup_{\{\nu_{\alpha} \}} 
 \left( \langle \rho_{\alpha}  \vert \nu_{\alpha} \rangle - 
 \log \Xi[\{\nu_{\alpha} \}] \right) \; ,  \\
 \label{Lege2} 
 \log \Xi [\{\nu_{\alpha} \}] &=& \sup_{\{ \rho_{\alpha} \}} 
 \left( \langle \rho_{\alpha}  \vert \nu_{\alpha} \rangle -  
 \beta {\cal A} \left[\{ \rho_{\alpha}\}\right] \right)
 \; .
\end{eqnarray}
\end{mathletters}

\noindent
Recall that the Young inequalities which follow directly from 
eqs\ (\ref{Lege1}),\ (\ref{Lege2}) say that
 \begin{equation}
\label{Young}  
\beta {\cal A}[\{ \rho_{\alpha}\}] + \log \Xi [\{\nu_{\alpha} \}] \geq 
 \langle \rho_{\alpha}  \vert \nu_{\alpha} \rangle \; \; (\forall \{ 
 \nu_{\alpha}\}, \; \forall \{ \rho_{\alpha}\} ) 
\; .
\end{equation} 
We have seen in sec.\ (\ref{MFfree}) that the MF functionals 
$\log\Xi_{MF}[\{ \mu_{\alpha} \}]$ and  ${\cal A}_{MF}[\{ \rho_{\alpha} 
\}]$  are also two convex functionals linked by a Legendre 
transform. We establish below some rigorous inequalities between the 
exact and  MF free energies and grand potentials which extend to 
ionic fluids  results  recently obtained  for neutral 
fluids.\cite{Caillol}

First, we notice that the fundamental eq.\ (\ref{basic}) rewritten as
\begin{equation}
    \Xi[ \{\nu_{\alpha}\}] =
    \left \langle \exp\left(
    \log \Xi_{HS}[ \{\nu_{\alpha}+\nu^{S}_{\alpha} + i \phi_{\alpha}\}] 
    \right) \right \rangle_{v_{c}}\; \; (\forall \{ 
 \nu_{\alpha}\} ) \; ,
\end{equation}     
can be inverted in order to give 
\begin{equation}
    \label{inv}
   \Xi_{HS}[ \{\nu_{\alpha}\}] =
    \left \langle \exp\left(
    \log  \Xi[ \{\nu_{\alpha}- \nu^{S}_{\alpha} + \phi_{\alpha}\}] 
    \right) \right \rangle_{v_{c}}\; \; (\forall \{ 
 \nu_{\alpha}\} ) \; .
\end{equation} 
Then, we apply Young inequality\ (\ref{Young}) to the RHS of eq.\ 
(\ref{inv}) which yields
\begin{eqnarray}\label{io}
   \Xi_{HS}[ \{\nu_{\alpha}\}] & \geq&
    \exp\left(  - \beta {\cal A}[\{ \rho_{\alpha}\}] +
     \langle \rho_{\alpha}  \vert \nu_{\alpha}- \nu^{S}_{\alpha} \rangle
    \right) \times \nonumber \\
    &\times &  \left \langle \exp\left( 
     \left \langle j \vert \varphi \right \rangle 
    \right) \right \rangle_{v_{c}} 
    \; \; (\forall \{ 
 \nu_{\alpha}\}, \; \forall \{ \rho_{\alpha}\} ) 
    \; , 
 \end{eqnarray}  
 where it follows from the definition\ (\ref{Phi}) of the smeared 
 field $\phi_{\alpha}$ that
\begin{equation}
    j(1)= \beta^{1/2} q_{\alpha} \tau_{\alpha}(1,1') 
    \rho_{\alpha}(1') \; .
 \end{equation}  
 The field average in the RHS of eq.\ (\ref{io}) is computed by 
 making use of Wick's theorem which yields
 \begin{equation}
 \left \langle \exp\left( 
     \left \langle j \vert \varphi \right \rangle 
    \right) \right \rangle_{v_{c}}  =
    \exp \left(  
    \frac{\beta}{2}  \left \langle \rho_{C} \vert
    v_{c}
    \vert \rho_{C} \right \rangle
    \right)  \; .
\end{equation} 
By inserting the above result into  eq.\ (\ref{io}) we obtain 
\begin{equation}
    \log \Xi_{HS}[ \{\nu_{\alpha}\}]  \geq
  - \beta {\cal A}[\{ \rho_{\alpha}\}]  
  + \langle \rho_{\alpha}  \vert \nu_{\alpha}- \nu^{S}_{\alpha} 
  \rangle +
  \frac{\beta}{2}  \left \langle \rho_{C} \vert
    v_{c}
    \vert \rho_{C} \right \rangle 
    \; \; (\forall \{ 
 \nu_{\alpha}\}, \; \forall \{ \rho_{\alpha}\} ) 
\; ,
\end{equation}   
which implies
\begin{eqnarray}
    \label{tuy}
 \beta {\cal A}[\{ \rho_{\alpha}\}]   & 
  \geq   & \sup_{\{\nu_{\alpha} \}} 
 \left[  \langle \rho_{\alpha}  \vert \nu_{\alpha}  \rangle
 -  \log \Xi_{HS}[ \{\nu_{\alpha}\}] 
 \right] \nonumber \\
 &+&  \frac{\beta}{2}  \left \langle \rho_{C} \vert
    v_{c}
    \vert \rho_{C} \right \rangle 
 -\langle \rho_{\alpha}  \vert \nu^{S}_{\alpha} 
  \rangle     \; \; (\forall \{ \rho_{\alpha}\} )  \; .
\end{eqnarray}
The $\sup$ in the RHS of eq.\ (\ref{tuy}) is the Legendre transform 
of the exact grand potential $\log \Xi_{HS}[ \{\nu_{\alpha}\}] $, 
i.e. the free energy  $\beta {\cal A}^{HS}[\{ 
\rho_{\alpha}\}] $.  It follows then from the expression\ (\ref{AMF}) 
of $\beta {\cal A}_{MF}$ that eq.\ (\ref{tuy}) may be 
recast under the form
\begin{equation}
    \label{bound1}
  \beta {\cal A}[\{ \rho_{\alpha}\}]    \geq   
\beta {\cal A}_{MF} [\{ \rho_{\alpha}\}]   \; \; (\forall \{ \rho_{\alpha}\} )  \; .
 \end{equation}   
 Therefore the MF free energy $\beta {\cal A}_{MF}[\{ 
 \rho_{\alpha}\}]$ is a rigorous lower bound of the exact GC free energy 
  $\beta {\cal A}[\{ \rho_{\alpha}\}]$. Moreover since   $\beta 
 {\cal A}_{MF}$ and $\log \Xi_{HS}$ are linked by a Legendre 
 transform (see eqs.\ (\ref{Leg1}) and (\ref{Leg2})) we also have
 \begin{eqnarray}
  \log \Xi_{MF}[ \{\nu_{\alpha}\}] &=&  \sup_{\{\rho_{\alpha} \}} 
  \left[ \langle \rho_{\alpha}  \vert \nu_{\alpha}  \rangle -
  \beta {\cal A}_{MF}[\{ 
 \rho_{\alpha}\}]  \right]
 \; \; (\forall \{ \nu_{\alpha}\} ) 
 \nonumber \\
 & \geq & 
 \sup_{\{\rho_{\alpha} \}} 
  \left[\left \langle \rho_{\alpha}  \vert \nu_{\alpha}  \right\rangle -
  \beta {\cal A}\left[\{ 
 \rho_{\alpha}\}\right]
  \right]
   \; \; (\forall \{ \nu_{\alpha}\} ) \; ,
\end{eqnarray}  
which can be, with help of eq.\ (\ref{Lege2}), rewritten as  
\begin{equation}
    \label{bound2}
 \log \Xi_{MF}[ \{\nu_{\alpha}\}] \geq
 \log  \Xi[ \{\nu_{\alpha}\}]
  \; \; (\forall \{ \nu_{\alpha}\} ) \; . 
   \end{equation}  
In other words $\log \Xi_{HS}[ \{\nu_{\alpha}\}]$ is a {\em rigorous 
upper bound} of the grand potential  $\log  \Xi[ \{\nu_{\alpha}\}]$.
Recall that, for a homogeneous system in the point-like limit,
the Debye-H\"{u}ckel approximation
for the pressure constitutes a {\em rigorous lower bound} for the pressure. This
result can also be proved with the help of 
the Sine-Gordon transform.\cite{Kennedy}

Let us specialize now to the homogeneous case. We denote by $\beta f(\{ 
\rho_{\alpha}\})$ the exact free energy per unit volume of the system. 
It is bounded from below by $\beta f_{MF}(\{ \rho_{\alpha}\})$ and 
 we have therefore
\begin{equation}
    \label{supf}
 \beta f(\{ \rho_{\alpha}\})  \geq \beta f_{HS}(\{ \rho_{\alpha}\})-
 \rho_{\alpha}  \nu^{S}_{\alpha} \; ,
\end{equation}
where the electroneutrality condition $\rho_{C}=0$ has been imposed. 
It follows from eq.\ (\ref{supf}) 
that the "best" or "optimized" MF free energy is obtained by minimizing the 
functional $K[\{ \tau_{\alpha}\}]= \rho_{\alpha}  
\nu^{S}_{\alpha} = \beta \rho_{\alpha} q_{\alpha}^{2} w_{\alpha 
\alpha}(0)/2 $ with respect to the variations of the smearing 
functions $\{ \tau_{\alpha}(\vec{r}) \}$. Since these distributions 
are normalized to unity we introduce $M$ Legendre parameters 
$\lambda_{\alpha}$ and define the functional
\begin{eqnarray}
    \overline{K}[\{ \tau_{\alpha}\}] &=& \frac{1}{2}
    \beta \rho_{\alpha} q_{\alpha}^{2} w_{\alpha 
\alpha}(0) - \lambda_{\alpha} Q_{\alpha}[\tau_{\alpha}]  \nonumber 
\; \\
Q_{\alpha}[\tau_{\alpha}] &=& \int_{r<\overline{\sigma}_{\alpha}}d^3 
\vec{r} \; \tau_{\alpha}(\vec{r}) \; .
 \end{eqnarray}   
 Now, we minimize $ \overline{K}[\{ \tau_{\alpha}\}] $ for a given 
 set of $\lambda_{\alpha}$. The constraints $Q_{\alpha}=1 \; \; 
( \forall \alpha $ ) will utimately serve to determine the values of the Legendre 
 parameters  $\lambda_{\alpha}$ .
 The conditions for an extremum of  $\overline{K}$ are 
 \begin{equation}
     \label{staK}
     \frac{\delta  \overline{K}[\{ \tau_{\alpha}\}] }
     {\delta \tau_{\gamma}(\vec{r})} = 0 \; \; (\forall \gamma \in (1, 
     \ldots,M), \; 
     \| \vec{r} \| \leq \overline{\sigma}_{\gamma}) \; .
 \end{equation}
Since $w_{\alpha \alpha}(r_{12}) \equiv 
w_{\alpha \alpha}(1,2)=\tau_{\alpha}(1,1')  \tau_{\alpha}(2,2') 
v_{c}(1',2')$ we have
 \begin{equation}
      \frac{\delta w_{\alpha \alpha}(0) }{\delta \tau_{\gamma}(\vec{r})}
      =2 \delta_{\alpha \gamma} V_{\gamma}(\vec{r}) \; ,
 \end{equation} 
 where $ V_{\gamma}$ is the electrostatic potential created by the 
 charge distribution $\tau_{\gamma}$ (i.e. $V_{\gamma}(r_{12}) \equiv
 V_{\gamma}(1,2)=\tau_{\gamma}(1,2') v_{c}(2',2)$).
 The stationarity condition\ (\ref{staK}) takes thus the form
 \begin{equation}
     V_{\gamma}(\vec{r})= \lambda_{\gamma} \; \;
     (\forall \gamma \in (1, 
     \ldots,M), \; 
     \| \vec{r} \| \leq \overline{\sigma}_{\gamma}) \; .
 \end{equation}   
 The potential $ V_{\gamma}$ created by the distribution $ \tau_{\gamma}$ 
 must therefore be constant inside the sphere of radius 
 $\overline{\sigma}_{\alpha}$. From elementary electrostatics we 
 conclude that $ \tau_{\gamma}(\vec{r})$ must be an uniform surface 
 distribution of charge of radius  $\overline{\sigma}_{\alpha}$. In 
 order to satisfy the constraint $Q_{\gamma}=1$ one must have
 \begin{equation}
     \label{tauopti}
     \tau_{\gamma}(\vec{r}) = \delta( \| \vec{r} \| - \overline{\sigma}_{\gamma})
     \frac{1}{\pi \sigma_{\gamma}^2} \; .
\end{equation}    
Note that the solution\ (\ref{tauopti}) is indeed a minimum of 
$\overline{K}$ since its second order functional derivative with 
respect to $\tau_{\alpha}$ is the Coulomb potential $v_{c}(1,2)$ which 
is a positive operator. 
The simple form of the Fourier transform of  $\tau_{\gamma}(\vec{r})$ 
which is found to be
\begin{equation}
    \label{tauk}
    \widetilde{\tau}_{\gamma}(k)= \frac{\sin(k 
    \overline{\sigma}_{\gamma})}
    {k  \overline{\sigma}_{\gamma}}
\end{equation}   
allows us to compute explicitly  the pair potentials $w_{\alpha \beta}(r)$ 
as the inverse Fourier transforms of the functions $\widetilde{w}_{\alpha 
\beta}(k)= 4 \pi  \widetilde{\tau}_{\alpha}(k) 
\widetilde{\tau}_{\beta}(k)/k^2$ with the result
\begin{mathletters}
    \label{wMF}
\begin{eqnarray}
    r > \overline{\sigma}_{\alpha} +  \overline{\sigma}_{\beta} 
    &\Rightarrow& w_{\alpha \beta}(r)= \frac{1}{r} \\
\overline{\sigma}_{\alpha} - \overline{\sigma}_{\beta} 
<r< \overline{\sigma}_{\alpha} +  \overline{\sigma}_{\beta}
& \Rightarrow & w_{\alpha \beta}(r)= 
\frac{\sigma_{\alpha} + \sigma_{\beta} -r 
}{\sigma_{\alpha}\sigma_{\beta}} +\frac{1}{2r}
- \frac{1}{4 r} \frac{\sigma_{\alpha}^{2} + \sigma_{\beta}^{2}}
{\sigma_{\alpha}\sigma_{\beta}} \\
0 < r < \overline{\sigma}_{\alpha} - \overline{\sigma}_{\beta} 
& \Rightarrow & w_{\alpha \beta}(r)= \frac{2}{\sigma_{\alpha}} \; ,
\end{eqnarray}    
\end{mathletters}

\noindent
where it was assumed that $\sigma_{\alpha} \geq \sigma_{\beta}$. As a 
subproduct of these equations we obtain the self-energies 
$\nu_{\alpha}^{S}=\beta q_{\alpha}^{2}/\sigma_{\alpha}$ and thus the 
expression of the optimized MF free energy
\begin{equation}
    \beta f_{MF}(\{ \rho_{\alpha}\})=
    \beta f_{HS}(\{ \rho_{\alpha}\}) - \frac{\beta 
    q_{\alpha}^{2} }{ \sigma_{\alpha}} \rho_{\alpha} \; . 
\end{equation}    
Note that the corresponding excess internal energy per unit volume 
$\beta
u_{MF} = \partial(  \beta f_{MF})/\partial \beta = - \beta \rho_{\alpha}
q_{\alpha}^2/\sigma_{\alpha}$ coincides with the Onsager lower
bound.\cite{Onsager} 
We note that the MF approximation is thermodynamically inconsistent in 
the sense that the MF energy $u_{MF}$ is obviously not equal to that which can 
be obtained by the integration of $g_{MF, \; \alpha \beta}(r) w_{\alpha \beta}(r)$.

Reporting the expressions\ (\ref{wMF}) of the pair potentials $w_{\alpha 
\beta}(r)$ in the equations\ (\ref{CMF}) one obtains the MF direct 
correlation functions. We shall discuss only the case of the SPM 
where all the radii $\sigma_{\alpha}$ are equal to the same $\sigma$. 
In this case one finds
\begin{mathletters}
\label{cbest}
\begin{eqnarray}
   r > \sigma & \Rightarrow & c_{MF, \; \alpha \beta}(r)=c_{HS}(r) 
   -\beta  q_{\alpha} q_{\beta}/r    \\
   0<r<\sigma & \Rightarrow & c_{MF, \; \alpha \beta}(r)=c_{HS}(r) 
   -\beta  q_{\alpha} q_{\beta} \frac{2 \sigma -r }{\sigma^2} \; , 
\end{eqnarray}
\end{mathletters}

\noindent
where $c_{HS}(r) $ is the exact direct correlation function of the HS 
fluid at the density $\rho=\sum_{\alpha} \rho_{\alpha}$.
The expressions\ (\ref{cbest}) of the $c_{\alpha \beta}$ are very similar 
to that obtained in the framework of the MSA approximation i.e.
\cite{Blum,Waisman}
\begin{mathletters}
 \label{cMSA}
\begin{eqnarray}
   r > \sigma & \Rightarrow & c_{MSA, \; \alpha \beta}(r)=c_{HS}^{PY}(r) 
   -\beta  q_{\alpha} q_{\beta}/r    \\
   \label{cMSAb}
   0<r<\sigma & \Rightarrow & c_{MSA, \; \alpha \beta}(r)=c_{HS}^{PY}(r) 
   -\beta  q_{\alpha} q_{\beta} \frac{2 B \sigma - B^{2}r }{\sigma^2} \; , 
\end{eqnarray}
\end{mathletters}

\noindent
 where $c_{HS}^{PY}(r) $ denotes the direct correlation function of the 
HS fluid in the Percus-Yevick approximation which is an excellent 
approximation of the exact $c_{HS}(r) $ especially at low or moderate 
densities. \cite{Hansen} The main difference between eqs.\ 
(\ref{cbest}) and.\ (\ref{cMSA}) is the occurrence of a parameter 
$B$ in the MSA solution.
Clearly the MF and MSA solutions coincide for $B=1$. Since we have
\cite{Blum,Waisman}
\begin{equation}
    B= \frac{\kappa^2 \sigma^2}{ (\kappa \sigma)^2 + \kappa \sigma 
    -\kappa \sigma (1 + 2 \kappa \sigma )^{1/2}} \; ,
 \end{equation}  
 where $\kappa$ is the Debye number, 
 it happens only in the limit $\kappa \to \infty$. For a finite 
 $\kappa$, $B(\kappa)$ is positive  and comprised between $0$ and 
 $1$. Obviously the electrostatic contribution to $c_{\alpha 
 \beta}^{MSA}(r)$ in the core (i.e. the second term in the RHS of 
 eq.\ (\ref{cMSAb})) may be seen as the interaction energy 
 $w_{\alpha \beta}(r)$ of two surface distributions of charge of 
 equal radii $\overline{\sigma}'= \overline{\sigma}/B $.
 However, since 
$ B <1 \Rightarrow \overline{\sigma}' >\overline{\sigma}$, this 
interaction energy is not equal to $1/r$ outside the core. 
Consequently the MSA solution cannot be interpreted as a MF KSSHE theory 
except in the limiting case $B=1$.

\section{The Gaussian approximation}
\label{Gauss}
\subsection{The general case}
\label{GaussI}
Let us  define the Gaussian approximation of the KSSHE theory
in the following way. \cite{Ma} We write $
 \varphi= \overline{\varphi} + \delta \varphi $ where 
 $\overline{\varphi} $ is the mean field solution  and $\delta \varphi$ a real 
 scalar field and we expand functionally
the action ${\cal H}[\varphi]$ (cf eq.\ (\ref{H}) ) up to second order 
in $\delta \varphi$ around the MF solution.
In this way the exact action ${\cal H}[\varphi]$ is 
replaced by an approximate action ${\cal H}_{G}[\varphi]$ given by
\begin{equation}
    \label{HG}
  {\cal H}_{G}[\varphi] = {\cal H}[\overline{\varphi}] +
  \frac{1}{2} \langle \delta \varphi \vert \Delta^{-1} \vert \delta 
  \varphi \rangle  \; ,
\end{equation}
where the terms linear in $\delta \varphi $ are absent  as a 
consequence of the stationarity condition\ (\ref{statio}), and
the inverse of 
the propagator $\Delta$ is given by:
\begin{equation}
    \label{Delta}
    \Delta^{-1}(1,2)=v_{c}^{-1}(1,2) + \beta G^{T}_{HS, \;C}(1,2) \; .
\end{equation}
 A comparison of eq.\ (\ref{Delta})
with eq.\ (\ref{GphiMF}) yields the expected result \cite{Ma}
\begin{equation}
    \label{Delta2}
    \Delta(1,2)= G^{T}_{MF, \; \varphi}(1,2)\; .
\end{equation}
In the Gaussian approximation, the grand partition 
function  is therefore given by
\begin{equation}
    \Xi_{G}\left[ \left\{ \nu_{\alpha} \right\} \right]=   
    \exp \left(  -{\cal H}[\overline{\varphi}]
    \right)
    \frac{\int {\cal D} \varphi \; \exp \left( -\frac{1}{2}
    \langle \varphi  \vert \Delta^{-1} \vert \varphi \rangle 
    \right)}
    {\int {\cal D} \varphi \; \exp \left( -\frac{1}{2}
    \langle \varphi  \vert v_{c}^{-1} \vert \varphi \rangle 
    \right)
    }\; .
\end{equation}    
From now we specialize to the case of a homogeneous system for which 
an explicit expression of $\Xi_{G}$ can be obtained, i.e.
\begin{equation}
    \label{yj1}
    \Xi_{G}\left[ \left\{ \nu_{\alpha} \right\} \right]=  
 \exp \left(  -{\cal H}[\overline{\varphi}]
    \right)   \frac{{\cal N}_{\Delta}}{{\cal N}_{v_{c}}} \; ,
 \end{equation}    
 where the normalization constants ${\cal N}_{\Delta}$ and ${\cal 
 N}_{v_{c}}$ are defined according to eq.\ (\ref{moyv}). A well known 
 property of functional integrals of Gaussian functionals gives us
 \cite{Ma,Dowrick,Parisi}
 \begin{eqnarray}
     \label{yj2}
     \log \frac{{\cal N}_{\Delta}}{{\cal N}_{v_{c}}}&=& \frac{\Omega}{2}
     \int d^{\vee}k
     \log \frac{\widetilde{\Delta}(k)}
     {\widetilde{v}_{c}(k)}\;  \nonumber \\
     &=& -  \frac{\Omega}{2}
      \int d^{\vee}k
      \log\left(  1 + \frac{4 \pi \beta}{k^2} \widetilde{G}^{T}_{HS, \; 
      C}(k) \right) \; .
  \end{eqnarray}     
   Taking the logarithm of eq.\ (\ref{yj1}) and taking into account the 
   relation\ (\ref{yj2}) we thus obtain the pressure of the ionic 
   solution at the Gaussian level:
   \begin{eqnarray}
       \label{PG}
       \beta P_{G}\left[ \left\{ \nu_{\alpha} \right\} \right] &=&
       \beta P_{HS}\left[ \left\{ \overline{\nu}_{\alpha}+ i 
       \beta^{1/2} \overline{\varphi} \right\} \right]  \nonumber \\
  & -&  \frac{1}{2}
      \int d^{\vee}k  
       \log\left(  1 + \frac{4 \pi \beta}{k^2} 
       q_{\alpha} q_{\beta} \widetilde{\tau}_{\alpha}(k)
       \widetilde{\tau}_{\beta}(k) 
       \widetilde{G}^{T}_{HS, \; \alpha \beta}(k)\right) \; ,
 \end{eqnarray}      
where we have expanded the HS charge correlation function in oder to 
make explicit the dependence of the result upon the smearing 
functions $  \widetilde{\tau}_{\alpha}(k) $. Note that the HS 
truncated pair 
correlation functions $\widetilde{G}^{T}_{HS, \; \alpha \beta}(k)$ must be 
evaluated at the chemical potentials $\left\{ \overline{\nu}_{\alpha}+ i 
       \beta^{1/2} \overline{\varphi} \right\}$. As a short examination 
reveals, the integral in the RHS of eq.\ (\ref{PG}) is convergent 
at large $k$ for any reasonable smearing function $\tau_{\alpha}$ 
(i.e. surface, volume distributions etc) but diverges for point 
charge distributions ($ \widetilde{\tau}_{\alpha}(k)=1$).

Some words on the field correlation functions at the Gaussian level. 
Following Ma \cite{Ma} we define
\begin{equation}
    G^{(n)}_{G, \; \varphi}(1, \ldots,n)=
    \frac{\int {\cal D} \varphi \; \exp
    \left( 
    - {\cal H}_{G}[\varphi] \right)
    \varphi(1) \ldots \varphi(n) }{\int {\cal D} \varphi \; \exp \left( 
    - {\cal H}_{G}[\varphi] \right)}
\end{equation}    
which entails the relations
\begin{mathletters}
    \begin{eqnarray}
        \label{Ga}
 \langle \delta \varphi \rangle_{{\cal H}_{G}} &=&0  \; ,  \\
        \label{Gb}
 G^{(2)\; T}_{G, \; \varphi}(1,2)&=& \Delta(1,2) 
 \equiv  G^{(2)\; T}_{MF, \; \varphi}(1,2)    
 \; , \\        
        \label{Gc}
 G^{(n)\; T}_{G, \; \varphi}(1, \ldots,n)&=&0 \text{ for } n\geq 3 \; . 
    \end{eqnarray}
\end{mathletters} 

\noindent   
In particular, eq.\ (\ref{Ga}) says that the number densities in the Gaussian 
approximation coincide with their MF values. The Legendre transform 
of $\beta P_{G}$ is therefore easily obtained and reads as
   \begin{equation}
       \label{fG}
       \beta f_{G}\left[ \left\{ \rho_{\alpha} \right\} \right] =
       \beta f_{HS}\left[ \left\{ \rho_{\alpha} \right\} \right]  
       -\rho_{\alpha}\nu_{\alpha}^{S}
   + \frac{1}{2}
      \int d^{\vee}k  \; 
       \log\left(  1 + \frac{4 \pi \beta}{k^2} 
       \widetilde{G}^{T}_{HS, \; C}(k)\right) \; ,
 \end{equation}     
 where the truncated charge correlation function $G^{T}_{HS, \; 
 C}$ of the reference HS fluid must be evaluated at the
  densities $\left\{ \rho_{\alpha} \right\} $. 
 Replacing the self energy $\nu_{\alpha}^{S}$ by its expression\ 
 (\ref{nuS}), $\beta f_{G}$ can be recast under the form
  \begin{eqnarray}
    \label{fGbis}
  \beta f_{G}\left[ \left\{ \rho_{\alpha} \right\} \right] &=&
    \beta f_{HS}\left[ \left\{ \rho_{\alpha} \right\} \right] +
    \frac{\beta}{2} \rho_{\alpha} \rho_{\beta} 
    q_{\alpha} q_{\beta}\int d^{3}\vec{r} \; 
    h_{HS, \; \alpha \beta} (r) w_{\alpha \beta}(r) \; \nonumber \\
    &+& \frac{1}{2} \int d^{\vee}k  \; 
      \left\{ \log\left(  1 + \frac{4 \pi \beta}{k^2} 
       \widetilde{G}^{T}_{HS, \; C}(k)\right) 
      \;  -\;\frac{4 \pi \beta}{k^2} 
       \widetilde{G}^{T}_{HS, \; C}(k)
      \right\} \; ,
 \end{eqnarray} 
 which will be discussed in next section.
\subsection{Comparison with the RPA}
\label{RPA}
 As in the case of neutral fluids,\cite{Caillol} the expression\ 
 (\ref{fGbis}) of the free 
 energy at the Gaussian level coincides with that obtained in the 
 framework of the random phase approximation (RPA) of the theory of 
 liquids.
 The proof of this equivalence is however more tricky in the 
 present case.  
 The RPA theory can be summarized as follows.\cite{Hansen,Chand-RPA,Outhwaite}
 The connected two-body 
 correlation functions are given by
 \begin{equation}
     \label{GRPA}
     G_{RPA, \; \alpha \beta}^{T}(1,2)=
      G_{HS, \; \alpha \beta}^{T}(1,2)  + \rho_{\alpha} \rho_{\alpha} 
      {\cal C}_{\alpha \beta}(1,2) \; ,
 \end{equation}
 where the "renormalized" potentials $ {\cal C}_{\alpha \beta}(1,2) $ 
 are given in Fourier space by
 \begin{equation}
     \label{CRPA}
\rho_{\alpha} \rho_{\alpha} \widetilde{{\cal C}}_{\alpha \beta} (k)=  \left.
\widetilde{\underline{P}}(k)\; . \; \left( \underline{U} -
\widetilde{\underline{P}}(k) \right)^{-1} \; . \; 
\widetilde{\underline{G}}_{HS}^{T} (k)\right\vert_{\alpha \beta} \; ,
\end{equation}
where the matrix $\widetilde{\underline{P}}(k)$ the elements of which 
are given by
\begin{equation}
    \label{matPRPA}
    \widetilde{P}_{\alpha \beta}=-\beta \widetilde{G}_{HS,\; \alpha 
    \beta}^{T} (k) \widetilde{v}_{\alpha \beta}(k) \; 
\end{equation}    
turns out to coincide with that we defined at eq.\ (\ref{matP}).  
Note that, in the diagrammatic formulation of Chandler and 
Andersen, $\underline{G}_{HS}^{T}$ is the matrix of the so-called 
"hypervertices".\cite{Chand-RPA} 
It is easy to check that  the 
properties\ (\ref{prop}) and\ (\ref{prop2}) of the matrix 
$\widetilde{\underline{P}}(k)$ imply that 
\begin{eqnarray}
 \left.
\widetilde{\underline{P}}(k)\; . \; \left( \underline{U} -
\widetilde{\underline{P}}(k) \right)^{-1} \; . \; 
\widetilde{\underline{G}}_{HS}^{T} (k)\right\vert_{\alpha \beta}&=&
\frac{1}{1 + \beta \widetilde{v}_{c}(k) \widetilde{\underline{G}}_{HS, \; 
C}^{T} (k)} \left. \widetilde{\underline{P}}(k)\; . \; 
\widetilde{\underline{G}}_{HS}^{T} (k) \right\vert_{\alpha \beta} 
\; , \nonumber \\
&=& -\frac{\beta \widetilde{v}_{c}(k) \Gamma_{ \alpha}(k)\Gamma_{
 \beta}(k)}
{1 + \beta \widetilde{v}_{c}(k) \widetilde{\underline{G}}_{HS, \; 
C}^{T} (k)} \; ,
\end{eqnarray}
and therefore, cf eq.\ (\ref{sy}), that
\begin{equation}
    G_{RPA, \; \alpha \beta}^{T}(1,2)=G_{G, \; \alpha \beta}^{T}(1,2) \; .
\end{equation}  

 The expression found by Chandler and Andersen for the  RPA free energy 
 reads as \cite{Chand-RPA,Outhwaite}
 \begin{eqnarray}
    \label{fRPA}
  \beta f_{RPA}\left[ \left\{ \rho_{\alpha} \right\} \right] &=&
    \beta f_{HS}\left[ \left\{ \rho_{\alpha} \right\} \right] +
    \frac{\beta}{2} \rho_{\alpha} \rho_{\beta} \int d^{3}\vec{r} \; 
    h_{HS, \; \alpha \beta} (r) v_{\alpha \beta}(r) \; \nonumber \\
    &+&  \frac{1}{2} \int d^{\vee}k  \; 
    \left\{ \tr \widetilde{\underline{P}}(k) +
    \log \det \left(\underline{U}-\widetilde{\underline{P}}(k)\right)
    \right\} \; .
\end{eqnarray}  
We first note that it follows from the definition\ (\ref{matP}) of 
$\widetilde{\underline{P}}(k)$ that
\begin{equation}
    \label{tr}
   \tr \widetilde{\underline{P}}(k)=- \beta \widetilde{v}_{c}(k)  
\widetilde{\underline{G}}_{HS, \; 
C}^{T} (k) \; .
\end{equation}
The determinant in the RHS of eq.\ (\ref{fRPA}) can be explicitly 
computed from a classical formula of linear algebra.\cite{Smirnov}
 Recall that for 
a $n \times n$ matrix $\underline{a}$ we have
\begin{equation}
    \label{smy1}
   \det \left(a_{ij}+ x\delta_{ij} \right)= x^n + S_{1}x^{n-1}
    + \ldots + S_{n}x 
   +x^n \; ,
\end{equation}   
where $S_{k}$ denotes the sum of the minors of order $k$, i.e.
\begin{equation}
      \label{smy2}
  S_{k}= \sum_{q_{1}< \ldots <q_{k}} \det \left(
  a_{q_{i} q_{j}}
  \right) \; .
 \end{equation}
When applied to the matrix  $\widetilde{\underline{P}}(k)$ (with 
$x=1$) the 
formula\ (\ref{smy1}) yields the simple result that 
\begin{equation}
    \label{dettr}
  \det \left(\underline{U}-\widetilde{\underline{P}}(k)\right)  =1 
  -\tr \widetilde{\underline{P}}(k) \; .
\end{equation}    
Indeed, since $\widetilde{P}_{\alpha \beta}(k)$ is of the form 
$s_{\alpha}(k)t_{\beta}(k)$ then all the minors in eq.\ (\ref{smy1})
 vanish except 
$S_{1}$ which identifies with the trace of the matrix
$\widetilde{P}_{\alpha \beta}(k)$. It follows then
from the expression\ (\ref{fRPA}) and the intermediate 
results\ (\ref{tr}) and \ (\ref{dettr}) 
that the RPA  free energy is identical to the Gaussian KSSHE free 
energy:
\begin{equation}
 \beta f_{RPA}\left[ \left\{ \rho_{\alpha} \right\} \right] =   
 \beta f_{G}\left[ \left\{ \rho_{\alpha} \right\} \right] \; .
\end{equation}    
It must be stressed that in the RPA the pair potential $v_{\alpha 
\beta}$ between two ions is arbitrary in the core (i.e. for 
$0<r< \overline{\sigma}_{\alpha} +\overline{\sigma}_{\beta}$).
The simple forms 
that we have obtained obtained for the pair correlations and the 
free energy (cf eqs.\ 
(\ref{sy}) and\ (\ref{fG})) are valid {\it only} if $v_{\alpha 
\beta}$ may be identified with the interaction of two smeared 
distributions of charge. In that case, the relevant mathematical 
property which yields important simplifications 
is that $\widetilde{v}_{\alpha \beta}(k)$ may be written as 
$\widetilde{v}_{\alpha \beta}(k)=\beta \widetilde{v}_{c}(k) 
q_{\alpha}\widetilde{\tau}_{\alpha}(k) 
q_{\beta}\widetilde{\tau}_{\beta}(k)$ for 
all the pairs $(\alpha, \beta)$ 
, i.e. as the product of two functions $s_{\alpha}(k)$ and  $ s_{\beta}(k)  $.

 In the optimized RPA (ORPA) the pair potentials in the core are 
chosen in such a way to ensure that the radial pair correlations 
$g_{ORPA, \; \alpha \beta}(r)$ vanish for 
$0<r< \overline{\sigma}_{\alpha} +\overline{\sigma}_{\beta}$.\cite{Chand-RPA}
 Chandler and 
Andersen have shown that this condition corresponds to an extremum 
of $\beta f_{RPA}$ considered as a functional of the core potentials. 
There is no analytical solution to this variational problem as far as 
the author knows, and the solution must be seeked numerically. It is 
however quite certain that this solution cannot be interpreted as 
the interaction of two smeared charges as suggested by the discussion
on the MSA 
integral equation of sec.\ (\ref{Bound}) (the MSA is identical 
with the ORPA if the HS reference fluid is described in the framework 
of the PY theory). Therefore, with smeared electrostatic potentials one
 can never
ensure that the $g_{G, \; \alpha \beta}(r)$  vanish in the core.
\subsection{The SPM}
\label{SPM1}
We conclude this section by specializing to the case of the SPM. In this model
all the hard spheres have the same diameter $\sigma$ and 
all the smearing distributions $\tau_{\alpha}$ are 
equal to the same $\tau$. It 
follows from the  expression\ (\ref{GHSC2})
of $G_{HS, \;C}^{(2) \; T}$ that the Gaussian free 
energy of the SPM takes then the very simple form :
\begin{equation}
\label{freeSPM}
     \beta f_{G}\left[ \left\{ \rho_{\alpha} \right\} \right] =
    \beta f_{HS}\left[ \left\{ \rho_{\alpha} \right\} \right] 
   +\frac{1}{2} \int d^{\vee}k  \; 
      \left\{ \log\left(  1 + \frac{\kappa^{2}}{k^2} 
      \widetilde{\tau}^{2}(k) \right) 
      \;  -\frac{\kappa^{2}}{k^2} 
      \widetilde{\tau}^{2}(k)
      \right\} \; .
\end{equation} 
Several comments on eq.\ (\ref{freeSPM}) are in order at this point.

(a) As it is well known,\cite{Outhwaite} and as a direct calculation will show, eq.\
(\ref{freeSPM})
reduces to the Debye-H\"uckel  free energy $\beta f_{DH}= 
\beta f_{HS}-\kappa^{3}/12 \pi$ in the limit of 
point-like  distributions (i.e. when $ \widetilde{\tau}(k)=1$ for all $k$).
Recall that in the point-like limit, the MF free energy diverges. 
Therefore the first order term in the loopwise expansion of the free 
energy makes the final result finite.

(b) The expression\ (\ref{freeSPM}) shows no sign of a hidden "RPA catastrophe" 
\cite{Wheeler} which would
be the case if the argument of the logarithm happened to be negative
for some $k$. This happy circumstance is a consequence of the regularization
by "smearing" of the Coulomb potential which was adopted in this paper. 
Other types of regularization can however
lead to such a RPA catastrophe. For instance, adopting the
Weeks-Chandler-Andersen (WCA)
recipe\cite{Weeks} one could set the potential constant in the core,
i.e.
$v_{\alpha \beta}(r)=q_{\alpha}q_{\beta}/\sigma$ for $0<r<\sigma$.
Clearly, one has in fact
$v_{\alpha \beta}(r)=q_{\alpha}q_{\beta} w(r) \; (\forall r)$ where
$w(r)$ can be interpreted as the electric potential created by the surface
distribution of charge $\tau_{S}(r)=\delta(r-\overline{\sigma})/\pi \sigma^2$.
Thus $w(1,2)=\tau_{S}(1,1')v_{c}(1',2)$.
As a consequence, one must replace the term
$\widetilde{\tau}^{2}(k)$ by 
$\widetilde{\tau}_{S}(k)=\sin(k \overline{\sigma})/k \overline{\sigma}$
({\em not} squared) in eq.\ (\ref{freeSPM}),
with the annoying consequence that the
argument of the $\log$ can become negative for some $k$ yielding a
"RPA catastrophe". 
The mathematical origin of this catastrophe is that,
within the WCA framework,  the
energy is not a positive definite quadratic form and the KSSHE transform is
ill-defined. It has been suggested that
this instability of the RPA theory could possibly be related to the
order-disorder transition of the lattice-version of the RPM.
\cite{Ciach-Stell,Oksanna} Note that this
transition is not present in the continuous version of the model although the
RPA catastrophe for the WCA-RPA free energy is still there. In all cases, 
one cannot be comfortable with a theory the ability of which to predict a 
possible transition depends on the way the regularization of
the interaction is performed.

\section{A two-loop order calculation }
\label{twoloop}
\subsection{The cumulant expansion}
\label{cumu}
We adopt the same decomposition $\varphi= \overline{\varphi} + \delta \varphi $ 
of the KSSHE field as that considered in the  section on the 
Gaussian approximation
but, this time, we do not 
truncate the action at the second order in $\delta \varphi $ and 
consider rather  the full  functional Taylor expansion of
${\cal H}[\varphi]$ about the saddle point
\begin{mathletters}
    \label{Hgene}
    \begin{eqnarray}
        {\cal H}[\varphi]&=& {\cal H}[\overline{\varphi}]
        +\frac{1}{2} \langle \delta\varphi\vert
        \Delta^{-1}_{\overline{\varphi}}\vert \delta\varphi\rangle
        + \Delta{\cal H}[\delta\varphi] \; , \\
\Delta{\cal H}[\delta\varphi]& =& \sum_{n=3}^{\infty}
\frac{1}{n!} 
 {\cal H} ^{(n)}[\overline{\varphi}](1, \ldots, n)
 \delta\varphi(1) \ldots \delta\varphi(n)  \; ,
    \end{eqnarray}
\end{mathletters}

\noindent    
where  the free propagator
$\Delta_{\overline{\varphi}}$ defined at eq.\ (\ref{Delta}) must be 
evaluated at the saddle point $\overline{\varphi}$  and 
the integral kernels are given by
\begin{eqnarray}
    \label{kernel}
{\cal H} ^{(n)}[\overline{\varphi}](1, \ldots, n) &=&- (i\beta^{1/2})^{n}
 q_{\alpha_{1}} \ldots  q_{\alpha_{n}} \;
 \tau_{\alpha_{1}}(1,1')  \ldots  \tau_{\alpha_{n}}(n,n')  \times 
 \nonumber \\
 &\times &
 G_{HS, \; \alpha_{1} \ldots \alpha_{n}}^{(n) \; T}\left[ \left\{ 
 \overline{\nu}_{\alpha} + i \overline{\phi}_{\alpha}
  \right\}\right](1', \ldots, n') \; , \nonumber \\
  &\equiv& - (i\beta^{1/2})^{n} 
  G_{HS, \; C}^{(n) \; T}\left[ \overline{\varphi}
 \right] (1, \ldots, n) \; .
 \end{eqnarray} 
With these notations, the grand partition function can be recast under 
the form
\begin{equation}
    \label{Xiloop}
    \Xi\left[ \left\{ \nu_{\alpha} \right\}\right] =
    \exp \left( - {\cal H}\left[\overline{\varphi}\right] \right) \; 
 \frac{{\cal N}_{\Delta_{\overline{\varphi}}}}{{\cal N}_{v_{c}}} \; 
 \langle  \exp \left(  - \Delta{\cal H}[\delta\varphi] \right)
 \rangle_{\Delta_{\overline{\varphi}}} \; ,
\end{equation} 
yielding the cumulant expansion\cite{Ma,Amit,Zinn}
\begin{equation}
    \label{cumul}
    \log  \Xi =
    -{\cal H}\left[\overline{\varphi}\right] 
    +\log  \frac{{\cal N}_{\Delta_{\overline{\varphi}}}}{{\cal 
    N}_{v_{c}}} \; 
    -  \langle \Delta{\cal H}
    \rangle_{\Delta_{\overline{\varphi}}}
    +\frac{1}{2!} \langle \Delta{\cal H}^{2}
    \rangle_{\Delta_{\overline{\varphi}}}^{T} \; -  
      \frac{1}{3!} \langle \Delta{\cal H}^{3}
    \rangle_{\Delta_{\overline{\varphi}}}^{T} \; +\ldots    
\end{equation}    
The two first terms in the RHS of eq.\ (\ref{cumul}) correspond 
the MF and Gaussian approximations respectively, the other terms are 
Gaussian averages which can be computed with the help of Wick's 
theorem. Reordering these additional terms in powers of some small 
parameter yields various types of expansions.
 Low fugacity and high temperature expansions of the pressure and the 
free energy of the RPM where  obtained in this way in papers I and II,
yielding to a 
rediscovery of old results derived years ago by Mayer graphs expansion
technics. \cite{Mayer,Stell-Lebo} In both cases the "small parameter" 
has a physical origin and thus each order of the expansion should be 
independent of the pair potentials in the core and therefore of the 
smearing functions $\tau_{\alpha}$. This point was 
checked carefully in I and II  for the RPM
by considering surface distributions of radius 
$\overline{a}<\overline{\sigma}$ and by showing that the first terms 
of the fugacity and temperature expansions were indeed independent 
of $\overline{a}$.

Here we consider  a loopwise expansion. The small parameter $\lambda$ 
of this expansion is defined by rewriting \cite{Amit,Zinn}
\begin{eqnarray}
\label{lambda}
   \delta \varphi & \rightarrow & \lambda^{1/2}  \delta \varphi \; , 
   \nonumber \\
 {\cal H}[\varphi]  & \rightarrow & \frac{1}{\lambda}  
 {\cal H}[ \lambda^{1/2} \varphi]
 \; , \nonumber\\
 \log \Xi & \rightarrow & \lambda \log \Xi  \; .
 \end{eqnarray}
 An expansion of eq.\ (\ref{cumul}) in powers of $\lambda$ yields the loopwise 
 expansion of $\log \Xi$; at the end of the calculation one set  
 $\lambda=1$. 
 The "small" parameter $\lambda (\equiv \hbar\; !!)$ 
 is not a physical parameter but serves mainly to keep track of 
 different classes of Feynman diagrams; therefore each term of the 
 loop expansion can {\em a priori} depend on the smearing functions
 $\tau_{\alpha}$. One can only hope that the larger the number of terms is 
 retained in the expansion the "smaller"  this dependence will be. In a 
 sense, the fact that the MF free energy diverges for point charge 
 distributions but remains finite in the Gaussian (one-loop) 
 approximation confirms this hope. Henceforth we shall retain for 
  $\tau_{\alpha}$
 the distributions\ (\ref{tauk}) which lead to the best MF free energy. 
 Finally, we note that it follows from eqs.\ (\ref{basic}),\ (\ref{Phi})
 and\ (\ref{lambda}) that
 a high-temperature expansion of the loop-expansion  of $\log \Xi$
 at order $\lambda^{k}$ obviously yields  the correct high-temperature
 expansion  of $\log \Xi$ at order $\beta^{k}$.
\subsection{The two-loop expansion of $\log \Xi$}
\label{twoloopXi}
\begin{figure}
\begin{center}
\epsfxsize=3.7 truein
\epsfbox{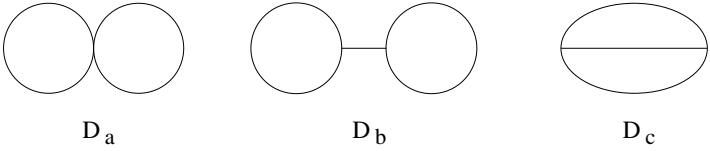}
 \caption{Two-loop diagrams contributing to $\log \Xi$. The lines 
 represent the bare propagator $ \Delta_{\overline{\varphi}}$.}
\label{D1}
\end{center}
\end{figure}

The loop expansion of $\log \Xi$ for the general Hamiltonian\ 
(\ref{Hgene}) can be found in the literature.\cite{Amit,Zinn} Taking 
into account the form\ (\ref{kernel}) of the kernels ${\cal H} ^{(n)}$ one 
finds, at the two-loop order
\begin{eqnarray}
    \label{Xi2loop}
    \log \Xi \left[ \left\{ \nu_{\alpha} \right\}\right] &=&
    \log \Xi_{MF} \left[ \left\{ \nu_{\alpha} \right\}\right] +
    \lambda \log \frac{{\cal N}_{\Delta_{\overline{\varphi}}}}{{\cal 
    N}_{v_{c}}} \nonumber \\
    &+&
    \frac{\lambda^2\beta^2}{8} G_{HS, \;C}^{(4) \; 
    T}[\overline\varphi]
    (1,2,3,4) \; 
    \Delta_{\overline{\varphi}}(1,2)  \Delta_{\overline{\varphi}}(3,4) 
  \; \;\; ( \equiv D_{a}) \; \nonumber \\
    &- &  \frac{\lambda^2\beta^3}{8} 
   G_{HS, \;C}^{(3) \; T}[\overline\varphi](X_{1}, X_{2},X_{3}) \; 
   G_{HS, \;C}^{(3) \; T}[\overline\varphi](Y_{1}, Y_{2},Y_{3}) \;
    \nonumber \times \\
   &\times&
     \Delta_{\overline{\varphi}}(X_{1}, X_{2})  
     \Delta_{\overline{\varphi}}(Y_{1}, Y_{2}) 
     \Delta_{\overline{\varphi}}(X_{3}, Y_{3})  \;\;\; 
       ( \equiv D_{b}) \; \nonumber \\
     &-&\frac{\lambda^2\beta^3}{12} 
  G_{HS, \;C}^{(3) \; T}[\overline\varphi](X_{1}, X_{2},X_{3}) \; 
   G_{HS, \;C}^{(3) \; T}[\overline\varphi](Y_{1}, Y_{2},Y_{3}) \; 
   \nonumber \times \\
   &\times&
 \Delta_{\overline{\varphi}}(X_{1}, Y_{1})  
     \Delta_{\overline{\varphi}}(X_{2}, Y_{2}) 
     \Delta_{\overline{\varphi}}(X_{3}, Y_{3})  \;\;\;  ( \equiv D_{c}) 
     \;  \nonumber \\
     &+&
     {\cal O}(\lambda^3)  ,
\end{eqnarray}
where the diagrams corresponding to the two-loop order terms of the RHS of 
eq.\ (\ref{Xi2loop}) have been sketched in fig.\ (\ref{D1}).

Some remarks.
In general the loop-expansion\ (\ref{Xi2loop}) does not converge
notably in the critical region as a consequence of the singular
behavior of the correlation functions.\cite{Amit,Zinn}
Here however, we deal with a non critical
field $\varphi$ the correlation functions of which were shown
to be short range in section\ \ref{SL};
eq.\ (\ref{Xi2loop}) is thus expected to be an asymptotic expansion.
The second remark is that 
the expression \ (\ref{Xi2loop}) is horribly intricate in the general case since, 
{\em a priori},
it involves  the $2,\;3$-body and $4$-body density 
correlation functions of the reference HS fluid. However it turns out that 
eq.\ (\ref{Xi2loop}) can be considerably simplified for the 
SPM since, in this case, only the $2$-body correlation functions  survive, 
yielding tractable expressions.  

Before specializing to the case of the SPM let us introduce some notations. All 
the thermodynamic quantities of interest will be expanded in powers 
of $\lambda$. We shall note, for intance for the density of species 
$\alpha$
\begin{equation}
    \rho_{\alpha}=\sum_{n=0}^{\infty}\lambda^n  \rho_{\alpha}^{(n)} 
    \; ,
\end{equation}  
with, of course $\rho_{\alpha}^{(0)} \equiv \rho_{MF, \; \alpha}
=\rho_{HS, \; \alpha}[\{ \overline{\nu}_{\alpha} + i q_{\alpha}
\overline{\varphi}\}]$. 
As it is well known, the two-loop order expansion of the free energy requires the 
densities at the one-loop order only. \cite{Amit,Zinn} We have checked 
in the general case that the one-loop order correction to the densities 
preserves the electroneutrality; the detailed calculation is made in 
appendix\ (\ref{apC}).

\begin{figure}
\begin{center}
\epsfxsize=3.7 truein
\epsfbox{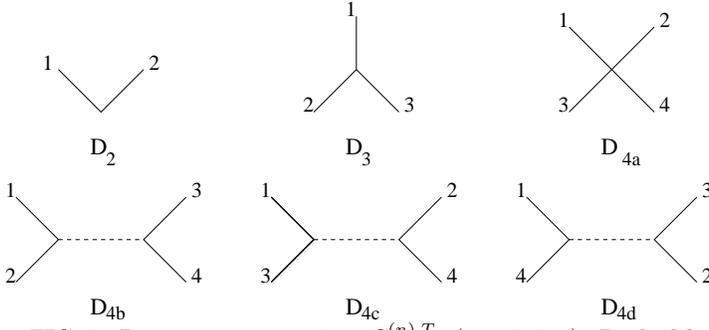}
 \caption{Diagrams representing $G_{HS,\; C}^{(n) \; T}$ ($n=2,3,4)$. 
 Dashed lines represent $h_{HS}(1,2)$ and solid lines the smearing function
  $\tau(1,2)$.}
\label{D2}
\end{center}
\end{figure}

Henceforth we consider only the homogeneous SPM. All the ions have the same 
diameter $\sigma$ and all the smearing functions are equal to 
$\widetilde{\tau}(k)=\sin(k \overline{\sigma})/k \overline{\sigma}$. 
We show in app.\ \ref{apD} that the charge neutrality condition at 
the MF level ($\rho_{C}^{(0)}=0$) kills most of the terms involved in the 
expressions of  $G_{HS, \;C}^{(n) \; T}[\overline{\varphi}]$ 
 and that the only surviving terms are
\begin{eqnarray}
    \label{kernelSPM}
    G_{HS,\; C}^{(2) \; T}[\overline{\varphi}](1,2) &=&
    \rho_{\alpha}^{(0)} q_{\alpha}^2 \; \tau(1,1')\tau(2,1') \; \; 
    \; (\equiv D_{2}) \nonumber \\   
 G_{HS,\; C}^{(3) \; T}[\overline{\varphi}](1,2,3)   &=&
 \rho_{\alpha}^{(0)} q_{\alpha}^3\; \tau(1,1')\tau(2,1') 
    \tau(3,1') 
    \; (\equiv D_{3}) \nonumber \\   
  G_{HS,\; C}^{(4) \; T}[\overline{\varphi}](1,2,3,4)   &=&   
  \rho_{\alpha}^{(0)}q_{\alpha}^4\; \tau(1,1')\tau(2,1') 
    \tau(3,1')   \tau(4,1')    \; (\equiv D_{4a}) \nonumber \\  
    &+&  \left( \rho_{\alpha}^{(0)} q_{\alpha}^2 \right)^{2} 
    \times \{ 
    \nonumber \\
  &\tau&(1,1')\tau(2,1') h_{HS}(1',2') \tau(2',4)\tau(2',3) 
   \; (\equiv D_{4b}) + \nonumber \\  
  &\tau&(1,1')\tau(3,1') h_{HS}(1',2') \tau(2',4)\tau(2',2) 
   \; (\equiv D_{4c}) + \nonumber \\  
    &\tau&(1,1')\tau(4,1') h_{HS}(1',2') \tau(2',2)\tau(2',3) 
   \; (\equiv D_{4d}) \} \; ,
\end{eqnarray}
where $h_{HS}(r)\equiv g_{HS}(r)-1$ is the pair distribution of the 
reference HS 
fluid at the total MF density $\rho^{(0)}=\sum_{\alpha} \rho_{\alpha}^{(0)}$.
Diagrammatic representations of  the various terms of the RHS of 
eq.\ (\ref{kernelSPM}) have been sketched in fig.\ (\ref{D2}).

Replacing either algebraically or graphically the kernels
 $G_{HS,\; C}^{(n) \; T}[\overline{\varphi}]$ by 
their expressions\ (\ref{kernelSPM}) in eq.\ (\ref{Xi2loop}) yields 
the expression for the pressure $\beta P=\log \Xi /\Omega$ at the 
two-loop order, i.e.  $P=P^{(0)} + \lambda P^{(1)} + \lambda^2 
P^{(2)}+{\cal O}( \lambda^3)$ with of course
\begin{equation}
      \label{P0loop}
    \beta P^{(0)} \left[ \left\{ \nu_{\alpha} \right\} \right]=  
    \beta P_{MF}\left[ \left\{ \nu_{\alpha} \right\} \right] \; ,
\end{equation}
and 
\begin{equation}
      \label{P1loop}
    \beta P^{(1)} \left[ \left\{ \nu_{\alpha} \right\} \right]= -
  \frac{1}{2} \int d^{\vee}k \; \log\left(1 + 
     \frac{\kappa^{(0) \;2}}{k^{2}} \widetilde{\tau}^{2}(k) \right) \; , 
\end{equation}
( where $\kappa^{(0) \;2}\equiv \kappa_{MF}^{2}=4 \pi \beta \rho_{\alpha}^{(0)}\; q_{\alpha}^2$ 
is the squared Debye number at the 0-loop order)  
as we already know, and 

\begin{eqnarray}
    \label{P2loop}
     \beta P^{(2)}\left[ \left\{ \nu_{\alpha} \right\} \right] &=&
     \frac{\beta^{2}}{8} \Big\{ \rho_{\alpha}^{(0)} q_{\alpha}^4 \; 
     \left[ \Delta_{\tau}^{(0)}(0) \right]^2  \;\;\; (\equiv 
     D_{1}) \nonumber \\
     &+& \left[ \Delta_{\tau}^{(0)}(0) 
     \right]^2\left[\rho_{\alpha}^{(0)}q_{\alpha}^2 \right]^2  
     \int d^{3}\vec{r} \; 
     h_{HS}(r)
     \;\;\; (\equiv 
     D_{2}) \nonumber \\
     &+& 2 \left[\rho_{\alpha}^{(0)} q_{\alpha}^2 \right]^2
     \int d^{3}\vec{r} \; h_{HS}(r)  \left[\Delta_{\tau}^{(0) }(r)\right]^{2}
     \Big\} 
   \;\;\; (\equiv 
     D_{3})\; \} \nonumber \\   
     &-&   \frac{\beta^{3}}{8} \left[\rho_{\alpha}^{(0)}q_{\alpha}^3 
     \right]^{2}  \left[\Delta_{\tau}^{(0)}(0) \right]^{2} 
     \widetilde{\Delta}_{\tau}^{(0)} (0)  \;\;\; (\equiv 
     D_{4})\;  \nonumber \\   
     &-&   \frac{\beta^{3}}{12} \left[\rho_{\alpha}^{(0)} q_{\alpha}^3 
     \right]^2  
     \int d^{3}\vec{r} \;  \left[\Delta_{\tau}^{(0) } (r) 
     \right]^{3}
     \;\;\; (\equiv 
     D_{5})\; ,
\end{eqnarray}  
where the "smeared" 
propagator $ \Delta_{\tau}^{(0)}$ which enters the RHS of eq.\ (\ref{P2loop})
is given by the convolution
\begin{equation}
    \label{Deltataur}
 \Delta_{\tau}^{(0)}(1,2)= \tau(1,1')    
 \Delta_{\overline{\varphi}}(1',2') \tau(2',2) 
 \end{equation}
 or, in Fourier space
 \begin{equation}
    \label{Deltatauk}
 \widetilde{\Delta}_{\tau}^{(0)} (k)= \frac{4 \pi 
 \widetilde{\tau}^{2}(k)}{
 k^{2} + \kappa^{(0) \; 2} \;  \widetilde{\tau}^{2}(k)} \; .
 \end{equation}
We have already met the function  $\Delta_{\tau}^{(0)}(r)$ (under the name
$X_{\tau}(r)$) in papers I 
and II. Recall its high temperature, or low $\kappa^{(0)}$,  
behavior 
  \begin{mathletters}
  \label{Xtau}
  \begin{eqnarray}
  r\leq\sigma &\Rightarrow &  
  \Delta_{\tau}^{(0)}(r)= \frac{2}{\sigma} -\frac{r}{\sigma^2}
  -\kappa^{(0)} + {\cal O}(\kappa^{(0) \;2})
    \; ,\\
   r\geq\sigma &\Rightarrow &  
    \Delta_{\tau}^{(0)}(r)=  
    \frac{\sinh^2(\kappa^{(0)}\overline{\sigma})}
    { (\kappa^{(0)}\overline{\sigma})^2} \frac{\exp( - \kappa^{(0)}r)}{r}
    + {\cal O}(\kappa^{(0) \;2})
    \; .
  \end{eqnarray}
  \end{mathletters}
  
\begin{figure}
\begin{center}
\epsfxsize=3.7 truein
\epsfbox{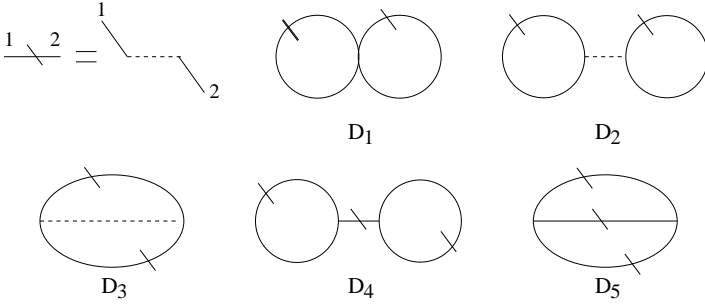}
 \caption{Two-loop diagrams contributing to the pressure (cf eq.\
 (\ref{P2loop})). The smeared
 propagator $\Delta_{\tau}^{(0)}(1,2)$ has been represented by a solid line plus
 a slash. As for the  diagrams of fig.\ (\ref{D2})
 the dashed lines represent $h_{HS}(1,2)$.}
\label{D3}
\end{center}
\end{figure}

\noindent  
Diagrammatic representations of  the various terms of the RHS of 
eq.\ (\ref{P2loop}) have been sketched in fig.\ (\ref{D3}). 

\subsection{The two-loop expansion of the free energy}
\label{twoloopf}
As it is well known from statistical field theory, the reducible 
diagram $D_{b}$ of fig.\ 
(\ref{D1}) should disappear after a Legendre transform of the pressure
yielding a simple expression for the free energy. \cite{Amit,Zinn} However 
in statistical field theory one is interested with the Legendre 
transform of $\beta P$ with respect to the mean value of the field 
$<\varphi>_{{\cal H}}$; by contrast, in our case, the quantity of interest 
is rather
the Helmoltz free energy $\beta f$ defined as the Legendre Transform of the 
pressure with respect to the $M$ chemical potentials $\nu_{\alpha}$ 
(cf eqs.\ (\ref{Lege1}), (\ref{Lege2})). These two Legendre 
transforms do not coincide  {\em a priori} and a direct calculation of
 $\beta f$ at the two-loop order is therefore necessary.

We have first to compute the densities at the one-loop order since the terms  
$\lambda^{2} \rho_{\alpha}^{(2)} $ and of higher order do not 
contribute to 
$\beta f$ at order ${\cal O}(\lambda^{2})$ (included) as a consequence of the 
stationarity condition  (this point will emerge
in subsequent developments).
Eqs.\ (\ref{P0loop}) and\ (\ref{P1loop}) entail that 
\begin{eqnarray}
    \label{ro2loop}
    \rho_{\alpha}&=& \frac{\partial \beta P\left[ \left\{ \nu_{\alpha} 
    \right\}\right] }{\partial \nu_{\alpha}} \nonumber \\
    &=&
    \rho_{\alpha}^{(0)} -\frac{\lambda}{2} \frac{\Delta_{\tau}^{(0)}(0)}
    {4 \pi} \frac{\partial \kappa^{(0) \; 2} }{\partial \nu_{\alpha}} +
    {\cal O}(\lambda^2)
    \; ,
\end{eqnarray}
where we recall that $  \rho_{ \alpha}^{0}= \rho_{HS, \; \alpha}
[\{ \overline{\nu}_{\alpha} + i \beta^{1/2}q_{\alpha} 
\overline{\varphi}  \}]$ is the 0-loop order (or MF) density of 
species $\alpha$. We note that
\begin{equation}
    \frac{\partial \kappa^{(0) \;2} }{\partial \nu_{\alpha}}=
    4 \pi \beta \; q_{\gamma}^{2} \left( \partial
    \rho_{\gamma}^{(0)}/ \partial \nu_{\alpha}\right)=
    4 \pi \beta \; q_{\gamma}^{2} \widetilde{G}_{MF, \; \alpha 
    \gamma}^{T} (k=0) \; ,
\end{equation}    
where $\widetilde{G}_{MF, \; \alpha \gamma}^{T} $ can be computed from 
eq.\ (\ref{tttt}). After some algebra  one finds finally
\begin{equation}
    \label{ro1}
  \rho_{\alpha}^{(1)}= -\frac{\beta \Delta_{\tau}(0)}{2} \left(
    \left[\rho_{\gamma}^{(0)} q_{\gamma}^{2}\right]  \; 
 \widetilde{h}_{HS}(0) \;   \rho_{\alpha}^{(0)} +
  \rho_{\alpha}^{(0)} q_{\alpha}^{2} 
 - \frac{ \left[\rho_{\gamma}^{(0)} q_{\gamma}^{3} \right] }
 { \left[\rho_{\gamma}^{(0)} q_{\gamma}^{2} \right] }\; 
  \rho_{\alpha}^{(0)} q_{\alpha} 
  \right) \; .
\end{equation} 
(no summation over $\alpha$ in the RHS).
We have already pointed out that the electroneutrality is satisfied 
at the MF level (i.e. $ \rho_{\alpha}^{(0)} q_{\alpha}=0$) and we 
note with satisfaction that eq.\ (\ref{ro1}) implies that it is 
also satisfied at the one-loop order (i.e.
 $ \rho_{\alpha}^{(1)} q_{\alpha}=0$; see also appendix C for a thorough
 treatment in the general case).
As a subproduct of eq.\ (\ref{ro1}) we obtain the one-loop order 
corrections to the total density and  Debye number
\begin{mathletters}
\begin{eqnarray}
  \rho^{(1)}   &=&  -\frac{\beta \Delta_{\tau}^{(0)}  (0)}{2}\;
   \left[\rho_{\gamma}^{(0)} q_{\gamma}^{2}\right] 
  \left(  1 +  \rho^{(0)} \widetilde{h}_{HS}(0) \right) \; \\
  \kappa^{2 \; (1)}&\equiv & 4 \pi \beta \rho_{\alpha}^{(1)} 
  q_{\alpha}^{2} \nonumber \\
  \label{ka1}
  &=& -2 \pi \beta^{2} \Delta_{\tau}^{(0)}(0)
  \left(   \left[\rho_{\gamma}^{(0)} q_{\gamma}^{2}\right]^2 \;  \widetilde{h}_{HS}(0)
  + \left[\rho_{\gamma}^{(0)} q_{\gamma}^{4}\right] - \frac{\left[\rho_{\gamma}^{(0)}
  q_{\gamma}^{3}\right]^{2}}{\left[\rho_{\gamma}^{(0)} q_{\gamma}^{2}\right]} 
  \right) \; .
\end{eqnarray}  
\end{mathletters}

\noindent
The free energy at the two-loop order is obtained by reexpressing
\begin{equation}
    \beta f = \rho_{\alpha} \nu_{\alpha} -
    \beta P^{(0)} -\lambda \beta P^{(1)}- \lambda^{2} \beta P^{(2)}
    +{\cal O}(\lambda^{3})
\end{equation}
in terms of the densities
\begin{eqnarray}
    \rho_{\alpha}&= &\rho_{\alpha}^{(0)} + \Delta \rho_{\alpha} 
    \nonumber \\
    \Delta \rho_{\alpha} &=& \lambda \rho_{\alpha}^{(1)} +
    \lambda^{2} \rho_{\alpha}^{(2)}  +{\cal O}(\lambda^{3}) \; .
\end{eqnarray}   
Let us first evaluate
\begin{equation}
    \label{bout1}
   \rho_{\alpha} \nu_{\alpha} -\beta P^{(0)}\left[
   \left\{\nu_{\alpha} \right\}\right] =
  \beta f^{(0)}\left[  \left\{\rho^{(0)}_{\alpha} \right\}\right] 
  + \nu_{\alpha} \Delta \rho_{\alpha}  \; .
\end{equation}  
We have already obtained the leading term of $\beta f$ (i.e. at the 
MF or 0-loop order) in sec\ (\ref{MFfree}) as a functions of the 
densities, i.e. for the homogeneous system considered here
\begin{equation}
   \beta f^{(0)}\left[  \left\{\rho_{\alpha} \right\}\right]  \equiv  
   \beta f_{MF}\left[  \left\{\rho_{\alpha} \right\}\right]=
   \beta f_{ HS}\left[  \left\{\rho_{\alpha} \right\}\right] - 
   \rho_{\alpha}\nu_{\alpha}^{S} \; .
\end{equation}      
At this point we remark that
\begin{equation}
     \label{hu0}
  \beta f^{(0)}\left[  \left\{\rho_{\alpha} +\Delta \rho_{\alpha} 
   \right\}\right]   =
  \beta f^{(0)}\left[  \left\{\rho_{\alpha}   \right\}\right]  + 
  \nu_{\alpha} \Delta \rho_{\alpha}  -\frac{1}{2\Omega} 
  \langle \Delta \rho_{\alpha} \vert \widehat{C}^{(0)}_{\alpha 
  \beta} \vert  \Delta \rho_{\beta} \rangle+ {\cal O}(\lambda^{3}) \; ,
\end{equation}    
where $\widehat{C}^{(0)}_{\alpha \beta}$ is the two-body MF proper vertex.
In the case of the SPM the expression\ (\ref{CMF2}) of 
$\widehat{C}^{(0)}_{\alpha 
\beta}$ takes the form
\begin{equation}
    \label{hu1}
 \widehat{C}^{(0)}_{\alpha \beta}(1,2)= c_{HS}^{(0)}(r_{12})- 
 \frac{\delta_{\alpha \beta}}{\rho^{(0)}_{\alpha}}   \delta(1,2) \; ,
\end{equation} 
where $c_{HS}^{(0)}$ is the usual direct correlation function of the 
HS fluid at the density $ \rho^{(0)}$. Reporting the expression\ 
(\ref{hu1}) of $\widehat{C}^{(0)}_{\alpha \beta}$ in eq.\ (\ref{hu0}) 
and taking into account the electroneutrality condition at the MF 
level one finds
\begin{equation}
 \beta f^{(0)}\left[  \left\{\rho_{\alpha} +\Delta \rho_{\alpha} 
  \right\}\right]   =
  \beta f^{(0)}\left[  \left\{\rho_{\alpha}   \right\}\right]  + 
  \nu_{\alpha} \Delta \rho_{\alpha}  -
  \frac{\lambda^{2}}{2} \; \widetilde{c}_{HS}^{(0)}(0) \rho^{(1) \; 2}
  +\frac{\lambda^{2}}{2} \sum_{\alpha}\frac{\rho_{\alpha}^{(1)\; 2}}
  {\rho_{\alpha}^{(0)}}
  + {\cal O}(\lambda^{3}) \; .
\end{equation} 
Reporting now the expression\ (\ref{ro1}) of $\rho_{\alpha}^{(1)}$ in the 
above equation one  obtains
\begin{eqnarray}
 \beta f^{(0)}\left[  \left\{\rho_{\alpha} +\Delta \rho_{\alpha}  
 \right\}\right]  & =&
  \beta f^{(0)}\left[  \left\{\rho_{\alpha}   \right\}\right]  + 
  \nu_{\alpha} \Delta \rho_{\alpha}  \nonumber \\
  &+&
  \frac{\lambda^2 \beta^{2}}{8} \; 
  \Delta_{\tau}^{(0)\; 2}(0)
  \left\{ \left[ \rho_{\alpha}^{(0)}q_{\alpha}^{2} \right]^{2}
   \widetilde{h}_{HS}^{(0)}(0)
  + \rho_{\alpha}^{(0)}q_{\alpha}^{4}  - 
  \frac{\left[ \rho_{\alpha}^{(0)}q_{\alpha}^{3} \right]^{2}}
  {\rho_{\alpha}^{(0)}q_{\alpha}^{2} }
  \right\}
  + {\cal O}(\lambda^{3}) \; ,
\end{eqnarray}
where we have made use of the OZ equation for the reference HS fluid 
$\widetilde{c}_{HS}^{(0)}(0)=
\widetilde{h}_{HS}^{(0)}(0)/(1+ \rho^{(0)}\widetilde{h}_{HS}^{(0)}(0))$. The 
first contribution\ (\ref{bout1}) of $ \beta f$ can thus be recast 
under the form
\begin{eqnarray}
    \rho_{\alpha} \nu_{\alpha} -\beta P^{(0)}\left[
   \left\{\nu_{\alpha} \right\}\right] &=& \beta f^{(0)}\left[  
   \left\{\rho_{\alpha}   \right\}\right]  \nonumber \\
   &- &  \frac{\lambda^2 \beta^{2}}{8} \; 
   \Delta_{\tau}^{2}(0)
 \left\{ \left[ \rho_{\alpha}q_{\alpha}^{2} \right]^{2} \widetilde{h}_{HS}(0)
  + \rho_{\alpha}q_{\alpha}^{4}  - 
  \frac{\left[ \rho_{\alpha}q_{\alpha}^{3} \right]^{2}}
  {\rho_{\alpha}q_{\alpha}^{2} }
  \right\}
  + {\cal O}(\lambda^{3}) \; ,  
\end{eqnarray}    
where the smeared propagator $ \Delta_{\tau}$ as well as the pair 
correlation $h_{HS}$ of the HS fluid  can be evaluated at the total 
density $\rho=\sum_{\alpha} \rho_{\alpha}$ (rather than $\rho^{(0)}$)
at this order in $\lambda$. 

The second step in the calculation of $\beta f$ is to reexpress $\beta 
P^{(1)}$ in terms of the densities $\rho_{\alpha}$. We note that 
\begin{equation}
    \lambda \beta P^{(1)} = 
    -\frac{\lambda}{2} \; \int d^{\vee}k \; 
    \log\left(
    1 + \frac{\kappa^{(0)\; 2}}{k^{2}} \widetilde{\tau}^{2}\left( k 
    \right)
    \right) \; 
\end{equation}    
which shows that $\beta P^{(1)}$ is a  function of the sole  squared Debye 
number 
$\kappa^{2}$. Defining the increment $\Delta \kappa^2=  \kappa^2-
\kappa^{(0)\; 2}$ and performing a Taylor expansion of 
$\beta P^{(1)}$ around $\kappa^{(0)\;2}$ yields

\begin{equation}
   \lambda \beta P^{(1)}= - \frac{\lambda}{2} \int d^{\vee}k \; 
    \log\left(
    1 + \frac{\kappa^{2}}{k^{2}} \widetilde{\tau}^{2}\left( k 
    \right)
    \right) 
    +\frac{\lambda}{8 \pi} \Delta_{\tau}(0) \Delta \kappa^2 
    + {\cal O}(\lambda^{3}) \; ,
\end{equation}   
which, after substitution of the expression of $\Delta \kappa^2$ at 
order ${\cal O}(\lambda)$ (cf eq.\ (\ref{ka1})  allows us to write
\begin{eqnarray}
  \lambda \beta P^{(1)}&=&   - \frac{\lambda}{2} \int d^{\vee}k \; 
    \log\left(
    1 + \frac{\kappa^{2}}{k^{2}} \widetilde{\tau}^{2}\left( k 
    \right)
    \right) \nonumber \\ 
    &-& \frac{\lambda^2 \beta^{2}}{4} \; 
   \Delta_{\tau}^{2}(0)
 \left\{ \left[ \rho_{\alpha}q_{\alpha}^{2} \right]^{2} \widetilde{h}_{HS}(0)
  + \rho_{\alpha}q_{\alpha}^{4}  - 
  \frac{\left[ \rho_{\alpha}q_{\alpha}^{3} \right]^{2}}
  {\rho_{\alpha}q_{\alpha}^{2} }
  \right\}
  + {\cal O}(\lambda^{3}) \; , 
\end{eqnarray}  
where, once again, the smeared propagator $ \Delta_{\tau}$ as well as the pair 
correlation $h_{HS}$ of the HS fluid  can be  evaluated at the total 
density $\rho=\sum_{\alpha} \rho_{\alpha}$ (rather than $\rho^{(0)}$)
at this order in $\lambda$. 
  
It remains to reexpress $\beta P^{(2)}$ in terms of the densities 
$\rho_{\alpha}$, which is done readily, and to gather all the 
intermediate results. After doing this job, one finds that the free energy
at the second order in the loop expansion reads as 
\begin{eqnarray}
    \label{jjjjo}
    \beta f \left[  \left\{\rho_{\alpha} \right\}\right] &=&
     \beta f_{HS} \left[  \left\{\rho_{\alpha} \right\}\right] +
  \frac{1}{2} \int d^{\vee}k \; 
   \left( \log\left(
    1 + \frac{\kappa^{2}}{k^{2}} \widetilde{\tau}^{2}\left( k 
    \right) \right) -  \frac{\kappa^{2}}{k^{2}} \widetilde{\tau}^{2}\left( k 
    \right) \right) \nonumber \\
   & -& \frac{\beta^{2}}{4} \left[ \rho_{\alpha}q_{\alpha}^{2} \right]^{2}
    \int d^{3}\vec{r} \; h_{HS, \; \rho}(r) \Delta_{\tau, \; 
    \kappa}^{2}(r) \nonumber \\
    &+& \frac{\beta^{3}}{12} \left[ \rho_{\alpha}q_{\alpha}^{3} \right]^{2}
 \int d^{3}\vec{r} \; \Delta_{\tau, \; 
    \kappa }^{3}(r)    \; ,
\end{eqnarray} 
where a subscript $\rho$ has been added to the function $h_{HS, \; 
\rho}(r) $ to emphasize that it has to be computed at the density 
$\rho=\sum_{\alpha}\rho_{\alpha}$. 
There are good theories for the HS pair correlations $h_{HS}(r) $, 
for instance one could try $h_{HS}(r)= h_{PY}(r) $, which makes the 
two-loop expression for the free energy manageable.
Similarly the smeared propagator in the RHS of 
eq.\ (\ref{jjjjo}) must be computed from\ 
(\ref{Deltatauk})  with $\kappa^{2}=4 \pi \beta \rho_{\alpha}q_{\alpha}^{2}$.

Several comment are in order at this point.

(a) We note that only the irreducible diagrams D3 and D5 of fig.\ 3 have
survived to the Legendre transform; note that, moreover, the symmetry
factor of D3 has changed as the result of many compensations of equal terms.
Diagram D3 is interesting because it can be interpreted as the contribution
of an effective attractive interaction $-\Delta_{\tau, \; \kappa}^{2}(r)$
to the free energy. Such a term should play an important role in the
description of the liquid-vapor transition of the SPM.

(b) Digging the literature we have found that the expression\ 
(\ref{jjjjo}) when specialized to the case of the RPM (i.e. with the last term
of the right hand side set to zero)
is not new and
has been obtained more than thirty years ago 
by Chandler 
and Andersen\cite{Chand-Ander} 
in the framework of the mode expansion formalism.
According to the numerical study performed by these authors 
eq.\ (\ref{jjjjo}) gives reasonable results in the liquid regime.

(c) Recent developments of the theory of collective variables (in its 
modern formulation\cite{NewYuko1,NewYuko2}) yield, 
in the case of the RPM, to
an expression for $\beta f$  that disagrees with eq.\ (\ref{jjjjo}). 
As far as I understand the details of ref. \cite{Oksanna} one can 
reconciliate the two points of view by replacing $h_{HS, \; \rho}(r)$ by 
its Fourier transform at $k=0$ in eq.\ (\ref{jjjjo}). 
This approximation is correct if the 
correlation length associated to charge fluctuations (i.e. 
$\xi_{C}\sim 1/\kappa$) is large compared to the correlation length 
associated with the HS fluid which does not seem to be fully justified in 
general.

(d) We have checked in the case of the RPM that,
 expanding the expression\ (\ref{jjjjo}) of $\beta 
f $ in powers of $\beta$ with the help of eq.\ (\ref{Xtau})
gives back the high-T expansion discussed in 
refs \cite{Raim-Cai}, i.e. in reduced units ($x=r/\sigma$, 
$\beta^{*}=\beta q^2/\sigma$, and $\rho^{*}=\rho \sigma^{3}$)
\begin{eqnarray}
    \beta^{*} f_{RPM}(\rho^{*}) 
    &= &\beta^{*} f_{HS}(\rho^{*}) -\rho^{*} \log 2 -\frac{2 
    \pi^{1/2}}{3} (\rho^{*} \beta^{*})^{3/2} -
    \frac{(\rho^{*} \beta^{*})^{2}}{4}\int d^{3}\vec{x} \; 
    \frac{h_{HS, \; \rho^{*}}(x)}{x^{2}} \nonumber \\
    &+& \pi^{1/2} (\rho^{*} \beta^{*})^{5/2} \int d^{3}\vec{x} \; 
    \frac{h_{HS, \; \rho^{*}}(x)}{x} 
    +{\cal O}( \beta^{*\; 3})
    \; .
\end{eqnarray}    
This result corroborates the conclusions of Stell in his study of the relation
between the $\gamma$-ordering and the mode expansion.\cite{StellT}
\section{Conclusion}
The KSSHE field theoretical representation of liquids provides a general 
framework for studying either neutral atomic liquids\cite{Caillol}
or ionic solutions, as in this series of papers. It is taylor made 
for building perturbation theories
with respect to a reference fluid chosen conveniently in general
as the HS fluid. The technics is roughly always the same, i.e. an ordering of
the cumulant expansion of the grand potential in ascending powers of some small
parameter. The latter can be either a physical parameter such as the fugacity
or the inverse temperature as considered in papers I and II or, as in the
present work, an abstract one related to the numbers of loops of the Feynman
diagrams retained in the expansion.  
The salient features of this loop expansion  can be summarized as follows.
\begin{itemize}
\item
The zero-loop order approximation of the free energy  $\beta {\cal A}_{MF}$
constitutes a rigorous lower bound for the exact GC free energy
$\beta {\cal A}$. An optimized MF theory can be obtained by
maximizing  $\beta {\cal A}_{MF}$ with respect to the smearing distribution
functions in the cores. The MF grand potential $\log \Xi_{MF}$ constitutes a rigorous 
upper bound for the exact grand potential. Both functionals 
$\beta {\cal A}_{MF}[\{\rho_{\alpha}\}]$ and $\log \Xi_{MF}[\{\nu_{\alpha}\}]$ are strictly 
convex in the fluid phase which rules out a fluid-fluid phase transition
at the MF level.
\item
At the MF level (or in the Gaussian
approximation) the pair correlation functions coincide with those of the RPA
theory. The direct correlation functions of 
the PM in the optimized MF theory are very similar to those considered
in the ORPA (or MSA) approximation of the theory of liquids, 
however the pair correlation
functions $g_{MF,\; \alpha\beta}(r)$, which have  simple analytical
expressions, do not vanish in the cores, except in limit cases.
\item
The one-loop free energy is identical with that obtained in the RPA theory of
liquids.\cite{Hansen,Chand-RPA,Outhwaite}
\item
An explicit and manageable expression of the two-loop order free energy can be written for the
SPM; in the case of the RPM
it coincides with a result obtained by  Chandler and Andersen in the
framework of the mode expansion theory.\cite{Chand-Ander} However the expression
\ (\ref{P2loop})
of the pressure  seems to be a new result even for the RPM.
\end{itemize}

 The homogeneous specific free energy $\beta f$ of the PM can be used to
study the critical point (CP) of the PM at the MF level. 
Recently, the one-loop order expression of $\beta f$ in the WCA scheme was
considered to study the CP of the RPM.\cite{Ciach-Stell,Oksanna} We have
performed a similar study
for the SPM with the one- and two-loop order expressions of $\beta f$ 
derived in this paper. These results will be discussed elsewhere. 

\acknowledgements
I acknowledge  B. Jancovici, E. Trizac and D. Levesque for 
discussions and comments. 
\newpage
\appendix
\section{Hierarchies}
\label{apA}
In this appendix we establish the hierarchies for the   various 
distribution functions 
introduced in sec.\ (\ref{def-corre}). 
\subsubsection{Hierarchy for the $G^{(n)\; T}_{C}$}
As a trivial consequence of the definition\ (\ref{theta}) of the 
generator $\Theta(1)$ and of eq.\ (\ref{defcorreCT}) we have
\begin{equation}
    \label{H1}
    \label{hie1}
    \Theta(n+1) \; G^{(n) \; T}_{C} (1, \ldots, n) =
    i \beta^{1/2} \; G^{(n+1) \; T}_{C} (1, \ldots, 
    n+1) \; , 
 \end{equation}  
 which is valid for all $n \geq 0$ with the convention that $G^{(n=0) \;
 T}_{C} \equiv \log \Xi$. 
\subsubsection{Hierarchy for the $G^{(n)}_{C}$}
Let us apply the linear operator $\Theta(1)$ to eq.\ 
(\ref{defcorreC3}). This gives
\begin{equation}
    \label{H2} 
    \label{hie2}
  \Theta(n+1) \; G^{(n) }_{C} (1, \ldots, n) =    i \beta^{1/2}
 \; \left[ G^{(n+1) }_{C} (1, \ldots,n+1)-
 \rho_{C}(n+1) \; G^{(n) }_{C} (1, \ldots, n) \; ,
 \right]
 \end{equation}  
where $ \rho_{C}\equiv  G^{(n=1) }_{C}$ is the equilibrium charge 
density. The relations\ (\ref{H2}) are valid for all $n \geq 0$ with 
the convention that $G^{(n=0)}_{C} \equiv 1 $.
\subsubsection{Hierarchy for the $G^{(n)}_{\varphi}$}
Let us first rewrite the definition\ (\ref{gphi}) of $G^{(n) }_{\varphi}$ 
more explicitly, i.e.
\begin{eqnarray}
    \label{eq1}
 G^{(n) }_{\varphi}(1,\ldots,n)&=& \Xi^{-1}\; {\cal N}_{v_{c}}^{-1}
 \; \int {\cal D} \varphi \;  \varphi(1) \ldots \varphi(n) \; \times 
 \nonumber \\
 & \times &
 \Xi_{HS}\left[ \{\overline{\nu}_{\alpha} +i 
\phi_{\alpha}\}\right] \; \exp \left(-\frac{1}{2}  \left< 
\varphi \vert  v_{c}^{-1} \vert \varphi \right>\right) \; .
 \end{eqnarray}  
Then we apply $\Theta(n+1)$ to  both sides of eq.\ (\ref{eq1}), 
which yields
 \begin{eqnarray}
  \Theta(n+1) \; G^{(n) }_{\varphi}(1,\ldots,n) &=& 
  -G^{(n) }_{\varphi}(1,\ldots,n) \; \Xi^{-1} 
  \Theta(n+1) 
  \Xi\left[ \{\nu_{\alpha}\}\right] \;  
  \nonumber \\
  &+& \Xi^{-1} {\cal N}_{v_{c}}^{-1} \; 
  \int {\cal D} \varphi \;  \varphi(1) \ldots \varphi(n) \; \times \nonumber \\
  &\times &
  \exp \left(-\frac{1}{2}  \left< 
\varphi \vert  v_{c}^{-1} \vert \varphi \right>\right) 
\Theta(n+1) \; \Xi_{HS}\left[ \{\overline{\nu}_{\alpha} +i 
\phi_{\alpha}\}\right]  \; .
 \end{eqnarray} 
Taking advantage of eqs.\ (\ref{defcorreC3}) (for $n=1$),\
(\ref{neq1}),  and of the relation\ (\ref{rela}) 
one concludes that
\begin{eqnarray}
  \Theta(n+1) \; G^{(n) }_{\varphi}(1,\ldots,n) &=& G^{(n) }_{\varphi}(1,\ldots,n)
  \frac{1}{4 \pi } \Delta_{n+1} <\varphi (n+1) >_{\cal{H}} \; + 
  \nonumber \\
  &+&  \Xi^{-1}\; {\cal N}_{v_{c}}^{-1}
 \; \int {\cal D} \varphi \;  \varphi(1) \ldots \varphi(n) \;
 \exp \left(-\frac{1}{2}  \left< 
\varphi \vert  v_{c}^{-1} \varphi \right>\right)
\frac{ \delta \Xi_{HS}}{\delta \varphi(n+1)}  \; .
 \end{eqnarray} 
 Now we perform a functional integration by parts which gives us
\begin{eqnarray}
    \label{H3}
    \label{hie3}
 \Theta(n+1) \; G^{(n) }_{\varphi}(1,\ldots,n) &=&
  \frac{-\Delta_{n+1}}{4 \pi }  \left[
  G^{(n+1) }_{\varphi}(1,\ldots,n+1) -G^{(n) }_{\varphi}(1,\ldots,n) 
   <\varphi (n+1) >_{\cal{H}}\right]  \; \nonumber \\
   &-& \sum_{j=1}^{n} \; \delta(n+1,j) 
   G^{(n-1) }_{\varphi}(1,\ldots,j-1,j+1,\ldots,n) 
   \; ,
\end{eqnarray}  
which is valid for all $n\geq 1$
with the convention $G^{(0)}_{\varphi} \equiv 1$.
\subsubsection{Hierarchy for the $G^{(n) \; T}_{\varphi}$}
Let us apply eq.\ (\ref{H3}) in the case $n=1$; we get
\begin{eqnarray}
    \label{t}
 \Theta(2) \; G^{(1) }_{\varphi}(1) &=&    
 \frac{\Delta_{2}}{4 \pi } \left[ <\varphi(1)>_{\cal{H}}<\varphi(2)>_{\cal{H}}
 -G^{(2) }_{\varphi}(1,2) \right] -\delta(1,2) \; , \nonumber \\
 &=& \frac{-\Delta_{2}}{4 \pi } G^{(2) \; T}_{\varphi}(1,2)
-\delta(1,2) \; .
\end{eqnarray}   
A similar brut force calculation yields
\begin{equation}
    \label{tt}
 \Theta(3) \; G^{(2) \; T}_{\varphi}(1,2) =
 \frac{-\Delta_{3}}{4 \pi } G^{(3) \; T}_{\varphi}(1,2,3) \; .
\end{equation}    
We want to prove that eq.\ (\ref{tt}) is valid for all $n \geq 3$. The 
proof will be  by induction. 
Suppose that, for all $2 \leq m\leq n$ with $n\geq 3$ we have either\ 
(\ref{t}) or\ (\ref{tt}) if $m=2$ or $m=3$ respectively or, in other 
cases ($m\neq 2,3$)
\begin{equation}
\label{hie4}
    \label{ttt}
   \Theta(m) \; G^{(m-1) \; T}_{\varphi}(1,\ldots,m-1) =  
 \frac{-\Delta_{m}}{4 \pi } G^{(m) \; T}_{\varphi}(1,\ldots,m) \; .   
\end{equation} 
We want to prove\ (\ref{ttt}) for $m=n+1$. From the definition
\begin{equation}
    \label{tt1}
    G_{\varphi}^{(n) \; T}=
    G_{\varphi}^{(n) } - \sum_{m<n} \prod  G_{\varphi}^{(m) 
    \; T} \; ,
\end{equation}    
where the sum runs over all the partitions of the set $\{1, \ldots,n \}$.
Now apply $\Theta(n+1)$ to both sides. One obtains
\begin{equation}
 \Theta(n+1)   G_{\varphi}^{(n) \; T}(1,\ldots,n) = I - II \; ,
\end{equation} 
where $I \equiv  \Theta(n+1)  \;  G_{\varphi}^{(n) }(1,\ldots,n) $ is 
given by eq.\ (\ref{H3}) and II results from the application of  
$\Theta(n+1)$ upon the sum of products of the RHS of eq.\ (\ref{tt1}).
Since only functions $G_{\varphi}^{(m) \; T}(1,\ldots,m)$ of order 
$m<n$ are involved in these products, we can apply eq.\ (\ref{ttt}). Therefore
 $II$ is made of all the terms of

\begin{equation}
    -\frac{\Delta_{n+1}}{4 \pi} \sum_{m<n+1} \prod  G_{\varphi}^{(m) 
    \; T} =  -\frac{\Delta_{n+1}}{4 \pi} \left[ 
    G_{\varphi}^{(n+1) }(1,\ldots,n+1) -G_{\varphi}^{(n+1) \; T}(1,\ldots,n+1)
    \right]
\end{equation}   
except : 
\begin{itemize}
    \item{(a)} the term 
    \begin{equation}
 -\frac{\Delta_{n+1}}{4 \pi} <\varphi(n+1)>_{{\cal H}}
 G_{\varphi}^{(n) \; T}(1,\ldots,n)     
    \end{equation}      
    \item{(b)} the terms involving $<\varphi(n+1)>_{{\cal H}}$, the sum 
    of which is given by
    \begin{eqnarray}
&-\frac{\Delta_{n+1}}{4 \pi}\left[
<\varphi(n+1)>_{{\cal H}}\sum_{m<n} \prod  G_{\varphi}^{(m) \; T}
\right] & \nonumber \\
&=& \nonumber \\
&
-\frac{\Delta_{n+1}}{4 \pi}\left[   G_{\varphi}^{(n) }(1,\ldots,n)
- G_{\varphi}^{(n) \; T}(1,\ldots,n)
\right] 
&
    \end{eqnarray} 
\end{itemize}    
Moreover, additional terms are generated in $II$  in the 
case where $\Theta(n+1)$ acts on some $<\varphi(j)>$ ($1\leq j \leq 
n$), this correspond to the special case\ (\ref{t}).
The total contribution of these terms is given by
    \begin{eqnarray} 
&       -\sum_{j=1}^{n} \delta(n+1,j) \left[
\sum_{m<n \; \text{, no $j$}} \prod  G_{\varphi}^{(m) \; T}
+ G_{\varphi}^{(n-1) \; T}(1, \ldots,j-1,j+1,\ldots,n) 
        \right] & \nonumber \\
        &=& \nonumber \\
        &
-\sum_{j=1}^{n} \delta(n+1,j) \; G_{\varphi}^{(n-1) }
(1, \ldots,j-1,j+1,\ldots,n) \; .&
\end{eqnarray}     
Gathering the intermediate results, one finds for $II$
\begin{eqnarray} 
    II &=& -\frac{\Delta_{n+1}}{4 \pi}
    \big[ G_{\varphi}^{(n+1) }(1,\ldots,n+1) - G_{\varphi}^{(n+1) \; 
    T}(1,\ldots,n+1) \nonumber \\
   & - &  <\varphi(n+1)>_{{\cal H}}G_{\varphi}^{(n)}(1,\ldots,n) 
   \big] \nonumber \\
    &-& \sum_{j=1}^{n} \delta(n+1,j) \; G_{\varphi}^{(n-1) }
(1, \ldots,j-1,j+1,\ldots,n) \; . 
 \end{eqnarray}   
  Finally we conclude that
 \begin{equation}
   \Theta(n+1) \; G^{(n) \; T}_{\varphi}(1,\ldots,n) =  I -II=
 \frac{-\Delta_{n+1}}{4 \pi } G^{(n+1) \; T}_{\varphi}(1,\ldots,n+1) \; ,   
\end{equation} 
We have just checked that if eq.\ (\ref{ttt}) is valid for $m \leq n$ 
it is also valid for $m=n+1$, therefore it is valid for all $n$.
\section{A derivation of the first SL rule}
\label{apB}
We derive here the first SL rule\ (\ref{sla}) in the KSSHE formalism. Let 
us define the two-points function
\begin{equation}
    \label{AA}
    A(1,2)= \Xi^{-1} {\cal N}_{v_{c}}^{-1} 
    \int {\cal D}  \varphi \; 
\exp \left(-\frac{1}{2} \left<  \varphi \vert  v_{c}^{-1}
\vert  \varphi \right>\right)
\frac{\delta}{\delta \varphi(1)} \frac{\delta}{\delta \nu_{\alpha}(2)}
 \Xi_{HS}\left[ \{\overline{\nu}_{\alpha} +i \phi_{\alpha}\}\right] \; .
\end{equation}  
First, we integrate by parts; it gives 
\begin{eqnarray} 
    \label{AA1}
    A(1,2)&=& - \Xi^{-1} {\cal N}_{v_{c}}^{-1} 
     \int {\cal D}  \varphi \; 
 \frac{\delta}{\delta \varphi(1)} \left[ 
 \exp \left(-\frac{1}{2} \left<  \varphi \vert  v_{c}^{-1}
\vert  \varphi \right>\right)  \right]\Xi_{HS} \; \rho_{HS,\; \alpha}(2) 
\nonumber \\
&=& -\frac{\Delta_{1}}{4 \pi} \left< \varphi(1) \; \rho_{HS,\; \alpha}(2) 
\right>_{{\cal H}} \; .
 \end{eqnarray}  
 Another way to evaluate the function $A(1,2)$ is to make use of 
 the identity\ (\ref{rela}) in eq.\ (\ref{AA}). This gives 
\begin{eqnarray}
    \label{AA2}
    A(1,2) &=& \Xi^{-1} {\cal N}_{v_{c}}^{-1} 
    \int {\cal D}  \varphi \;  
 \exp \left(-\frac{1}{2} \left<  \varphi \vert  v_{c}^{-1}
\vert  \varphi \right>\right)
\Theta(1) \; \frac{\delta}{\delta \nu_{\alpha}(2)} \Xi_{HS}
\left[ \{\overline{\nu}_{\alpha} +i \phi_{\alpha}\}\right] \nonumber \\
&=& i \beta^{1/2} \Xi^{-1} {\cal N}_{v_{c}}^{-1} 
 \int {\cal D}  \varphi \;  
 \exp \left(-\frac{1}{2} \left<  \varphi \vert  v_{c}^{-1}
\vert  \varphi \right>\right)
\Xi_{HS} \; q_{\beta} \tau_{\beta}(1,1') G^{(2)}_{HS, \;  \beta \alpha }(1',2)
\nonumber \\
&=& i \beta^{1/2}\; q_{\beta} \; \tau_{\beta}(1,1')\;  G^{(2)}_{\beta \alpha 
}(1',2) \; ,
\end{eqnarray}     
where we have made use of eq.\ (\ref{klug}). Comparing eqs.\ 
(\ref{AA1}) and\ (\ref{AA2}) we thus have the identity
\begin{equation}
    \label{AA3}
    -\frac{\Delta_{1}}{4 \pi} \left< \varphi(1) \; \rho_{HS, \; \alpha}(2) 
\right>_{{\cal H}}=
   i \beta^{1/2}\; q_{\beta}\;  \tau_{\beta}(1,1')\;  G^{(2)}_{\beta \alpha }(1',2)   \; .
\end{equation}    
Moreover, we already know that 
\begin{mathletters}
\begin{eqnarray}
   -\frac{\Delta_{1}}{4 \pi} \left< \varphi(1)\right>_{{\cal H}} &=&
    i \beta^{1/2}\; q_{\beta} \;  \tau_{\beta}(1,1') \; \rho_{\beta}(1')  
    \label{f1} \\
  \left<  \rho_{HS, \; \alpha}(2)  
  \right>_{{\cal H}}&=& \rho_{\alpha}(2) 
  \label{f2} \; .  
 \end{eqnarray}
 \end{mathletters}
 
 \noindent
 Taking the product of\ (\ref{f1}) and\ (\ref{f2}) and substracting 
 the  eq.\ (\ref{AA3})  yields
\begin{equation}
    \label{AA4}
    -\frac{\Delta_{1}}{4 \pi} \left< \varphi(1) \rho_{HS, \; \alpha}(2) 
\right>_{{\cal H}}^{T}=
   i \beta^{1/2}\; q_{\beta} \; \tau_{\beta}(1,1') \; G^{(2)\; 
   T}_{\beta
\alpha}(1',2)   \; .
\end{equation}  
We now specialize to the case of a homogeneous system and 
consider this equation in Fourier space in the limit $k\to 0$. We 
assume that the field correlation in the LHS of eq.\ (\ref{AA4}) is 
regular in this limit which gives
\begin{equation}
    q_{\beta} \; \widetilde{G}^{(2) \; T}_{\beta \alpha}(k=0)=0 \; ,
\end{equation}
Which is equivalent, in terms of the pair distributions 
$\widetilde{h}_{\beta \alpha}$, to the first SL rule\ (\ref{sla}).

\section{The electroneutrality at the one-loop order.}
\label{apC}
In the general case, and at the one-loop order, the densities of 
species $\alpha$ reads as
\begin{eqnarray}
    \rho_{\alpha}(1)&=&\rho_{\alpha}^{(0)}(1) + \lambda 
    \rho_{\alpha}^{(1)}(1)+
    {\cal O}(\lambda^2) \; \nonumber \\
     \rho_{\alpha}^{(1)}(1)&= &\frac{\delta \log {\cal N}_{\Delta_
     {\overline{\varphi}}}}
     {\delta \nu_{\alpha}(1)} \; .
\end{eqnarray}   
We have shown (cf eq.\ (\ref{neutraMF})) that the MF densities 
$\rho_{\alpha}^{(0)}\equiv  \rho_{MF, \; \alpha}$ of the homogeneous 
system satisfy to the electroneutrality condition. In this appendix we 
prove that it remains true at the one-loop order.

We have 
\begin{eqnarray}
  \rho_{\alpha}^{(1)}(1)&=&  {\cal N}_{\Delta_
     {\overline{\varphi}}}^{-1}  \int {\cal D}\varphi \; 
     \frac{\delta}{\delta \nu_{\alpha}(1)} 
    \exp \left(  -\frac{1}{2}
  \langle \varphi \vert  \Delta_ {\overline{\varphi}}^{-1} 
  \vert \varphi 
  \rangle  \right) \nonumber \\
  &=&-\frac{1}{2} \Delta_ {\overline{\varphi}}(2,3) 
  \frac{\delta}{\delta \nu_{\alpha}(1)} 
  \Delta_ {\overline{\varphi}}^{-1}(2,3) \; ,
\end{eqnarray}    
where we have applied Wick's theorem $\langle \varphi(1) \varphi(2)
\rangle_{\Delta_ {\overline{\varphi}}} = \Delta_ 
{\overline{\varphi}}(1,2)$.
It follows from the definition\ (\ref{Delta}) of the propagator
$\Delta_ {\overline{\varphi}}$ that
\begin{equation} 
    \label{ff0}
 \rho_{\alpha}^{(1)}(1)= -\frac{\beta}{2} \Delta_ {\overline{\varphi}}(2,3) 
  \frac{\delta}{\delta \nu_{\alpha}(1)} G_{HS, \; C}^{(2)\; T}
  \left[ \left\{ \overline{\nu}_{\gamma} + i \overline{\phi}_{\gamma} \right\}\right]
  (2,3) \; .
\end{equation}    
The functional derivative of  $G_{HS, \; C}^{(2)\; T}$ can be 
recast under the form
\begin{eqnarray}
    \label{ff}
 \frac{\delta}{\delta \nu_{\alpha}(1)} G_{HS, \; C}^{(2)\; T}(2,3)&=&
 G_{HS, \; \alpha C}^{(3)\; T}(1,2,3) \nonumber \\
 &-&\beta  G_{HS, \; C}^{(3)\; T}(2,3,4) v_{c}(4,5)
 q_{\delta}\tau_{\delta}(5,6)
 G_{MF, \; \delta 
 \alpha}^{(2)\; T}(6,1) \; ,
\end{eqnarray} 
where we have defined the hybrid density-charge correlation function
\begin{equation}
     G_{HS, \; \alpha C}^{(3)\; T}(1,2,3)\equiv 
     q_{\beta}q_{\gamma}\tau_{\beta}(2,2')\tau_{\gamma}(3,3')
       G_{HS, \; \alpha \beta \gamma }^{(3)\; T}(1,2',3') \; .
\end{equation}   
Combining eqs.\ (\ref{ff0}) and\ (\ref{ff}) one finds that
\begin{eqnarray}
    \label{ff1}
  \rho_{\alpha}^{(1)}(1)&= & -\frac{\beta}{2} \Delta_ 
  {\overline{\varphi}}(2,3) \big\{G_{HS, \; \alpha C}^{(3)\; T}(1,2,3)
  \nonumber \\
  &- &\beta G_{HS, \; C}^{(3) \; T}(2,3,4) v_{c}(4,5)
  q_{\delta}\tau_{\delta}(5,6) G_{MF, \; \delta \alpha}(6,1)  
  \big\} \; .
\end{eqnarray}    
The one-loop correction to the charge density 
$ \rho_{C}^{(1)}(1)=q_{\alpha} \tau_{\alpha}(1,1') \rho_{\alpha}^{(1)}(1')$
is therefore given by
\begin{equation}
  \rho_{C}^{(1)}(1)  = -\frac{\beta}{2} 
  \Delta_ {\overline{\varphi}}(2,3) \left\{
  G_{HS, \; C}^{(3) \; T}(2,3,1) -\beta  G_{HS, \; C}^{(3) \; T}(2,3,4)
  v_{c}(4,5) G_{MF, \; C}^{(2) \; T}(5,1)
  \right\} \; .
\end{equation}    
Now it should be clear to the careful reader that
\begin{equation}
    v_{c}(4,5)G_{MF, \; C}^{(2) \; T}(5,1)=\langle \widehat{V}(4) \widehat{\rho}_{C}(1) 
    \rangle_{GC, \; MF}^{T} \; ,
\end{equation}    
where $\widehat{V}$ and $\widehat{\rho}_{C}$
are respectively  the microscopic electric potential 
and charge density in a given 
configuration. With this remark the one-loop charge density reads as
\begin{equation}
    \label{resuroC}
 \rho_{C}^{(1)}(1) = -\frac{\beta}{2} 
  \Delta_ {\overline{\varphi}}(2,3) \left\{
    G_{HS, \; C}^{(3) \; T}(2,3,1) -\beta  G_{HS, \; C}^{(3) \; T}(2,3,4)
  \langle \widehat{V}(4) \widehat{\rho}_{C}(1) 
    \rangle_{GC, \; MF}^{T}   
  \right\} \; . 
\end{equation} 
In sec\ (\ref{MF}) we made the remark that the SL rules were satisfied 
in the MF approximation, in particular the Carnie-Chan rule\ 
(\ref{carni}) is satisfied and one has
\begin{equation}
    \label{carnibis}
\beta  \int d(1) \;  \langle \widehat{V}(4) \widehat{\rho}_{C}(1) 
    \rangle_{GC, \; MF}^{T}    =1.  
\end{equation}     
Integrating eq.\ (\ref{resuroC}) over the volume of the  system and making 
use of the Carnie-Chan rule\ 
(\ref{carnibis}) gives us
\begin{equation}
    \int_{\Omega}d(1) \;  \rho_{C}^{(1)}(1) =0
\end{equation}    
leading, for a homogeneous system, to the local charge neutrality 
condition
$\rho_{C}^{(1)}=0$, as expected. 
\section{The functions $G_{HS, \; C}^{(n) \; T}$ for the SPM}
\label{apD}
The aim of this appendix is to determine the form of the functions 
$G^{(n)}_{HS, \; C} (1, \ldots, n) $ (cf. eq.\ (\ref{bidule})) for a 
homogeneous SPM. In this case all the smearing functions 
$\tau_{\alpha}$ are equal to the same $\tau$ and we have
\begin{equation}
    \label{C0}
G^{(n)}_{HS, \; C} (1, \ldots, n) =
q_{\alpha_{1}} \ldots q_{\alpha_{n }} \; 
\tau(1,1') \ldots \tau (n,n')  \;
G^{(n)}_{HS, \; \alpha_{1} \ldots \alpha_{n} } (1', \ldots, n') \; .
 \end{equation} 
The functions  $G^{(n)}_{HS, \; C} (1, \ldots, n) $
are to be computed at the saddle point where we have 
shown that, for a homogeneous system, the electroneutrality condition 
\begin{equation}
    \label{Celectro}
    \rho_{HS, \; \alpha} \; q_{\alpha}=0 
 \end{equation}   
holds.

The HS reference system of the SPM is thus a mixture of $M$ species of 
hard spheres with all the same diameter $\sigma$ . Two hard spheres of 
different species differ solely by their charge. However these charges do not 
contribute to the configurational energy and  can thus be seen as  mere
internal degrees of freedom  which allow to distinguish the various 
species. 
Denoting by  $z_{\alpha}=\exp(\nu_{\alpha})$ the activity of the 
species $\alpha$ we have therefore
\begin{equation}
    \label{C1}
    \Xi_{HS}\left[ \left\{ z_{\alpha} \right \} \right] =
    \Xi_{HS, \ 0} \left[  z= \sum_{\alpha}z_{\alpha}  
    \right]  \; ,
\end{equation}
where $\Xi_{HS, \; 0} $ is the GC partition function of the usual HS 
fluid. Note that  eq.\ (\ref{C1}) is valid for both homogeneous and 
inhomogeneous systems. Of course, the density correlation functions
of the mixture are also simply related with those of the pure fluid.

Let us first consider the following type of n-body correlation 
functions
\begin{equation}
    \rho_{HS, \; \alpha_{1} \ldots \alpha_{n}}^{(n) \; T
    }\left[ \left\{ z_{\alpha} \right \} \right] 
    (1,\ldots,n)=    \prod_{i=1}^{n}  z_{\alpha_{i}}(i) \; 
    \frac{ \delta^{n} \log \Xi_{HS}\left[ \left\{ z_{\alpha} \right \} \right]}
    {\delta  z_{\alpha_{1}}(1) \ldots  z_{\alpha_{n}}(n)} \; .
\end{equation}    
It follows from eq.\ (\ref{C1}) that
\begin{equation}
  \rho_{HS, \; \alpha_{1} \ldots \alpha_{n}}^{(n) \; T }
     \left[ \left\{ z_{\alpha} \right \} \right] 
    (1,\ldots,n)=    \prod_{i=1}^{n} \left[\frac{\rho_{HS, \; \alpha_{i}}(i)}
    {\rho_{HS, \; 0}(i)}\right] \; 
     \rho_{HS, \;  0}^{(n) \; T}[z] (1,\ldots,n) \; ,
\end{equation}  
with self explanatory notations. Defining as usual the functions 
$h^{(n)\; T}$ by the relations
\begin{equation}
    \label{hn0}
  \rho_{HS, \; \alpha_{1} \ldots \alpha_{n}}^{(n) \; T }
     \left[ \left\{ z_{\alpha} \right \} \right] 
    (1,\ldots,n)=  \rho_{HS, \; \alpha_{1}}(1) \ldots  
    \rho_{HS, \; \alpha_{n}}(n)  \;
  h_{HS, \; \alpha_{1} \ldots \alpha_{n}}^{(n) \; T }
     \left[ \left\{ z_{\alpha} \right \} \right] 
    (1,\ldots,n)  \; ,  
\end{equation} 
we thus have    simply
\begin{equation}
    \label{hn}
     h_{HS, \; \alpha_{1} \ldots \alpha_{n}}^{(n) \; T }
     \left[ \left\{ z_{\alpha} \right \} \right] 
    (1,\ldots,n)= h_{HS, \; 0}^{(n) \; T }
     \left[z \right] (1,\ldots,n) \; .
  \end{equation}   
 
  Unfortunately the relations between the
  $ G_{HS, \; \alpha_{1} \ldots \alpha_{n}}^{(n) \; T }$ (cf. eq.\ 
  (\ref{defcorre})) 
  of the mixture and those of the pure HS fluid are not as so simple 
  as eq.\ (\ref{hn}). They can be obtained  by first establishing the relation between 
  the $G^{(n) \; T}$ and the  $\rho^{(n) \; T}$ which can be done as 
  follows:
  we consider a simple fluid made of
  a single species of particles and write down the hierarchies for the $G^{(n) \; T}$
  and the  $\rho^{(n) \; T}$ 
  \begin{eqnarray}
      \label{C2}
      \frac{\delta G^{(n) \; T}(1,\ldots,n)}{\delta \nu(n+1)}&=&
      G^{(n+1) \; T}(1,\ldots , n, n+1) \; ,  \nonumber \\
  \frac{\delta \rho^{(n) \; T}(1,\ldots,n)}{\delta \nu(n+1)}     &=& 
  \rho^{(n+1) \; T}(1,\ldots, n ,  n+1)  + 
       \sum_{i=1}^{n}
       \delta(n+1,i) \; \rho^{(n) \; T}(1,\ldots,  n)  \; .
 \end{eqnarray}     
 Noting that $G^{(n=1) \; T} \equiv \rho ^{(n=1) \; T} \equiv \rho$ and applying 
 eqs.\ (\ref{C2}) for $n=1,2$, etc one finds 
 \begin{mathletters}
 \begin{eqnarray}
    G^{(2) \; T}(1,2)&=& \rho^{(2) \; T}(1,2) + \delta(1,2) \rho(1) 
    \; , \\
   G^{(3) \; T}(1,2,3)&=& \rho^{(3) \; T}(1,2,3)   +\left[ 
   \delta(3,1) +\delta(2,1) \right]  \rho^{(2) \; T}(1,2) \nonumber \\
   &+& \delta(1,2) \rho^{(2) \; T}(1,3) +  \delta(1,2) 
   \delta(1,3) \rho(1,3) \: , \\
   G^{(4) \; T}(1,2,3,4)&=& \ldots \; .
\end{eqnarray}  
 \end{mathletters}
 
 \noindent
(We have omitted to report the cumbersome expression of $G^{(4) \; T}$). 
The generalization to mixtures is straightforward 
 \begin{mathletters}
     \label{C3}
 \begin{eqnarray}
    G_{\alpha_{1}\alpha_{2} }^{(2) \; T}(1,2)&=&
   \rho_{\alpha_{1}\alpha_{2} }^{(2) \; T}(1,2)  +
   \delta(1,2)\delta_{\alpha_{1}\alpha_{2} } \rho_{\alpha_{1}}(1) 
    \; , \\  
  G_{\alpha_{1}\alpha_{2} \alpha_{3}}^{(3) \; T}(1,2,3)&=&  
   \rho_{\alpha_{1}\alpha_{2} \alpha_{3}}^{(3) \; T}(1,2,3) \nonumber 
   \\
   &+& \delta(1,2) \delta_{\alpha_{1}\alpha_{2} } 
   \rho_{\alpha_{2}\alpha_{3} }^{(2) \; T}(2,3) \nonumber \\
  &+& \delta(1,3) \delta_{\alpha_{1}\alpha_{3} } 
   \rho_{\alpha_{2}\alpha_{3} }^{(2) \; T}(3,2) \nonumber \\ 
  &+& \delta(2,3) \delta_{\alpha_{2}\alpha_{3} } 
   \rho_{\alpha_{3}\alpha_{1} }^{(2) \; T}(3,1) \nonumber \\  
   &+&  \delta(1,2) \delta_{\alpha_{1}\alpha_{2} } 
   \delta(1,3) \delta_{\alpha_{1}\alpha_{3} }
   \rho_{\alpha_{1}}(1) \; , \\
   G_{\alpha_{1}\alpha_{2} \alpha_{3} \alpha_{4}}^{(4) \; T}(1,2,3,4)&=&  
   \ldots \; .
 \end{eqnarray}
 \end{mathletters} 
 
 \noindent  
Returning to our mixture of hard spheres one applies the eqs.\ 
(\ref{C3}) by taking into account the relations\ (\ref{hn0}) and\ 
(\ref{hn}). For instance for $n=2$ this yields
\begin{equation}
    \label{aG2}
 G_{HS, \; \alpha_{1}\alpha_{2} }^{(2) \; T}(1,2)=
 \rho_{HS, \; \alpha_{1}} \rho_{HS, \; \alpha_{2}} \; h^{(2)\; T}_{HS,\; 0}(1,2)
 +   \rho_{HS,\;  \alpha_{1}}\delta(1,2) \delta_{\alpha_{1}\alpha_{2} } 
 \; .
 \end{equation}   
Reporting the expression\ (\ref{aG2}) of  $G_{HS, \; \alpha_{1}\alpha_{2} }^{(2) \; T}$
in the definition\ (\ref{C0}) of  $G_{HS, \; C}^{(2) \; T}$ one finds
\begin{equation}
    \label{GC2}
G_{HS, \; C}^{(2) \; T}(1,2)= \rho_{HS, \; \alpha} \; q_{\alpha}^{2} 
\delta(1,2)
\; ,
\end{equation}
where we have made use of the electroneutrality condition\ 
(\ref{Celectro}) and of the normalization condition 
$\widetilde{\tau}(0)=1$. The same job can be done for $G_{HS, \; C}^{(3) \; T}$
and $G_{HS, \; C}^{(4) \; T}$  leading,
after some tedious algebra, to
the relations\ (\ref{kernelSPM}) reported in the text.


\end{document}